\pdfoutput=1
\documentclass[aps,prd,superscriptaddress,11pt,showpacs,notitlepage,nofootinbib]{revtex4}

	\usepackage{graphicx}
	\usepackage{amsmath,amssymb,mathrsfs}
	\usepackage[colorlinks=true, a4paper=true, pdfstartview=FitV,
linkcolor=blue, citecolor=blue, urlcolor=blue]{hyperref}

\usepackage{subfigure}
\usepackage{epsf}
\usepackage{epsfig}
\usepackage[usenames,dvipsnames]{xcolor}
\usepackage{bbm}
\usepackage{color}
\usepackage{comment}
\usepackage{cleveref}

\usepackage{layouts}
\usepackage[titletoc]{appendix}

\newcommand{\be}{\begin{equation}}
\newcommand{\ee}{\end{equation}}
\newcommand{\bea}{\begin{eqnarray}}
\newcommand{\eea}{\end{eqnarray}}

\newcommand{\bb}{\bibitem}

\newcommand{\eqn}{\begin{eqnarray}}
\newcommand{\eqnx}{\end{eqnarray}}

\usepackage{physics}
\numberwithin{equation}{section}

\DeclareMathOperator{\sgn}{sgn}
\newcommand{\oldchi}{f}

\begin{document}

\title{Scattering of compact oscillons}
\author{F. M. Hahne}
\thanks{CNPq Scholarship holder -- Brazil}
\author{P. Klimas}
\email{pawel.klimas@ufsc.br}
\author{J. S. Streibel}
\affiliation{Departamento de F\'isica, Universidade Federal de Santa Catarina, Campus Trindade, 88040-900, Florian\'opolis-SC, Brazil}
\author{W. J.  Zakrzewski}
\affiliation{Department of Mathematical Sciences, Durham University, Durham DH1 3LE, U.K.}
\begin{abstract}

We study various aspects of the scattering of generalized compact oscillons in the signum-Gordon model in (1+1) dimensions. Using covariance of the model we construct traveling oscillons and study their interactions and the dependence of these interactions on the oscillons' initial velocities and their relative phases. The scattering processes transform the two incoming oscillons into two outgoing ones and lead to the generation of extra oscillons which appear in the form of jet-like cascades. Such cascades vanish for some values of free parameters and the scattering processes, even though our model is non-integrable, resemble typical scattering processes normally observed for integrable or quasi-integrable models.
 
Occasionally, in the intermediate stage of the process, we have seen the emission of shock waves and we have noticed  that, in general, outgoing oscillons have been more involved in their emission than the initial ones {\it i.e.} they have a border in the form of curved worldlines.

 The results of our studies of the scattering of oscillons suggest that the radiation of the signum-Gordon model has a fractal-like nature.
\end{abstract}

\maketitle
\section{Introduction}

Oscillons are localized, time dependent, quasi-periodic solutions observed in many  scalar field models. Their presence was first reported in  \cite{bogmukh}. In a vast majority of cases, in which physical models are non-integrable, oscillons radiate very slowly \cite{gleiser, gleiser1, gleiser2, shnir, salmi, fodor, roman, roman1, correa, fodor2, sakstein, correa2}. Oscillons can be created in some dynamical processes like, for instance, in kink-antikink collisions \cite{roman, adam, roman2, bazeia1, bazeia2, izquierdo}. On the other hand, oscillating structures seen in some integrable models, like ({\it e.g.} sine-Gordon model \cite{ablowitz,ablowitz2, LAFWJZ}, affine Toda models \cite{olive} and non-linear Schrodinger model \cite{tajiri, kedziora})  do not radiate at all and so they can live forever. Such infinitely long living objects are called  ``breathers'' which distinguishes them from standard oscillons. 

In this paper we describe our study of another interesting group of oscillons which share many properties of oscillons and breathers. A solution of such a type was first discovered in the signum-Gordon  model \cite{oscillon}. The signum-Gordon oscillon is an exact solution which, if not perturbed, behaves like a breather {\it i.e.} it can live forever without emitting any radiation. This is a very interesting and extremely rare behaviour for time dependent solutions in non-integrable field theories. Of course, due to the non-integrability of the model 
a perturbed signum-Gordon oscillon would emit some radiation. Such radiation often takes the form of emissions of smaller oscillon-like packages. So, this type of an oscillon can be thought of as being a stable (or perhaps metastable) time dependent non-topological solution of a non-integrable model. Moreover, some very special perturbations of such oscillons lead to more general, exact and infinitely long lived oscillons (generalized oscillons). Such oscillons were constructed in \cite{arodzswier, swierczynski}.

The signum-Gordon model \cite{signumGordon} is perhaps the simplest example of a wider class of scalar field-theoretic models with non-analytic potentials. A very important and characteristic property of such models is their possession of compactons \cite{topcomp, bas1, bas2, bas3, bas4, bas5} and scaling symmetry \cite{scalling}. This symmetry makes these models relevant in the description of dynamics of  fields in other models with approximate scaling symmetry in the limit of small amplitudes \cite{2bps, baby, kswz, kl}. In other words, the signum-Gordon model can be thought of as emerging, in this limit, from models containing non-analytic potentials. This shows that studies of solutions of this model are useful and can have relevance in the description of some aspects of solutions of other models with non-analytic potentials. Of course, due to an often encountered rich structure of minima of such more general models, the field configurations with larger amplitudes could also have some nontrivial topology (kinks, skyrmions etc)  \cite{shnir, topcomp, 2bps, adam2} and so their complete dynamics would be essentially different from the dynamics of the signum-Gordon compactons.

In what follows we present some basic notions about the signum-Gordon model. The model is defined by the Lagrangian density
\be
{\cal L}=\frac{1}{2}\partial_{\mu}\phi\partial^{\mu}\phi-|\phi|\label{sgmodel}
\ee
and its dynamics is described by solutions of the Euler-Lagrange equation
\be
\partial_{\mu}\partial^{\mu}\phi+{\rm sgn}(\phi)=0.\label{EL}
\ee
 The Euler-Lagrange equations contain a term ${\rm sgn}(\phi)=\frac{\partial}{\partial\phi}|\phi|=\pm1$ and so they do not include the vacuum solution $\phi=0$. In order to include explicitly the vacuum solution into the set of solutions of \eqref{EL} we require that ${\rm sgn}(0):=0$.
The model \eqref{sgmodel} has naturally appeared in the study of the behaviour of  scalar fields in the vicinity of minima of  V-shaped potentials {\it i.e.} potentials whose left and right side derivatives at minima are different from each other. Such models are perfectly well-defined from a physical point of view. Moreover, in some cases they can be seen as field-theoretic limits of certain mechanical models, which certainly admit experimental realizations. In fact, it was such  mechanical models that led to scalar field models with non-analytic potentials \cite{topcomp}.

The physical origin of models with non-analytic potentials is wider than the continuous limit of mechanical models. Quite recently it was reported in \cite{2bps} that models with V-shaped potentials may be obtained from other physical models when a parametrization associated with the symmetry reduction leads to new field variables that are restricted, {\it i.e} they cannot take arbitrarily large (or  small) values. In the case of models with a mechanical realisation such a restriction is {\it a priori} imposed on the system. The restrictions on values of fields lead to some inconvenience in the description of the dynamics of the system which, in such a case, is governed by both the Euler-Lagrange equations and the extra condition on the time derivative of the field. For instance, the mechanical model, from which the signum-Gordon model originates, has a continuous limit described by the field variable that satisfies $\tilde\phi\ge0$. Thus the model possesses the potential with an infinite barrier at $\tilde\phi=0$ {\it i.e.} $\tilde V(\tilde\phi)=\tilde\phi$ for $\tilde\phi\ge0$ and $\tilde V(\tilde\phi)=\infty$ for $\tilde\phi<0$.  The field must also satisfy the reflection condition $\partial_t\tilde\phi\rightarrow -\partial_t\tilde\phi$ at $\tilde\phi=0$. One can avoid such an inconvenient reflection condition by introducing an auxiliary model with a new field $\phi\in(-\infty,\infty)$ and the potential $V(\phi)=|\phi|$. \footnote{This potential generates the term ${\rm sgn}(\phi)=\frac{dV}{d\phi}$ in the field equations which justifies the name of the model.} This new model is so-called the {\it unfolded model}. The dynamics of this auxiliary field can be mapped onto the dynamics of the physical field  through the {\it folding transformation}, see \cite{signumGordon}. Thus the signum-Gordon model and other models  of this type can describe behaviour of physical systems with restrictions on the values of scalar fields.

The signum-Gordon model is certainly non-integrable. This conclusion can be drawn from the existence of radiation in numerical simulations of generic initial field configurations.  In fact, very little is known about the nature of this radiation. In this paper we describe results of our study of some aspects of the radiation in the signum-Gordon model. We pay particular attention to the exact time-dependent solutions of the model known as compact oscillons\footnote{In the literature the infinitely long lived oscillons are more often called breathers than oscillons. Here we keep the name oscillon following the original paper \cite{oscillon}.}  \cite{oscillon, arodzswier}, which are our principal candidates for constituents of the radiation. The signum-Gordon oscillons rely on  three principal properties of  models with $V$-shaped potentials: the existence of compact solutions, their scale invariance (exact or approximate) and the lack of linearization of small amplitude oscillations. The existence of compact solutions, like compact oscillons in particular, follows from the fact that models with standard kinetic and gradient terms in the Lagrangian approach vacuum in a quadratic manner if the potential has a  $V$-shaped form close to its minimum \cite{topcomp}. The scale invariance \cite{scalling} is a straightforward  consequence  of the form of the field equations. In the case of the signum-Gordon model the scale invariance is exact because ${\rm sgn}(\phi)$ is a scale invariant term. A very important consequence of this fact is the existence the self-similar solutions and oscillons of  all scales of energy and length. Thus the perturbed oscillons may lose energy by the emission of smaller (perturbed) oscillons.  We will demonstrate in this paper  that this is exactly what happens and so that the oscillons are main ingredients of the radiation of the signum-Gordon model. Finally, the absence of linear small amplitude oscillations follows from the fact that ${\rm sgn}(\phi)$ term cannot be linearized at $\phi=0$. This implies that, independently of their size, the signum-Gordon oscillons are fundamentally non-linear field configurations.

The existence of many exact solutions of the signum-Gordon model in (1+1) dimensions follows from the fact that equation \eqref{EL} reduces to a non-homogeneous wave equation on  segments of the $x$ axis, where the sign of the scalar field is fixed, and so, on each segment, it has the general solution
of the form:
\be
\phi(t,x)=F(x+t)+G(x-t)+\frac{1}{4}(x^2-t^2),\label{partsol}
\ee
where $F$ and $G$ are arbitrary functions.  The main point here is that this reduction is local {\it i.e.} the size and the localization of the segments of constant sign varies with time. This is a direct manifestation of a non-linear character of the model. Solutions like \eqref{partsol} are called {\it partial solutions}. The exact global solution of the model is given by the explicitly known set of properly patched partial solutions which must hold for arbitrary times. Determining  such a closed set of partial solutions usually generates some small technical difficulties. They correspond to the unpleasant side of finding solutions of models with $V$-shaped potentails.

The main aim of this paper is to describe the results of our study of interactions between  two oscillons.  Our motivation for such a study follows from the fact that the previous numerical simulations of models with $V$-shaped potentials \cite{kswz} have indeed observed collisions of oscillon-like structures in emerging radiation.  The study of systems containing colliding oscillons may throw new light on the nature of the radiation of the signum-Gordon model. In particular, we are interested in the scattering processes involving only two oscillons. Unfortunately, even such a `simple' scattering process is too complicated for a purely analytical investigation. For this reason we have used the numerical analysis as our principal tool and have restricted our attention to initial field configurations which possess certain symmetries. We hope that our results will find applications not only for models with a single minimum potential but also for models with multi $V$-shaped minima that can support the existence of compact kinks etc.

The paper is organized as follows. In Section II we  present some facts about generalized oscillons with vanishing total linear momentum. Then, making use of the Lorentz covariance of the model, we construct traveling oscillons. In Section III  we study scattering of two oscillons. Due to the compactness of the oscillons we construct the initial configurations 
of such oscillons by considering simple superpositions of non-overlapping oscillons. Further on in this section we also discuss shock waves and oscillons with non uniformly moving borders. The last section presents a short summary of our results and some comments.

\section{Generalized oscillons}



The {\it generalized oscillons} are exact compact solutions  of the signum-Gordon model which are distinguished by the fact that their borders move periodically from left to right and back again. The first particular solutions of this class, characterized by constant velocity of the motion, were  reported in \cite{arodzswier} and, due to this nature of the   motion, were called {\it swaying oscillons}.  A further generalization of the swaying oscillons to arbitrary periodic motions of the borders was discussed in \cite{swierczynski}. In order to simplify our initial discussions we first consider the field configurations describing oscilllons 
swaying with a constant velocity $v$. The more general oscillons are discussed later in Section \ref{nonuniform} where we also comment on the results of scattering of compact oscillons. 

Before we go further let us establish the terminology for different types of oscillons considered in this paper. The oscillons which are {\it a priori} exact solutions (like initial field configurations) will be called the {\it exact oscillons}  if their borders do not move periodically or, if they do move, the {\it generalized exact oscillons}.  The oscillons produced in the process of the scattering are numerical solutions  of the signum-Gordon equation, hence, they are not {\it a priori} exact.  They will be called {\it quasi-oscillons}  in the case of their strong similarity to the exact oscillons and {\it perturbed oscillons}  if such a similarity is only approximate {\it i.e.} when they are approximately periodic, emit radiation etc.

\subsection{Oscillons at rest}
In our work here we are particularly interested in scattering of generalised exact oscillons.  Such oscillons move uniformly (modulo a periodic motion given by $v$) in a certain reference frame $S$. We shall refer to this frame as to {\it the laboratory reference frame}.  Moreover, we preserve symbols $t$ and $x$ for coordinates exclusively in this reference frame. On the other hand, the reference frame of the oscillon is denoted by $S'$ and called the {\it rest frame of the oscillon}. The precise meaning of expression ``rest frame'' in the case of generalised exact oscillons is the following one: it is the inertial reference frame in which  the total linear momentum of the oscillon vanishes. The space time coordinates in $S'$ are denoted by $t'$ and $x'$.

\begin{figure}[h!]
\centering
\subfigure{\includegraphics[width=0.65\textwidth,height=0.8\textwidth, angle =0]{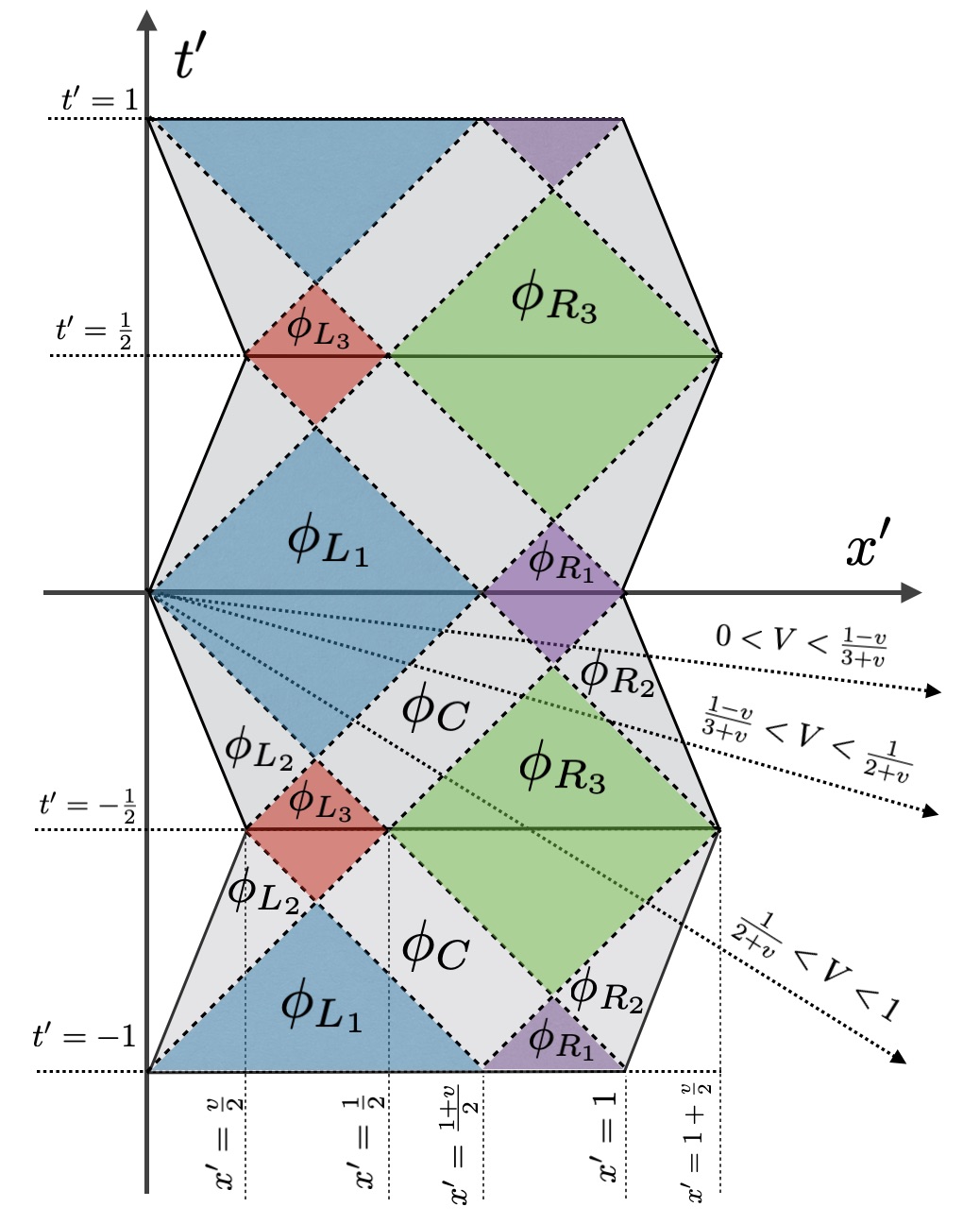}}
\caption{The world sheet of the generalized exact oscillon seen in its own rest frame $S'$. Only in the interval $t'\in[0,\frac{1}{2}]$  $\phi_{k}=\varphi_{k}$ holds; other partial solutions are given by \eqref{partialsolutions}. The axes $x$ of the laboratory reference frame seen in frame $S'$  have an inclination with respect to axis $x'$. Here we present three different cases of inclination corresponding to  the velocity $u'=-V$of the laboratory frame  $S$ that moves to the left on the diagram $S'$. The angle of this inclination between axes $x'$ and $x$ is given by $\arctan(V)$.}\label{worldsheet}
\end{figure}

The basic oscillon has period $T=1$. Smaller and bigger oscillons, which differ by their period, can be obtained from the basic one by  the scaling transformation which exploits the scaling symmetry of the signum-Gordon equation, see Section \ref{symmetries}. The Minkowski diagram presented in Fig.\ref{worldsheet} shows regions of validity of the partial solutions $\phi_{k}(t',x')$, $k=\{C,L_1,L_2, L_3, R_1, R_2, R_3\}$ that together describe the motion of the oscillon.  These partial solutions are given by quadratic polynomials in variables $t'$ and $x'$ in the rest frame of the oscillon. Physically relevant parts of the polynomial solutions are restricted to some intervals of $t'$ and $x'$.  In order to get a periodic solution for any $t'$ one has to replace $t'$ by a periodic function of $t'$. Below we discuss this process in detail.

We start our discussion with a set of partial solutions which are valid {\it only} in the interval $t'\in[0,\frac{1}{2}]$.  These solutions have been given in \cite{arodzswier} and they are denoted by the letter $\varphi$. Here, we present them in a new notation. In fact, there are seven partial solutions in the interval $t'\in[0,\frac{1}{2}]$. Among them, four solutions are essentially different, namely
\begin{align}
\varphi_{C}(t',x';v)&=\frac{(1+v-2x')^2-4t'(1+v-2vx')+4(2-v^2)t'^2}{8\,(1-v^2)},\label{phiC}\\
\varphi_{L_1}(t',x';v)&=\frac{t'^2}{2}-\frac{t'x'}{1+v},\label{phiL1}\\
\varphi_{L_2}(t',x';v)&=-\frac{(x'-vt')^2}{2\,(1-v^2)},\label{phiL2}\\
\varphi_{L_3}(t',x';v)&=\frac{1}{2}\left(t'-\frac{1}{2}\right)\left(t'+\frac{1}{2}+\frac{2x'-1}{1-v}\right)\label{phiL3}.
\end{align}
The other three partial solutions $\varphi_{R_j}(t',x';v)$, $j=1,2,3$ can be obtained from those shown above by performing the transformation
\be
x'\rightarrow 1-x', \qquad v\rightarrow -v\label{transformation1}
\ee
which gives
\be
\varphi_{R_j}(t',x';v)=\varphi_{L_j}(t',1-x';-v).\label{RL}
\ee
Note that the solution $\varphi_{C}(t,x;v)$ is invariant under this transformation. Note also that all solutions $\varphi_{k}(t',x';v)$ are negative-valued in their domains.

Each solution $\varphi_{k}(t',x';v)$ is valid only in a specific region of the Minkowski diagram. For this reason we define a few region step functions $\Pi_{k}(t',x';v)$ which are equal to unity only in the region in which the given partial solution holds and vanish outside this region:
\begin{align}
&{\textstyle \Pi_C(t',x';v)= \theta(x'-t')\theta(-x'-t'+1)\theta\left(x'+t'-\frac{1+v}{2}\right)\theta\left(-x'+t'+\frac{1+v}{2}\right),}\nonumber\\
&{\textstyle \Pi_{L_1}(t',x';v)=\theta(x'-t')\theta\left(-x'-t'+\frac{1+v}{2}\right),}\nonumber\\
&{\textstyle\Pi_{L_2}(t',x';v)=\theta(-x'+t')\theta\left(-x'-t'+\frac{1+v}{2}\right)\theta(x'-vt'),}\nonumber\\
&{\textstyle\Pi_{L_3}(t',x';v)=\theta\left(x'+t'-\frac{1+v}{2}\right)\theta(-x'+t'),}\nonumber\\
&{\textstyle\Pi_{R_j}(t',x';v)=\Pi_{L_j}(t',1-x',-v),\qquad j=1,2,3}\nonumber
\end{align}
where $\theta(z)$ is the unit step function $\theta(z)=0$ for $z<0$ and $\theta(z)=1$ for $z\ge 0$.
One can check that $\Pi_C(t',x';v)=\Pi_C(t',1-x';-v)$.

\begin{figure}[h!]
\centering
\subfigure{\includegraphics[width=0.4\textwidth,height=0.2\textwidth, angle =0]{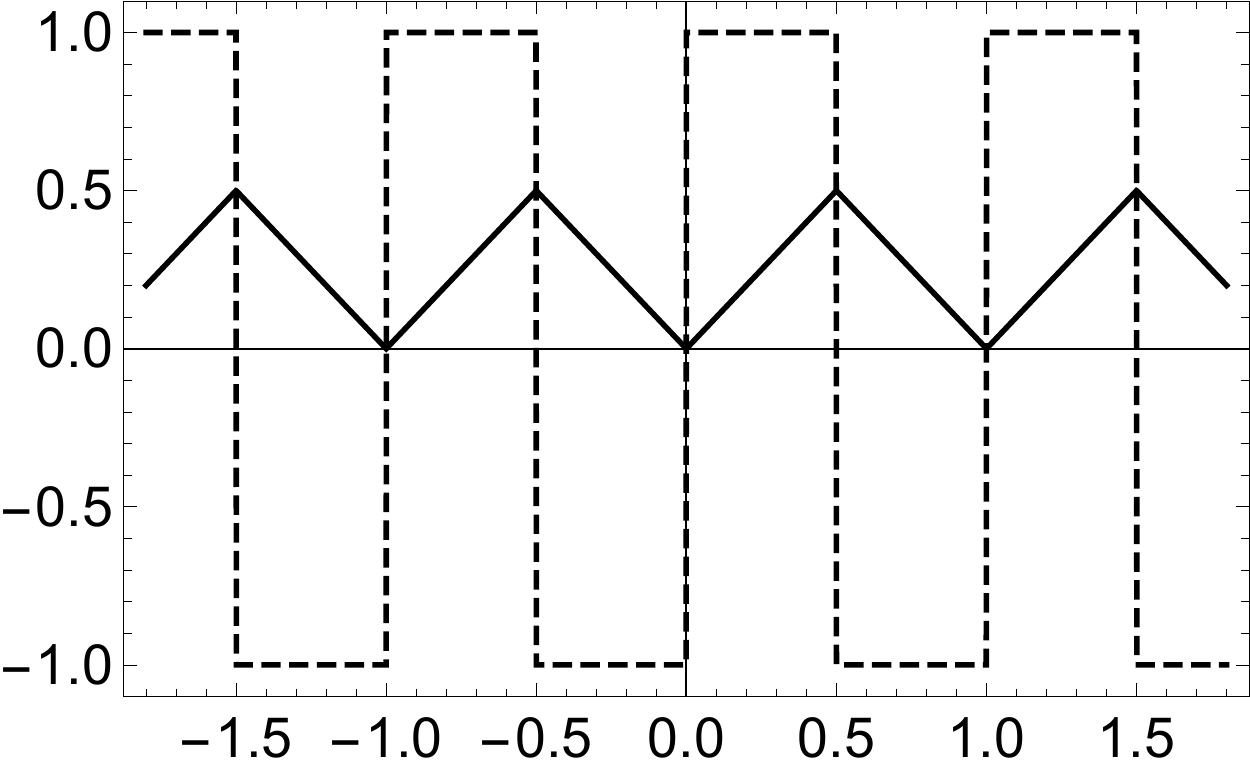}}
\caption{Function $\tau(z)$ (solid line) and $\sigma(z)$ (dashed line).}\label{tau}
\end{figure}

The parabolic functions like \eqref{phiC}-\eqref{phiL3} are not periodic whereas the oscillon is a periodic solution. In order to give formulas  which are valid for any $t'$, we define two periodic functions
\begin{align}
\tau(z)&:=\frac{1}{\pi}\arcsin(|\sin(\pi z)|),\label{funtau}\\
\sigma(z)&:={\rm sgn}(\sin(2\pi z))\label{fundertau},
\end{align}
where $\tau(z)$ maps any $t'$ onto the intarval $[0,\frac{1}{2}]$ and $\sigma(z)=\pm1$  agrees with classical derivative of $\tau(z)$. The functions \eqref{funtau} and \eqref{fundertau}  allow us to construct the periodic solutions. They are presented in Fig.\ref{tau}.
The function $\sigma(z)$ is needed to describe the changes of the sign of the partial solutions at $t'=\frac{n}{2}$.
Any partial solution (for arbitrary $t'$) can be written in terms of basic solutions $\varphi_{k}(t',x';v)$, where $k=\{C,L_1,\cdots, R_3\}$, that are valid only on the interval $t'\in[0,\frac{1}{2}]$. The partial solutions presented in the Minkowski diagram in Fig.\ref{worldsheet} have the form
\be
\phi_{k}(t',x';v)=\sigma(t')\varphi_{k}(\tau(t'),x';v)\Pi_{k}(\tau(t'),x';v).\label{partialsolutions}
\ee
The functions \eqref{funtau} and \eqref{fundertau} allow us to introduce a more compact notation than the one in \cite{kswz}. Note that here the functions $\phi^{-}_{L_1/R_1}$ and $\phi^{+}_{L_3/R_3}$ have been absorbed into the definitions of functions $\phi_{L_1/R_1}$. Similarly, $\phi^{-}_{L_3/R_3}$ and $\phi^{+}_{L_1/R_1}$ have been absorbed into the definitions of functions $\phi_{L_3/R_3}$.
The total solution is given by a continuous function which is a sum of non overlapping partial solutions \eqref{partialsolutions}. The derivative of the oscillon solution with respect to time is given by
\be
\partial_{t'}\phi_{k}(t',x';v)=\partial_{z}\varphi_{k}(z,x';v)|_{z=\tau(t')}\Pi_{k},(\tau(t'),x';v)\label{derivpartialsolutions}
\ee
where $\sigma^2(z)=1$. Note that  all the  derivatives of the region step functions $\Pi_{\alpha}$ can be ignored because the sum of  partial solutions is a continuous function so there is no reason to expect delta functions at the matching points.

\begin{figure}[h!]
\centering
\subfigure[]{\includegraphics[width=0.3\textwidth,height=0.15\textwidth, angle =0]{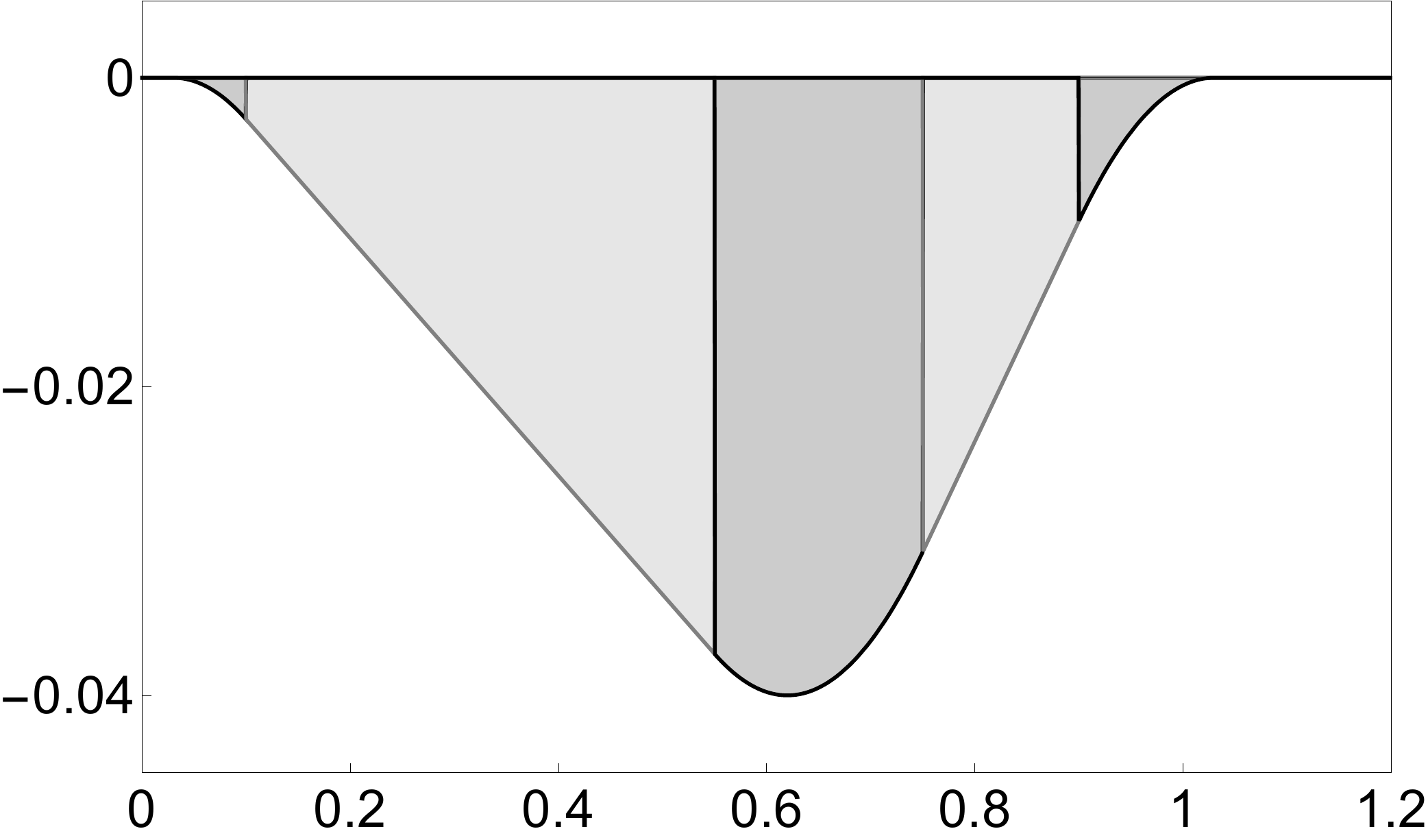}}
\subfigure[]{\includegraphics[width=0.3\textwidth,height=0.15\textwidth, angle =0]{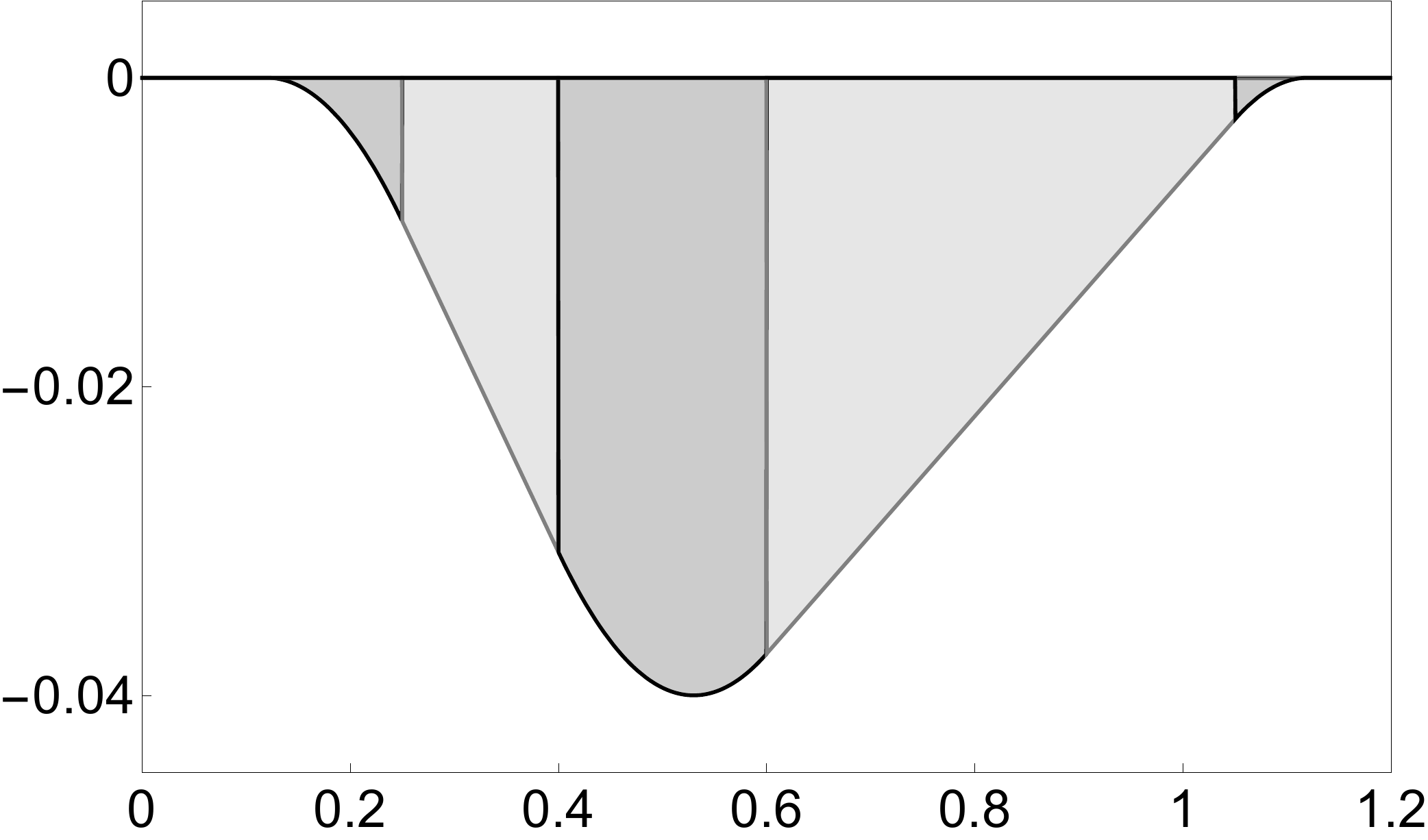}}
\subfigure[]{\includegraphics[width=0.3\textwidth,height=0.15\textwidth, angle =0]{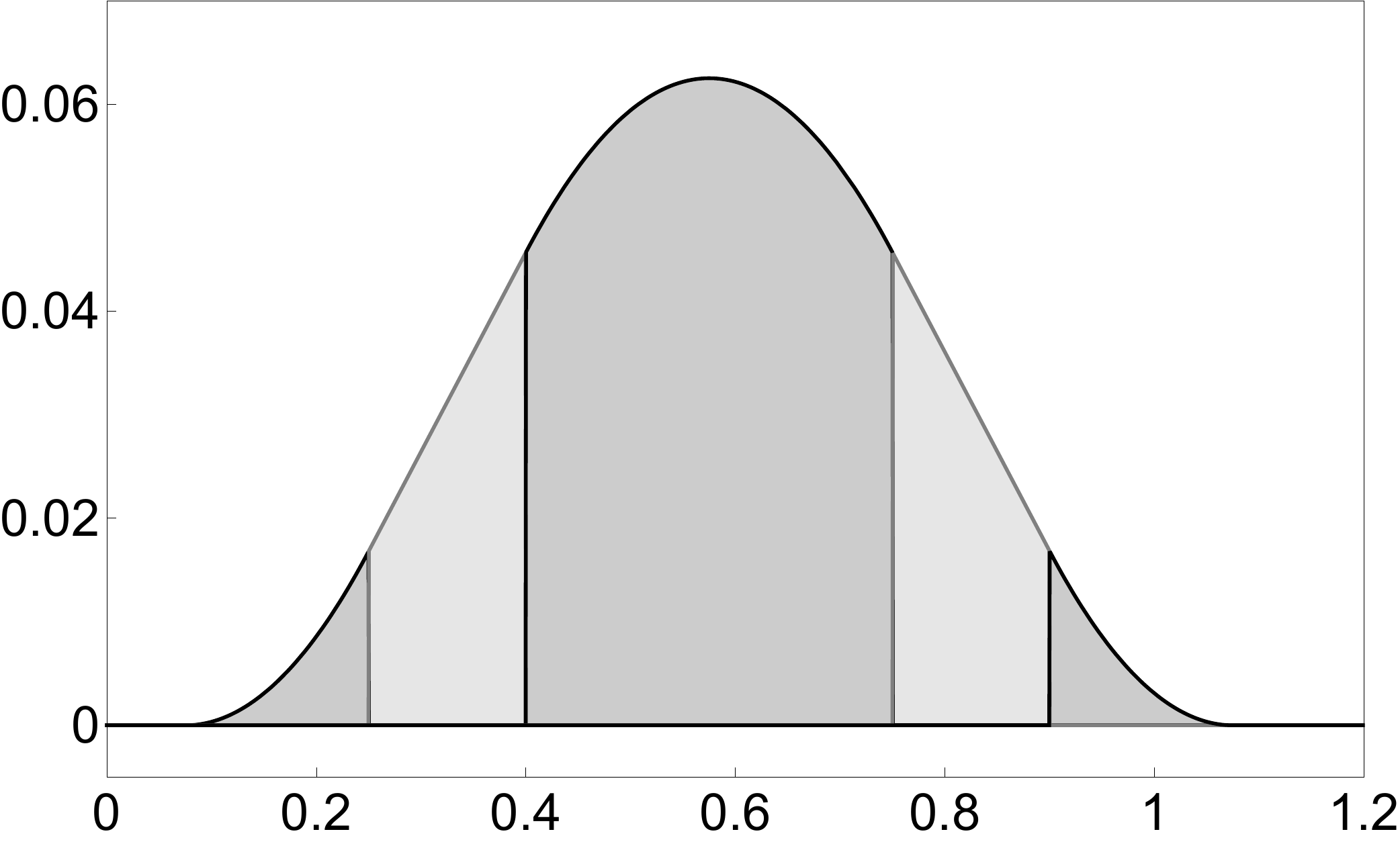}}
\subfigure[]{\includegraphics[width=0.3\textwidth,height=0.15\textwidth, angle =0]{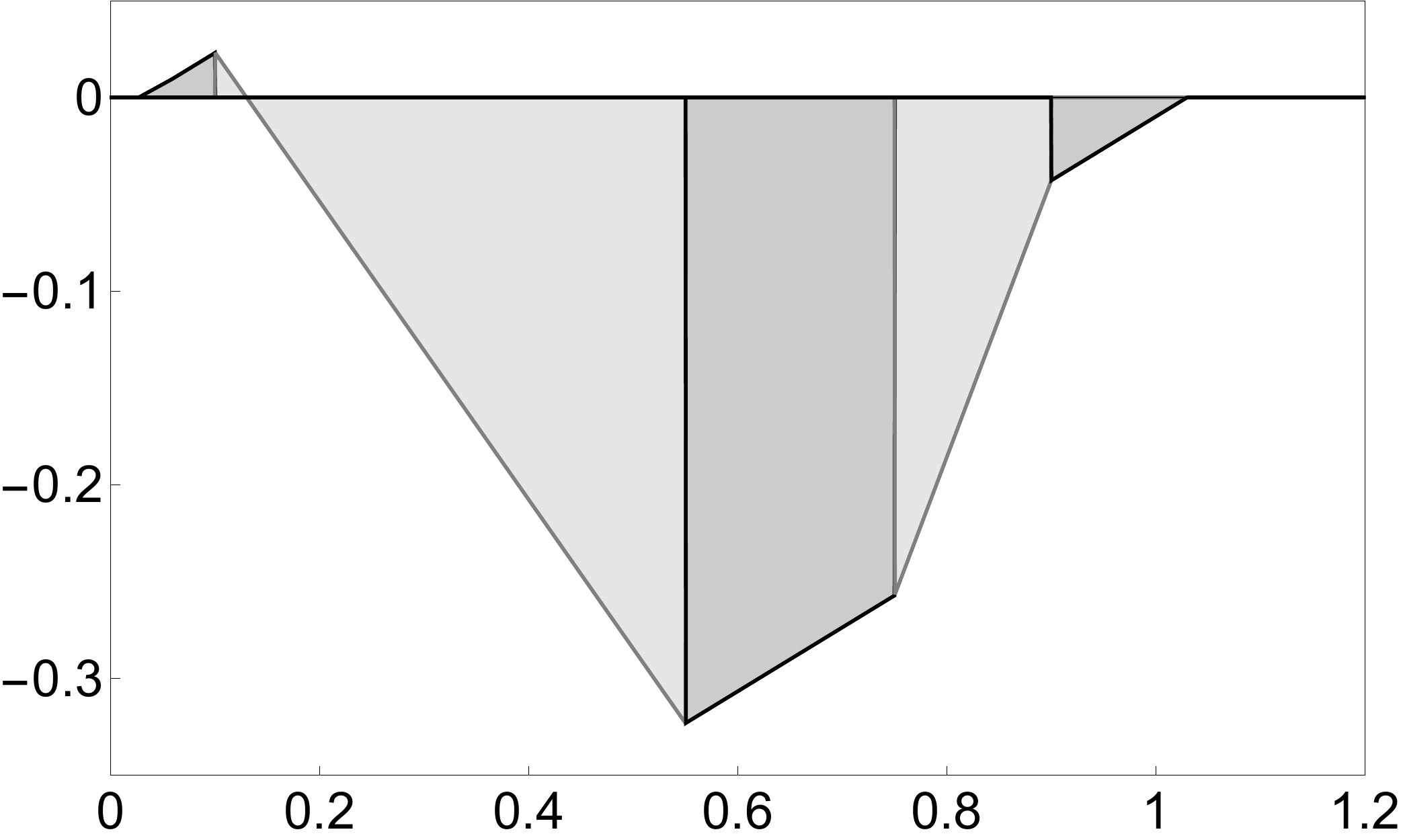}}
\subfigure[]{\includegraphics[width=0.3\textwidth,height=0.15\textwidth, angle =0]{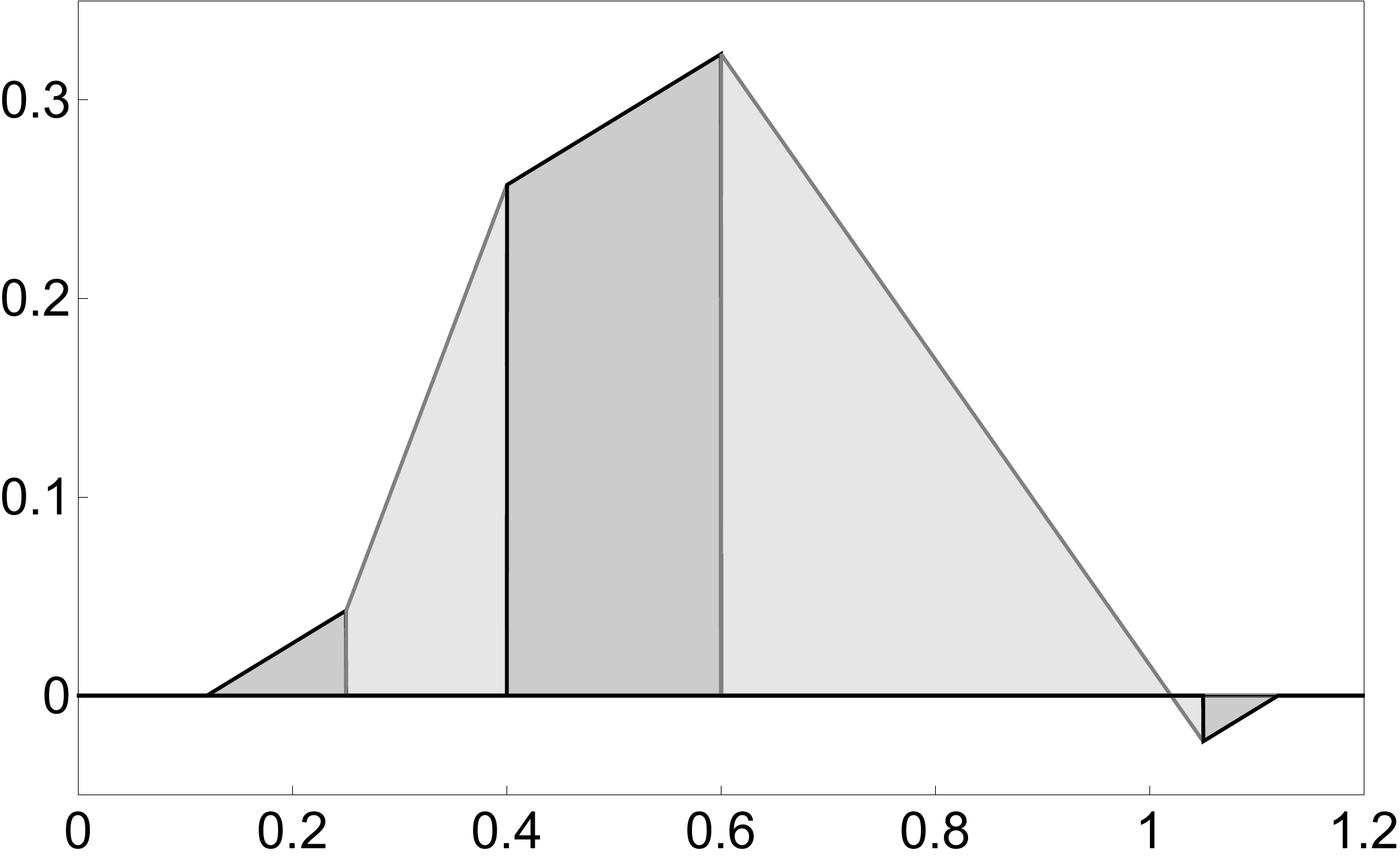}}
\subfigure[]{\includegraphics[width=0.3\textwidth,height=0.15\textwidth, angle =0]{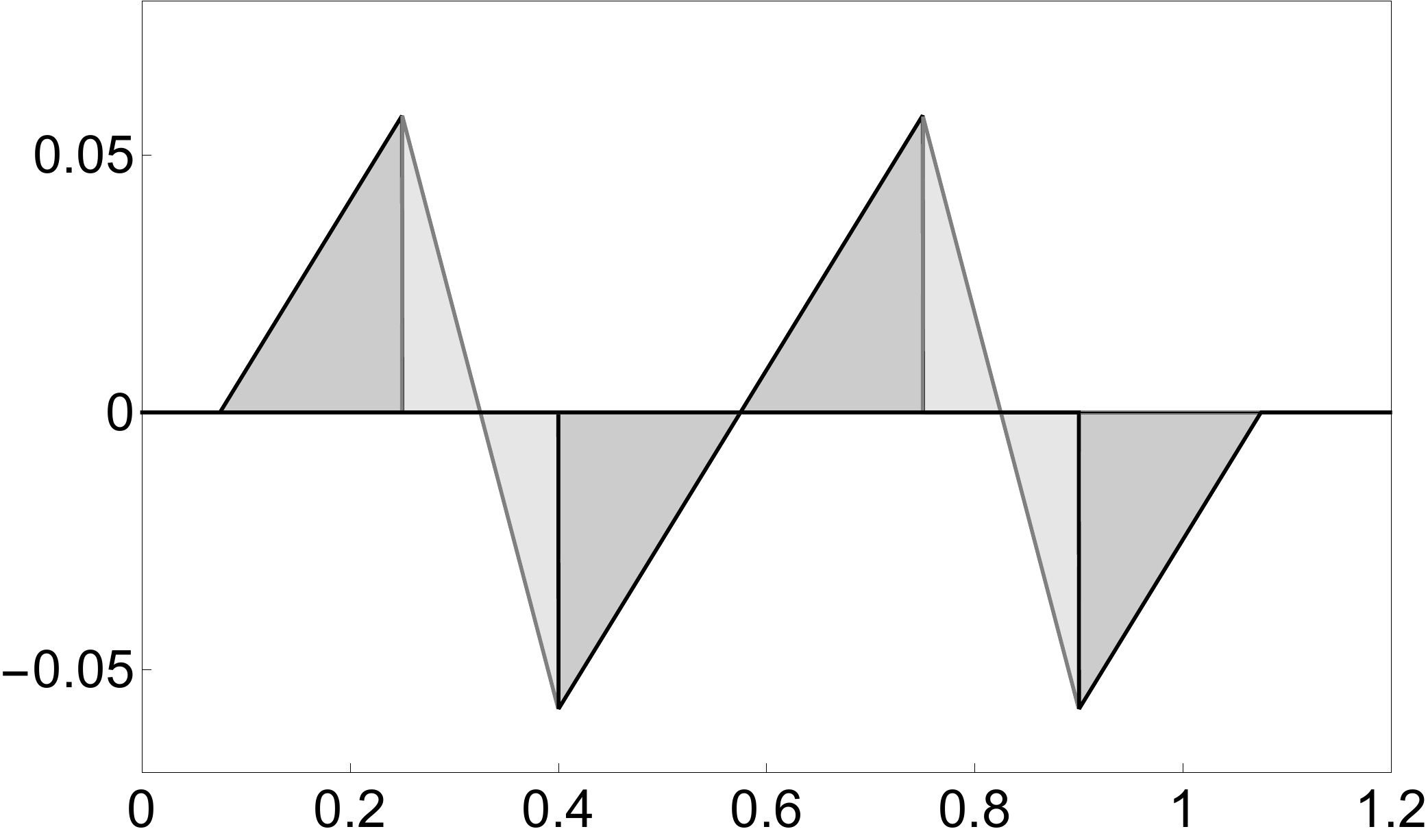}}
\caption{The profile function $\phi(t',x';v)$ of the generalized exact oscillon for $v=0.3$ at (a) $t'=0.1$, (b) $t'=0.4$ and (c)  $t'=0.75$. The corresponding time derivatives of $\partial_{t'}\phi(t',x';v)$ are shown in (d) at $t'=0.1$, in (e) at $t'=0.4$ and in (f) at $t'=0.75$.}\label{rest-brather}
\end{figure}

In Fig.\ref{rest-brather} we present three snapshots of the generalized exact oscillon solution $\phi(t',x';v)$ and its time derivative $\partial_{t'}\phi(t',x';v)$ at $t'=0.1$, $t'=0.4$ and $t'=0.75$.
Solutions presented in figures (a) and (d) consists of (from left to right) $\{\phi_{L_2}, \phi_{L_1},\phi_C,\phi_{R_1},\phi_{R_2}\}$, in figures (b) and (e) fn $\{\phi_{L_2}, \phi_{L_3},\phi_C,\phi_{R_3},\phi_{R_2}\}$
and, finally,  in figures (c) and (f) of $\{\phi_{L_2}, \phi_{L_1},\phi_C,\phi_{R_3},\phi_{R_2}\}$.

\subsection{Travelling oscillons}

The signum-Gordon equation
\be
(\partial_t^2-\partial_x^2)\phi(t,x)+{\rm sgn}\,\phi(t,x)=0\label{signumgordon}
\ee
is invariant under the Lorentz transformations. Thus, traveling compact oscillons  can be obtained from the non traveling ones by an appropriate Lorentz transformation. In particular, the oscillon with a non vanishing linear momentum is obtained from the generalized exact oscillon  by a Lorentz boost. In what follows we assume that the laboratory reference frame $S$ moves with velocity $u'=\mp V$ with respect to the rest frame of the oscillon $S'$.  Thus, the oscillon has velocity $u=\pm V$ in the laboratory frame $S$. The field configuration that describes  the traveling oscillon is a function of coordinates $t$ and $x$ and is given by performing the transformations  
\be
t'\rightarrow\xi:=\frac{t-ux}{\sqrt{1-u^2}},\qquad\qquad x'\rightarrow\zeta:=\frac{x-ut}{\sqrt{1-u^2}}
\ee
in \eqref{partialsolutions} and obtaining the  partial solutions
\be
\psi_{k}(t,x;v,u):=\sigma(\xi)\varphi_{k}\Big(\tau(\xi),\zeta; v\Big)\Pi_{k}\Big(\tau(\xi),\zeta; v\Big).\label{travel}
\ee
The derivatives of the fields with respect to time $t$ are given by the expression
\begin{align}
\partial_t\psi_{k}(t,x;v,u)=\left.\frac{1}{\sqrt{1-u^2}}\Big[\partial_{\bar\xi}\varphi_{k}(\bar\xi,\bar\zeta;v)-u\,\sigma(\xi)\partial_{\bar\zeta}\varphi_{k}(\bar\xi,\bar\zeta;v)\Big]\Pi_{k}(\bar\xi,\bar\zeta;v)\right|_{\bar\xi=\tau(\xi),\,\bar\zeta=\zeta},\label{dertravel}
\end{align}
where  $\frac{d}{dz}\tau(z)=\sigma(z)$ and $\sigma^2(z)=1$ at open supports of the partial solutions. Hence, the travelling oscillon in $S$ is a solution of \eqref{signumgordon} given by a sum of non overlapping partial solutions \eqref{travel}, namely,
\be
\psi(t,x;v,u)=\sum_{k}\psi_{k}(t,x;v,u),
\ee
where $k=\{L_1, L_2,L_3, C, R_1, R_2, R_3\}$.

Note, that the axis $x'$ {\it i.e} the line $t'=0$ in $S'$ is not a  simultaneity line in $S$. The scalar field $\phi(t',x';v)$ vanishes at the horizontal lines $t'=\frac{n}{2}$, $n=0,\pm1,\pm2,\ldots$ which are shown in Fig.\ref{worldsheet}. It shows that the oscillon seen in the laboratory reference frame $S$ has some isolated traveling zeros. Such isolated zeros are  given by points of intersection of lines parallel to the axis $x$ (given by $t=const$) with the lines $t'=\frac{n}{2}$. The number of isolated zeros and the composition of the  oscillon (types of partial solutions seen in $S$ at given instant of time $t$)  depend on the value of the velocity $u=\pm V$ which the oscillon has in $S$. The axes $x$ and $x'$ form an angle $\arctan(V)$ so for  $V<\frac{1}{2+v}$ there is only one point of intersection of straight lines parallel to $x'$ with the line  $t'=\frac{n}{2}$. For $\frac{1}{2+v}<V<1$, a second isolated zero arises at some value of the time interval.

According to the diagram presented in Fig.\ref{worldsheet}, the initial (at $t=0$) configuration of the field in $S$, obtained by the Lorentz boost of the oscillon given in $S'$, consists of different sets of partial solutions corresponding to different values of the velocity $V$. Let us consider the oscillon that moves with the velocity $u=+V$ in the laboratory reference frame. The oscillon configuration $\psi$ at $t=0$ consists of (from left to right)
\begin{align}
 &\{\psi_{L_1}, \psi_C, \psi_{R_1}, \psi_{R_2}\} &{\rm for} &&&0<V<\frac{1-v}{3+v},\label{case1}\\
&\{\psi_{L_1}, \psi_C, \psi_{R_3}, \psi_{R_2}\}& {\rm for} &&&\frac{1-v}{3+v}<V<\frac{1}{2+v},\label{case2}\\
& \{\psi_{L_1}, \psi_C, \psi_{R_3}, \psi_{R_2}\}&{\rm for}&&&\frac{1}{2+v}<V<1.\label{case3}
\end{align}
The cases \eqref{case2} and \eqref{case3} differ by a sign of partial solutions  $\psi_{R_3}$ and $\psi_{R_2}$.
\begin{figure}[h!]
\centering
\subfigure[]{\includegraphics[width=0.3\textwidth,height=0.15\textwidth, angle =0]{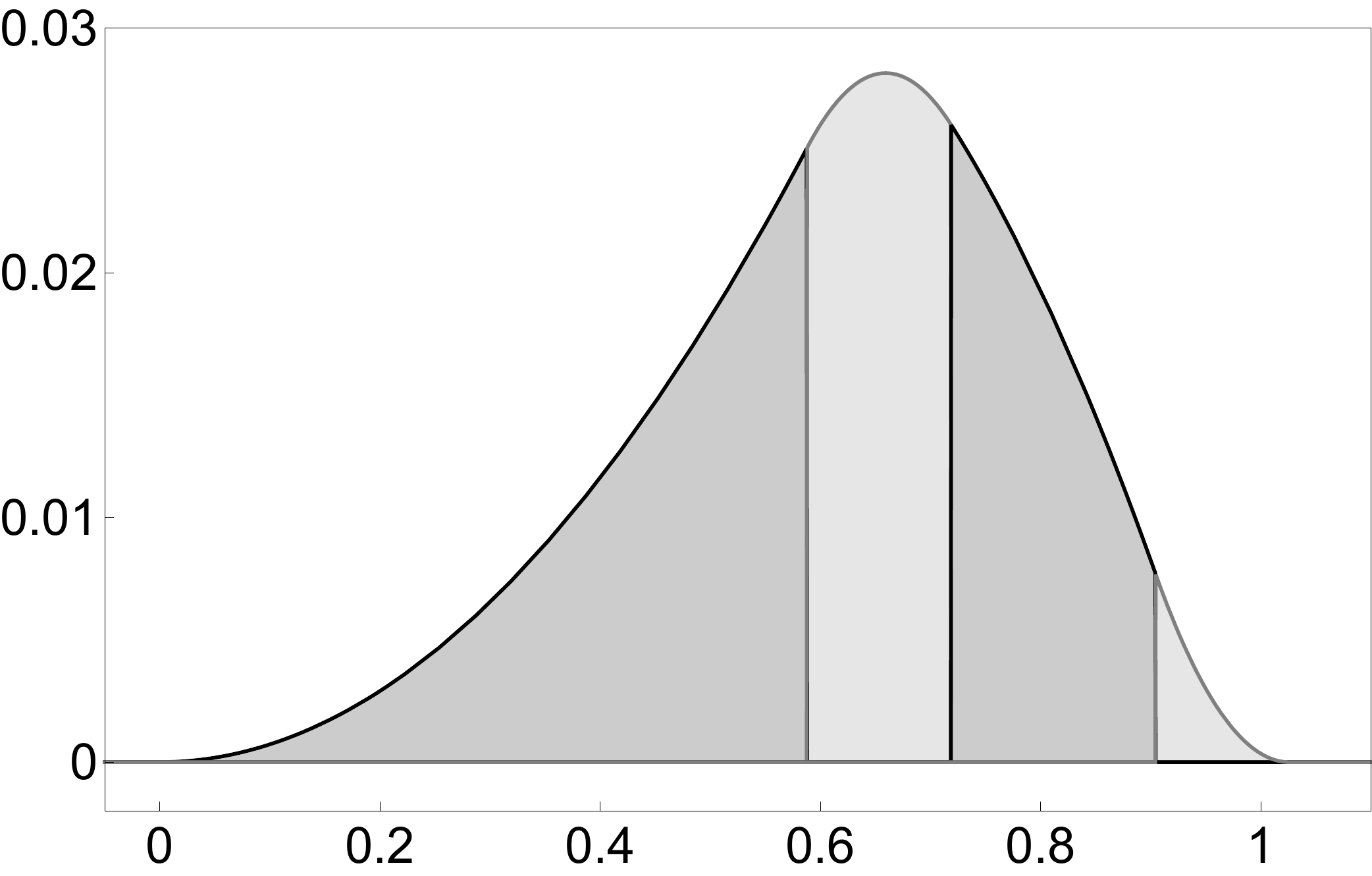}}
\subfigure[]{\includegraphics[width=0.3\textwidth,height=0.15\textwidth, angle =0]{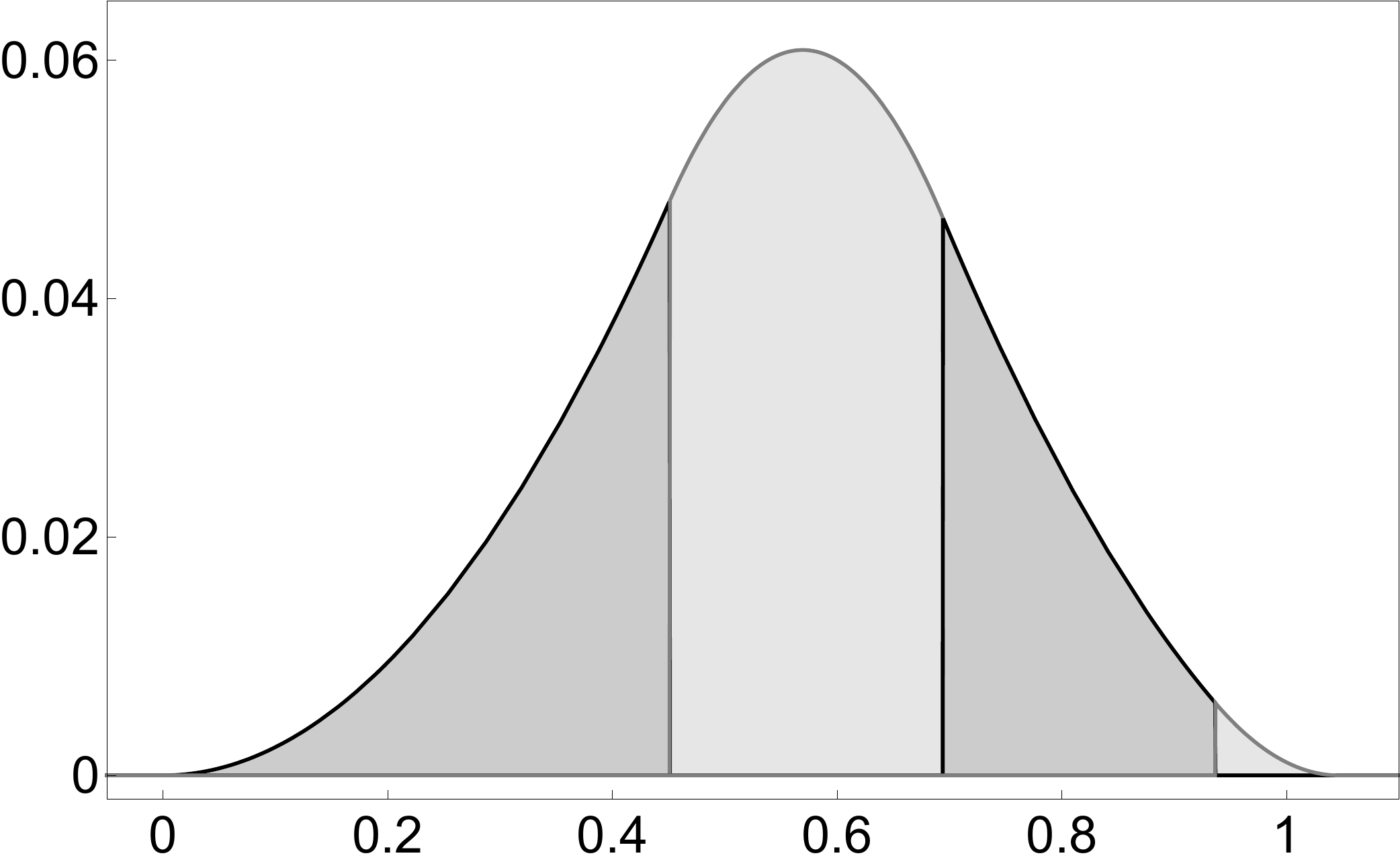}}
\subfigure[]{\includegraphics[width=0.3\textwidth,height=0.15\textwidth, angle =0]{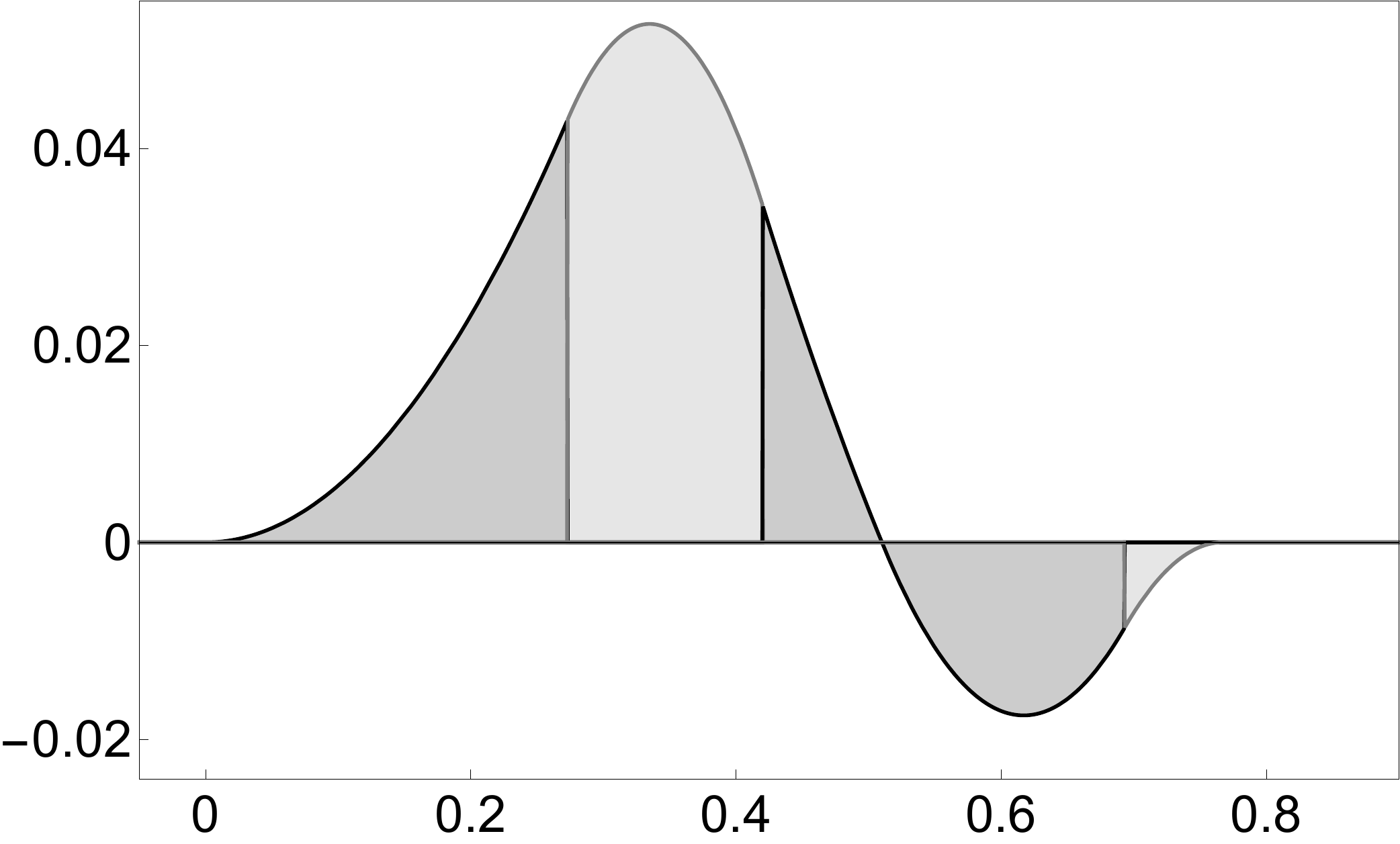}}
\subfigure[]{\includegraphics[width=0.3\textwidth,height=0.15\textwidth, angle =0]{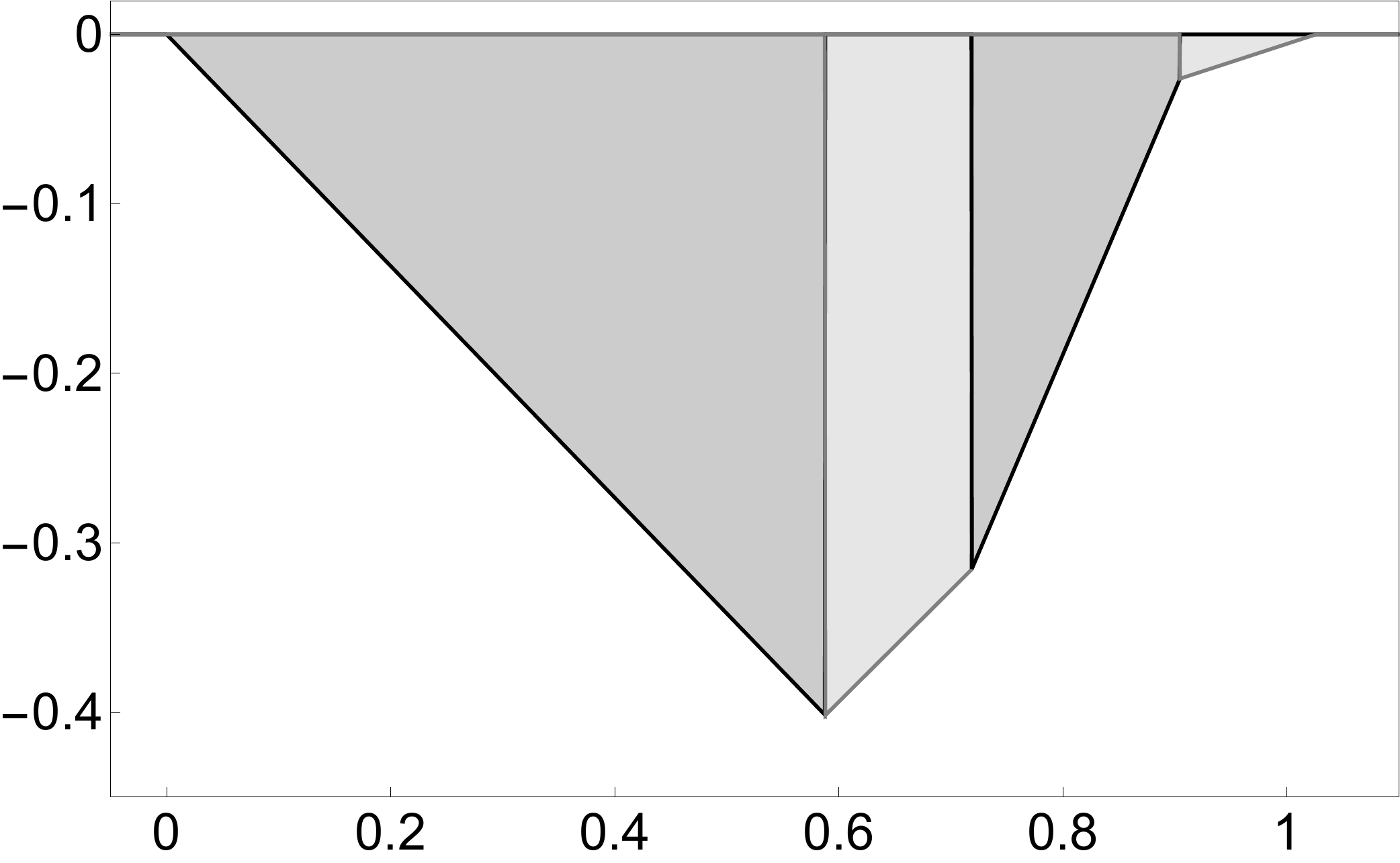}}
\subfigure[]{\includegraphics[width=0.3\textwidth,height=0.15\textwidth, angle =0]{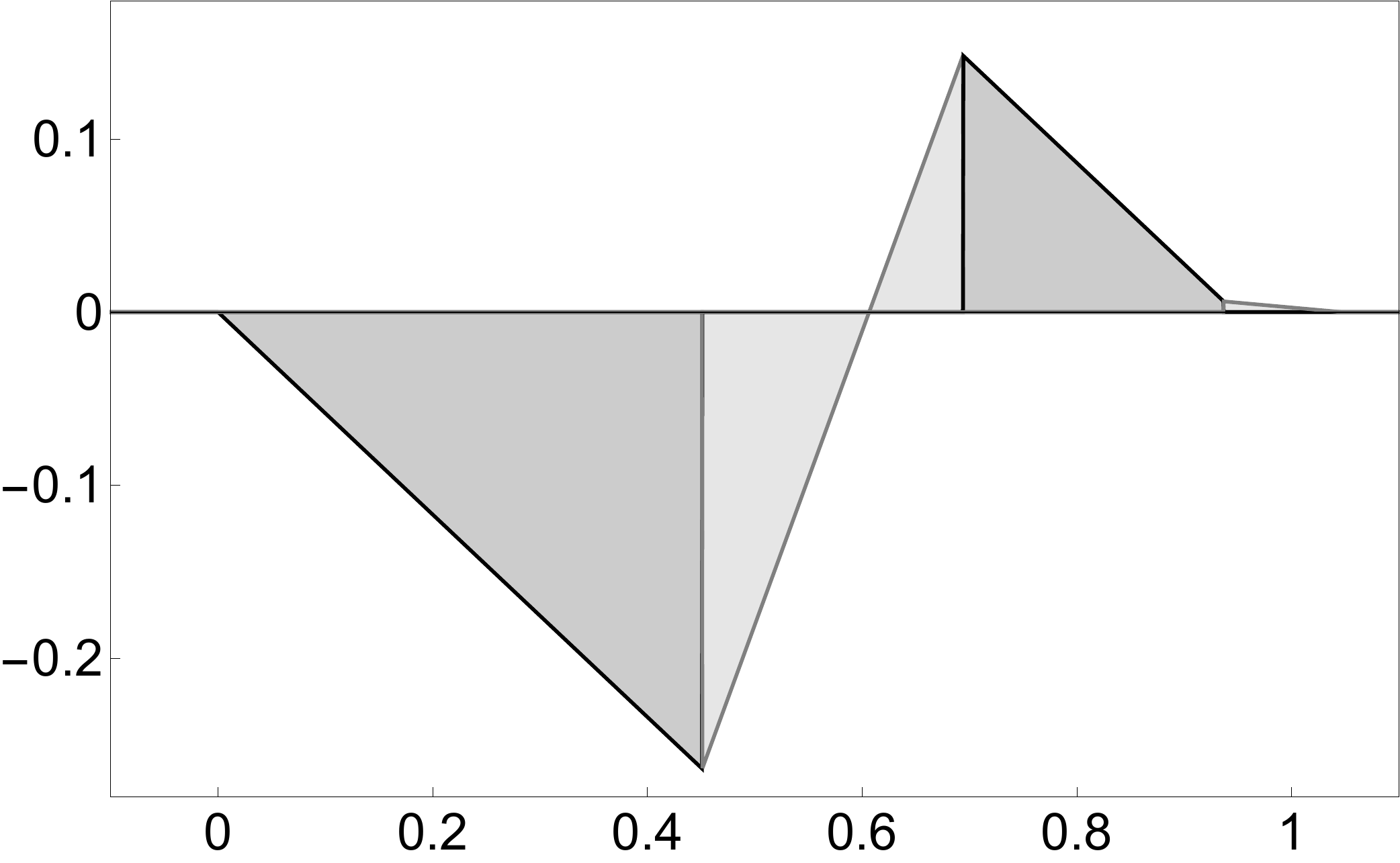}}
\subfigure[]{\includegraphics[width=0.3\textwidth,height=0.15\textwidth, angle =0]{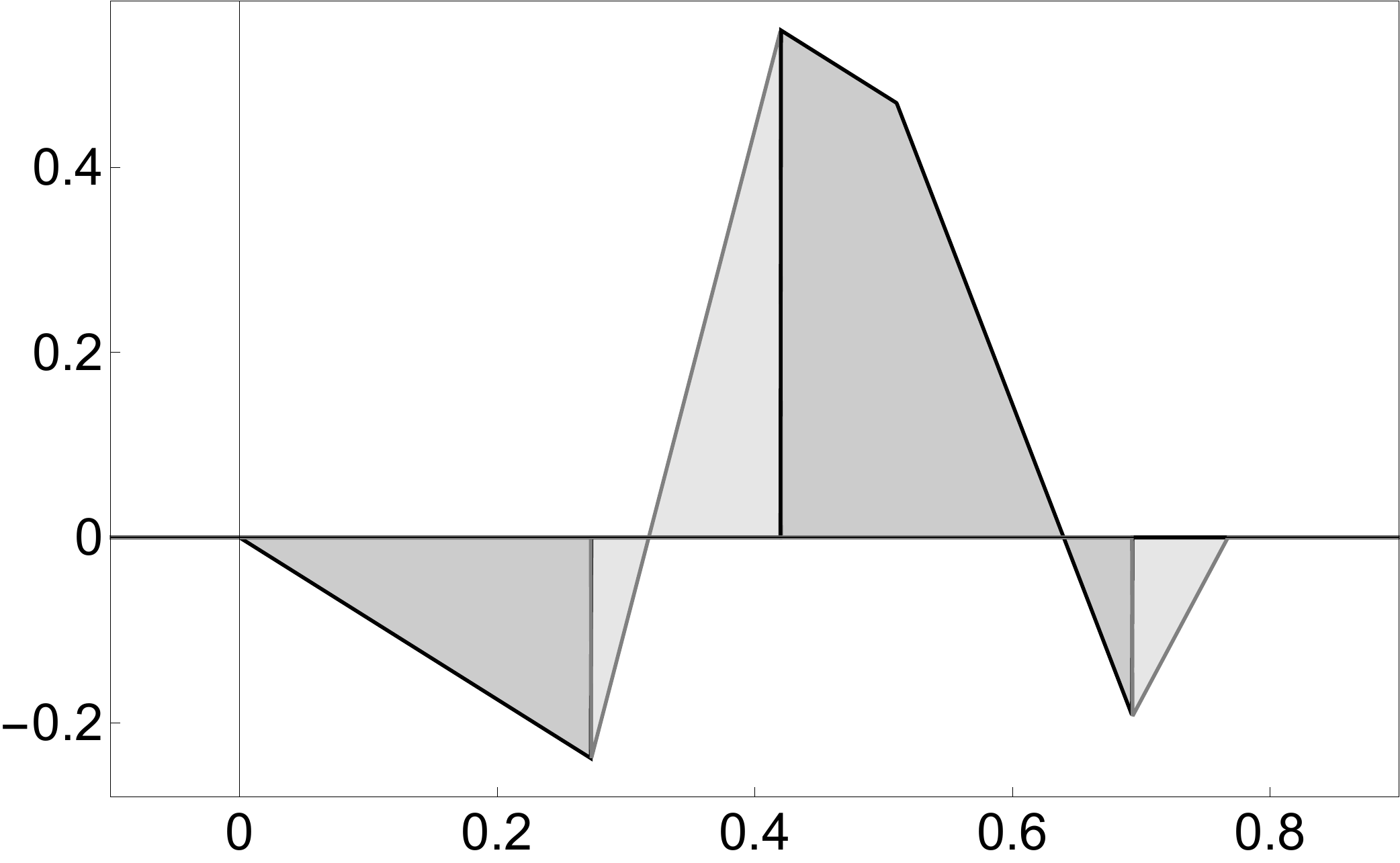}}
\caption{The generalized exact oscillon with the velocity $u=+V$ in the laboratory reference frame at $t=0$. The oscillon is parametrized by $v=0.3$. Figures (a), (b), (c) show the initial shape of the oscillon $\psi(t,x;vV)|_{t=0}$  whereas (d), (e), (f) show $\partial_t\psi(t,x;v,V)|_{t=0}$. Figures (a), (d) correspond with the oscillon velocity $V=0.1$, (b), (e) with $V=0.35$ and (c), (f) with $V=0.7$.}\label{initial1}
\end{figure}

Fig.\ref{initial1} shows three examples of such field configurations and their time derivatives at $t=0$. All three oscillons have $v=0.3$  (which gives $\frac{1-v}{3+v}\approx 0.212$ and $\frac{1}{2+v}\approx 0.434$) and they differ by the value of velocity $V$. Figures (a) and (d) were obtained for $V=0.1$ and they correspond to \eqref{case1},  (b) and (e) were obtained for $V=0.35$ and they correspond to \eqref{case2} and figures (c) and (f) show the case \eqref{case3} with $V=0.7$.

\section{Scattering of oscillons}

\subsection{Initial configurations for scattering process}

\subsubsection{Two-oscillon configurations}\label{symmetries}
The compactness of exact oscillons allows the construction of some multi-oscillon configurations which are exact solutions of the signum-Gordon equation. The only condition to satisfy is the non-overlapping of the supports of individual oscillons. A generic initial configuration $\{\Psi(x), \Psi_t(x)\}$ containing two travelling oscillons is given by the superposition of non-overlapping travelling oscillons obtained from generalized exact oscillons $\phi(t,x; v_1)$ and $\phi(t,x; v_2)$ by the transformations which are symmetries of the signum-Gordon equation, namely
\begin{itemize}
\item
Poincar\'e transformatons in (1+1) dimensions: boosts, spatial and temporal translations, spatial and temporal reflections,
\item
symmetry of the scale $\phi(t,x)\rightarrow\phi^{(\lambda)}(t,x)=\lambda^2\phi(\frac{t}{\lambda},\frac{x}{\lambda})$,
\item
sign flipping of the field $\phi\rightarrow-\phi$.
\end{itemize}
Two individual oscillons, $i=1,2$, are obtained by transformations
\[
\phi(t,x;v_i)\rightarrow \Psi_i(t,x):=\varepsilon_i\psi^{(\lambda_i)}(t+t_{0i},\epsilon_i x+x_{0i};v_i,u_i)
\],
where $u_i$ are boost velocities, $t_{0i}$ are temporal translations, $x_{0i}$ are spatial translations, $\lambda_i$ stand for scale parameters and $\varepsilon_i=\pm1$ and $\epsilon_i=\pm1$ for reflections. Let us fix the sign of $v_i>0$ because $\phi(t,x;-v)=\phi(t,1-x,v)$ and so the sign of $v$ can be absorbed into the combination of spatial reflexion and translation.

The initial configuration $\{\Psi(x), \Psi_t(x)\}$ is now  given by
\begin{align}
\Psi(x)&:=\Psi_1(0,x)+\Psi_2(0,x),\\
\Psi_t(x)&:=\partial_t\Psi_1(t,x)|_{t=0}+\partial_t\Psi_2(t,x)|_{t=0}
\end{align}
where the non-overlapping of their supports restricts the values of admissible spatial translations.

It turns out that some of left-over parameters can be omitted without any loss of generality. The only relevant parameters of the  initial configurations involving two oscillons are those which are not equivalent {\it i.e.} they cannot be related by a symmetry transformation. Thus the set of relevant free parameters contains relative scale, relative velocity etc. In some cases it will be more convenient fix one of them and so only study the dependence on the last one.

\begin{figure}[h!]
\centering
\subfigure[]{\includegraphics[width=0.45\textwidth,height=0.15\textwidth, angle =0]{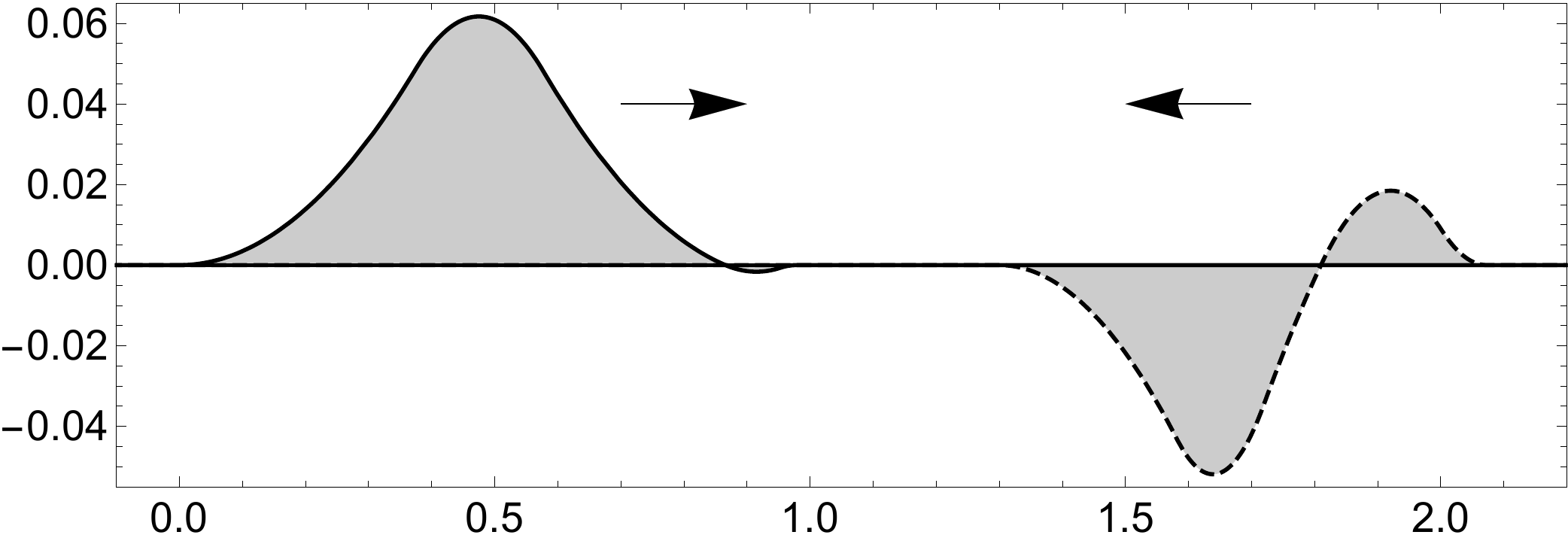}}
\subfigure[]{\includegraphics[width=0.45\textwidth,height=0.15\textwidth, angle =0]{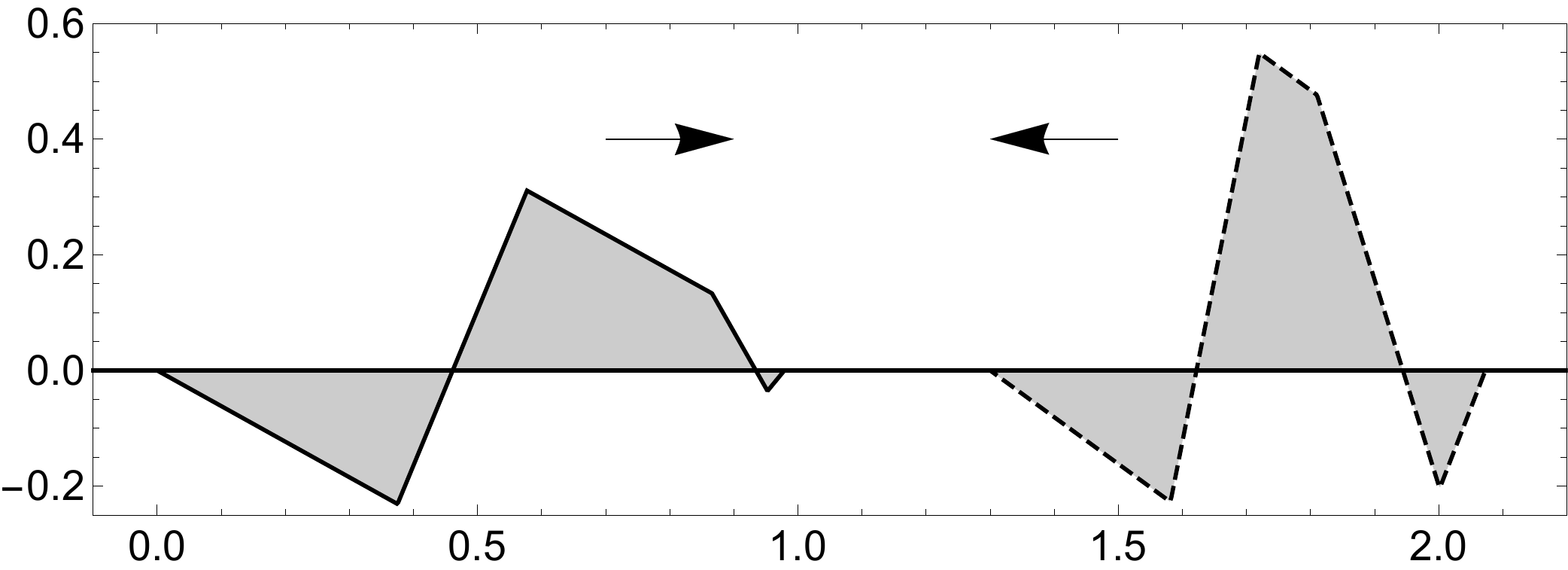}}
\caption{Initial field configuration (a) $\Psi(x)$ and (b) $\Psi_t(x)$ given by two compact oscillons at $t=0$. Parameters of oscillons $v_1=0.3$, $v_2=0.34$,  $u_1=0.5$, $u_2=-0.7$, $x_{01}=0$, $x_{02}=-1.3$ and $t_{01}=t_{02}=0$, $\lambda_i=1$, $\varepsilon_i=1=\epsilon_i$ where $i=1,2$. }\label{scat1}
\end{figure}

In Fig.\ref{scat1} we present an example of the initial configuration containing two  oscillons which move in opposite directions with velocities $u_1=0.5$ and $u_2=-0.7$. This configuration was obtained by taking the generalized exact oscillons with $v_1=0.3$ and $v_2=0.34$ and $\lambda_1=\lambda_2=1$. We have set $t_0=0$ and  $\varepsilon_i=1$ as well as $\epsilon_i=1$. The oscillons were shifted in space taking $x_{01}=0$ and $x_{02}=-1.3$.

\subsubsection{Symmetric configurations}
A numerical study of the scattering process shows that the process depends on many parameters like  $v_1$, $v_2$, the relative velocity of the oscillons, their initial distance, time shift and reflections. In order to simplify the set of parameters we have decided to restrict our considerations to {\it symmetric} and {\it anti-symmetric} initial configurations  which certainly reduces the number of free parameters. It has turned out that even with this restriction we have been left with a sufficiently rich set of physical systems. We think that more general configurations are even interesting; however, their systematic study would have required much more work so we have decided to put main effort on symmetric configurations.

In order to get a symmetric $\Psi^{(s)}$ or an antisymmetric $\Psi^{(a)}$ configuration we can take a single exact oscillon parametrized by $v$ and perform a sequence of symmetry transformations which leads to $\psi(t+t_0,x+x_{0};v,u)|_{t=0}$, where the boost velocity is chosen to be $u=+V$ with $V\ge 0$. The second oscillon can by obtained from this result by the spatial reflection $x\rightarrow-x$ and, optionally, by sign flipping. Naturally, $x_0$ must be chosen in a way that the support of $\psi$ lies on negative semiaxis $x$.  In such a case the support of the oscillon and its mirror image do not overlap. The resultant initial symmetric and antisymmetric configurations are given by
\be
\begin{split}
\Psi^{(s/a)}(x)&=\psi(t+t_0,x+x_{0};v,V)|_{t=0}\pm \psi(t+t_0,-x+x_{0};v,V)|_{t=0},\\
\Psi_t^{(s/a)}(x)&=\partial_t\psi(t+t_0,x+x_{0};v,V)|_{t=0}\pm \partial_t\psi(t+t_0,-x+x_{0};v,V)|_{t=0}.
\end{split}\label{inisymanti}
\ee
\begin{figure}[h!]
\centering
\subfigure[$\quad \Psi^{(s)}(x)$]{\includegraphics[width=0.45\textwidth,height=0.15\textwidth, angle =0]{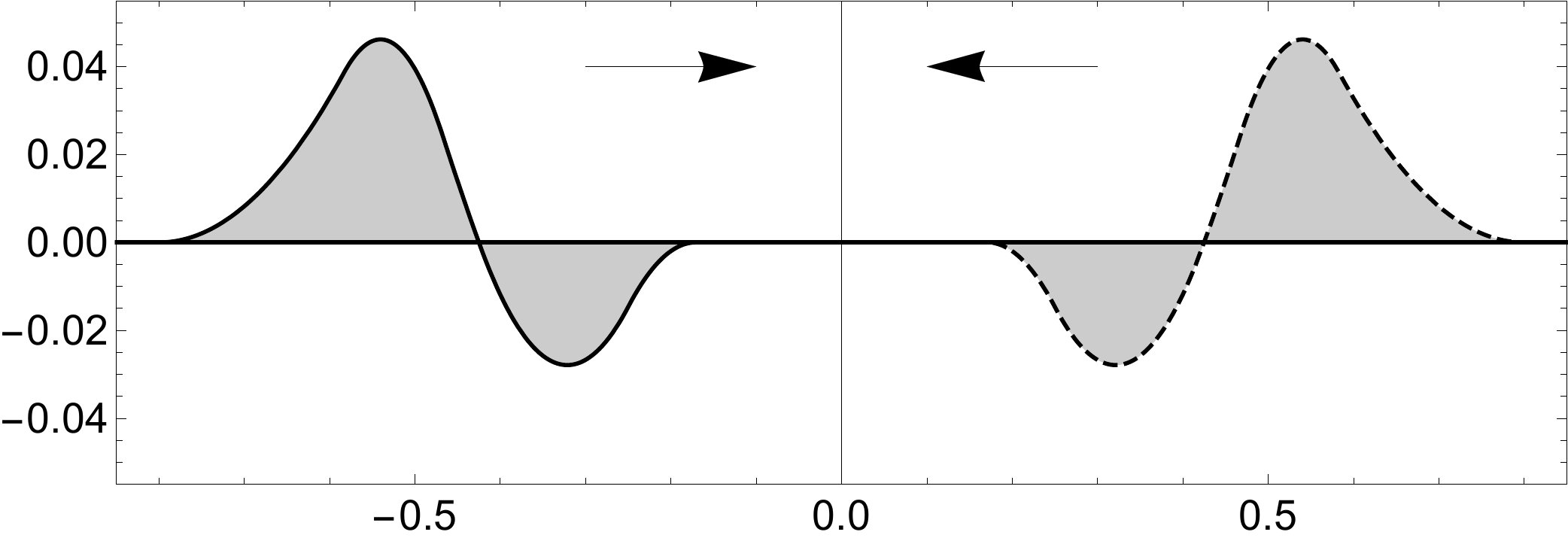}}
\subfigure[$\quad \Psi_t^{(s)}(x)$]{\includegraphics[width=0.45\textwidth,height=0.15\textwidth, angle =0]{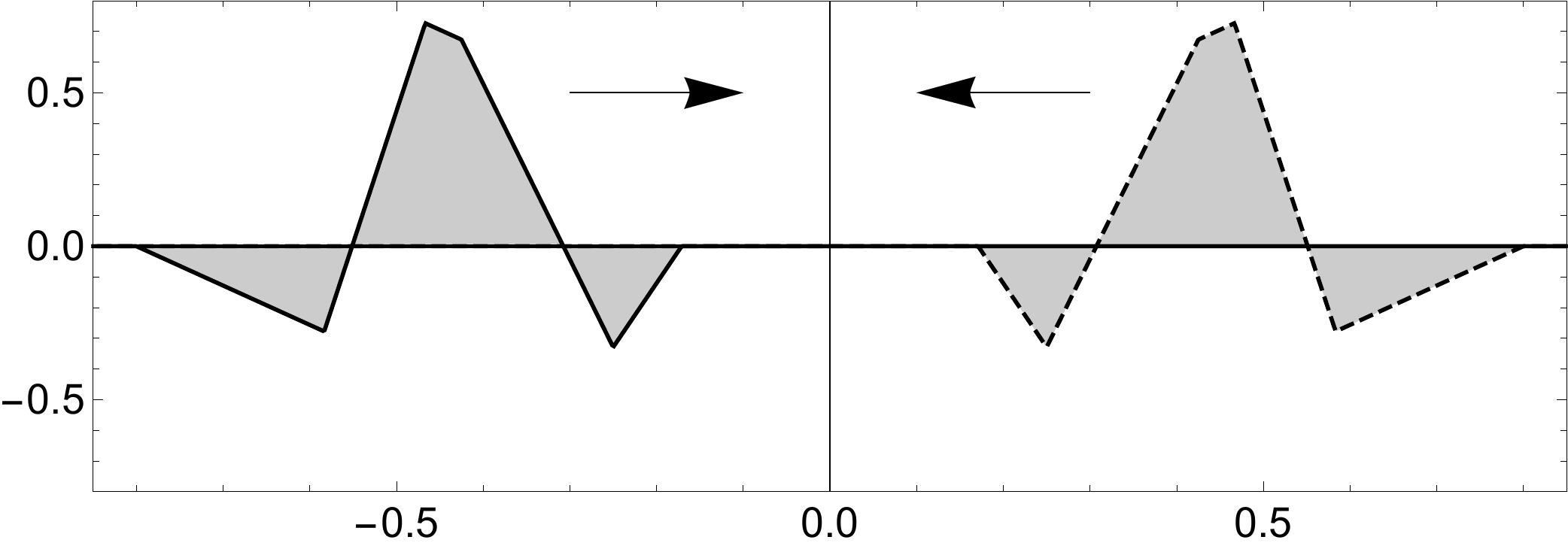}}
\caption{Symmetric initial configuration of two compact exact oscillons parametrized by $V=0.8$, $v=0.3$, $t_0=0$ and $x_0=0.8$.}\label{sym1}
\end{figure}
\begin{figure}[h!]
\centering
\subfigure[]{\includegraphics[width=0.45\textwidth,height=0.15\textwidth, angle =0]{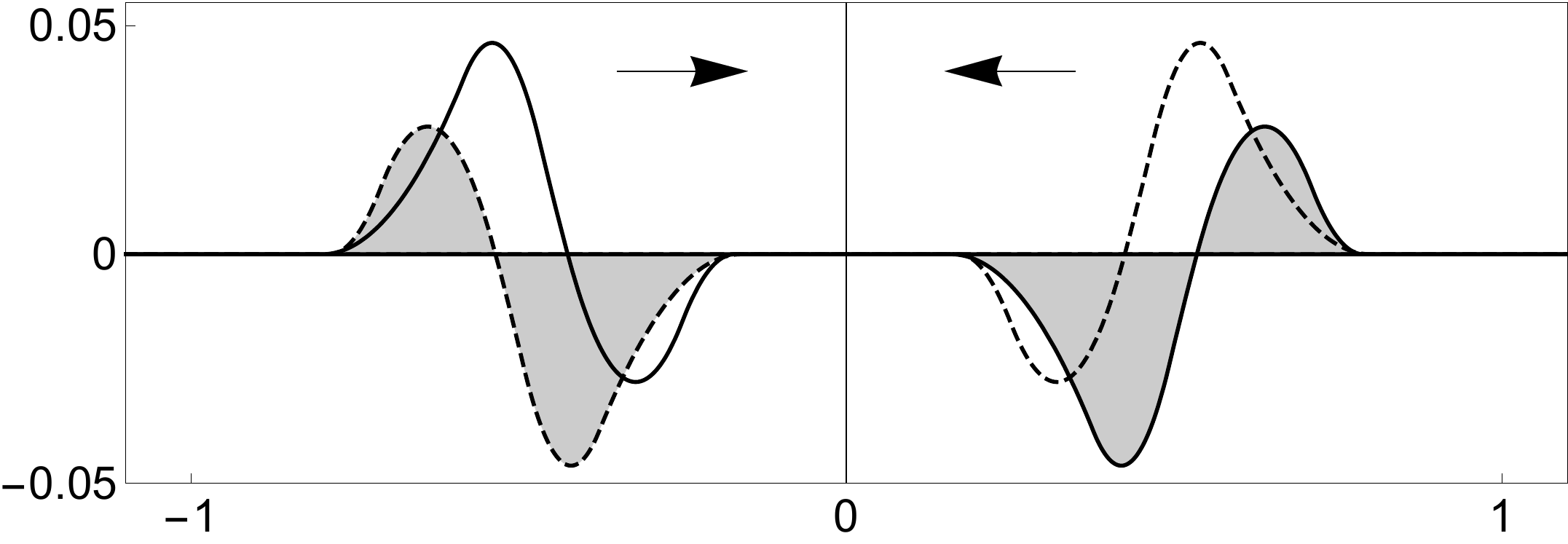}}
\subfigure[]{\includegraphics[width=0.45\textwidth,height=0.15\textwidth, angle =0]{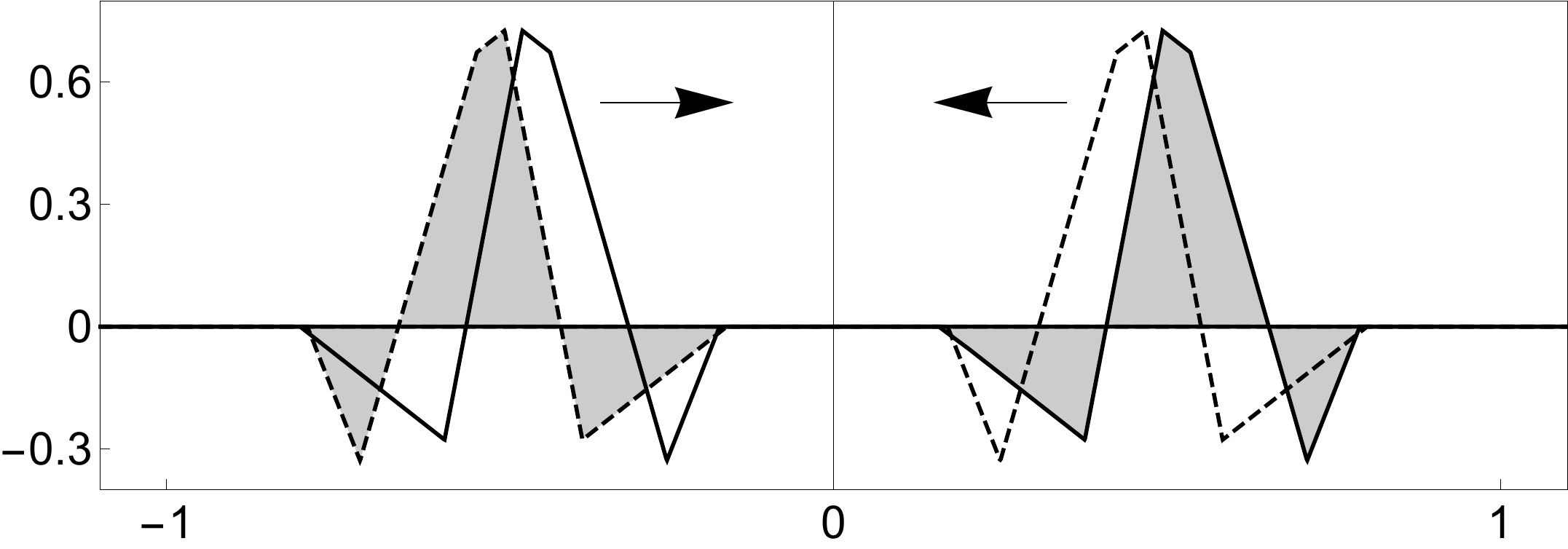}}
\caption{Another possible symmetric initial configuration of two compact oscillons parametrized by $V=0.8$, $v=0.3$, $t_0=0$ and $x_0=-0.16$.}\label{sym2}
\end{figure}

Fig.\ref{sym1} shows a symmetric initial configuration $\{\Psi^{(s)}(x),\Psi_t^{(s)}(x)\}$ involving two compact exact oscillons obtained this way. A similar antisymmetic configuration can obtained by flipping the sign of one of the oscillons.

Note that an alternative symmetric (antisymmetric) initial configuration can also be obtained by taking $u=-V$ and shifting the resulting oscillon to the right {\it i.e.} choosing $x_0<0$. The second oscillon is obtained by the  mirror reflection $x\rightarrow-x$ (and possibly the sign flip). An example of such an initial configuration is shown in  Fig.\ref{sym2}  (shadowed regions). The configuration shown in Fig.\ref{sym1} is marked in Fig.\ref{sym2} by curves without shadowing.

\subsubsection{Phase of the oscillon}

Unlike for the case of compact kinks, the scattering process of two oscillons depends on the initial distance  between them. This observation follows from the fact that the shape of the oscillons changes with time and the outcome of the scattering process depends strongly\footnote{This observation follows from our numerical investigation and it will be discussed later on in the paper.} on the  shapes of  oscillons  at the moment when their supports begin to collide. Thus the {\it phase} of the oscillon is another relevant parameter which must be taken into account in the analysis of the scattering of oscillons.
Two traveling oscillons that differ {\it exclusively} by the value of spatial translation are said to have  {\it the same phase}.
The phase of the oscillon, in its  own rest frame $S'$, is the number $\alpha\in[0,1)$, where the lowest value of $\alpha=0$ represents an oscillon configuration at $t'=0$ and  the upper limiting value $\alpha=1=T_{\rm rest}$ (for $\lambda=1$) is given by the period of the oscillon. The period of the oscillon in the laboratory reference frame $S$, in which the oscillon has a certain velocity $V$,  is given by $T_{\rm lab}=\gamma\equiv (1-V^2)^{-1/2}$ and the distance travelled during the period $T_{\rm lab}$ is $\gamma V$, thus
$
\psi(t+\gamma,x+\gamma V;v,V)=\psi(t,x;v,V).
$
The phase of the oscillon in the laboratory reference frame $S$ can be chosen again as the number $\alpha\in[0,1)$, whose  upper limit is given by $T_{\rm lab}/\gamma$. Note, that two oscillons with the same phase in two different inertial reference frames describe different field configurations. Below we describe in more detail the choice of the initial symmetric (antisymmetric) configurations containing two generalized exact oscillons.

The uniform motion of the oscillons from $t=0$ to the moment of the collision results in a variation of their individual phases or the variation of their common phase in the case of symmetric (antisymmetric) initial configurations. The variation of the phase depends on the initial distance between support of two oscillons. Since the oscillons do not interact until they supports begin to overlap one can eliminate this initial distance without any loss of generality. This can be done choosing properly the value of spatial translation.  The condition that oscillons begin to collide at $t=0$ (their supports touch each other) makes the parameter of spatial shift a function of the phase {\it i.e.} $x_0=x_0(\alpha)$.

\begin{figure}[h!]
\centering
\subfigure[]{\includegraphics[width=0.5\textwidth,height=0.3\textwidth, angle =0]{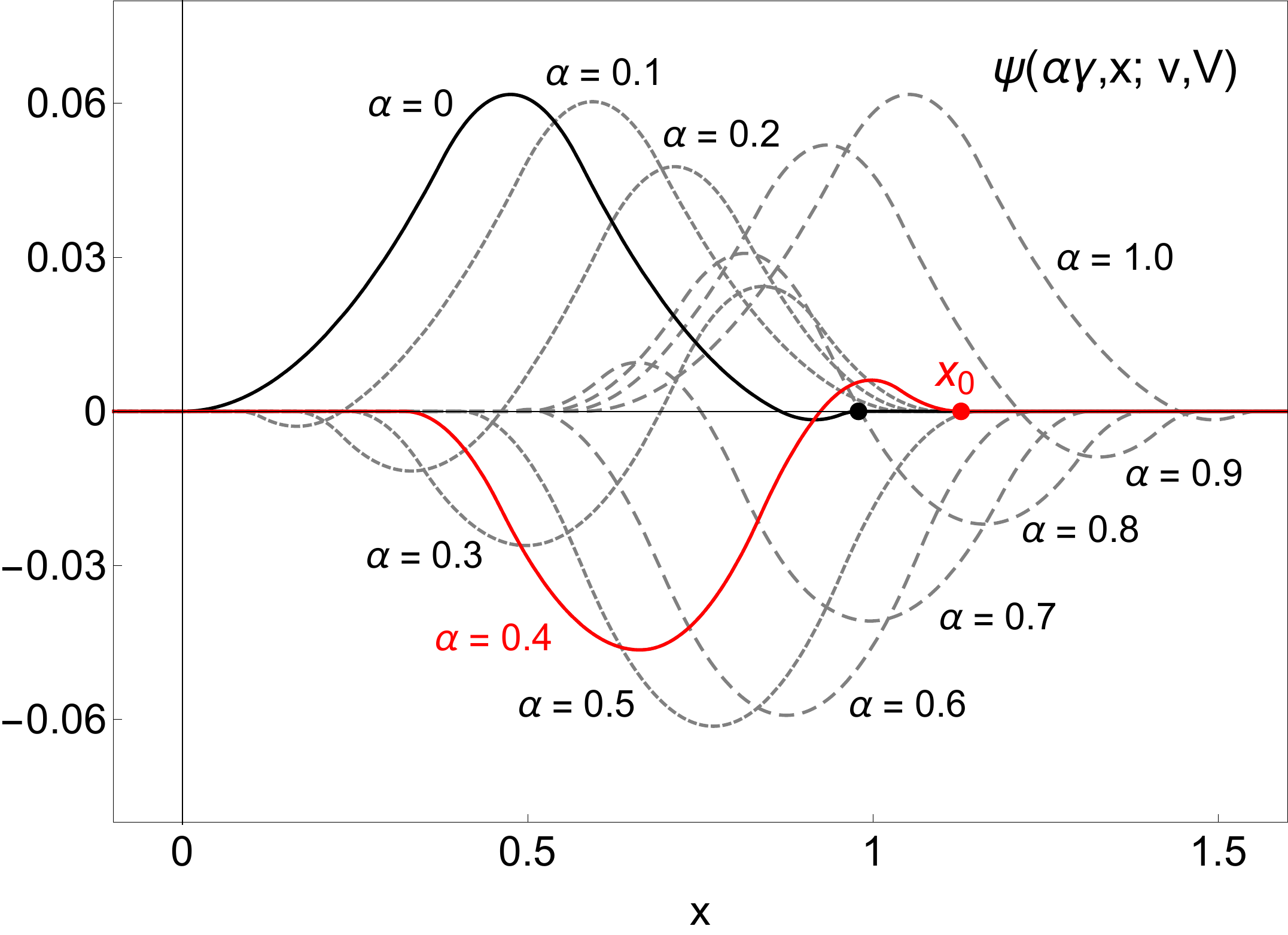}}
\hskip0.5cm
\subfigure[]{\includegraphics[width=0.4\textwidth,height=0.3\textwidth, angle =0]{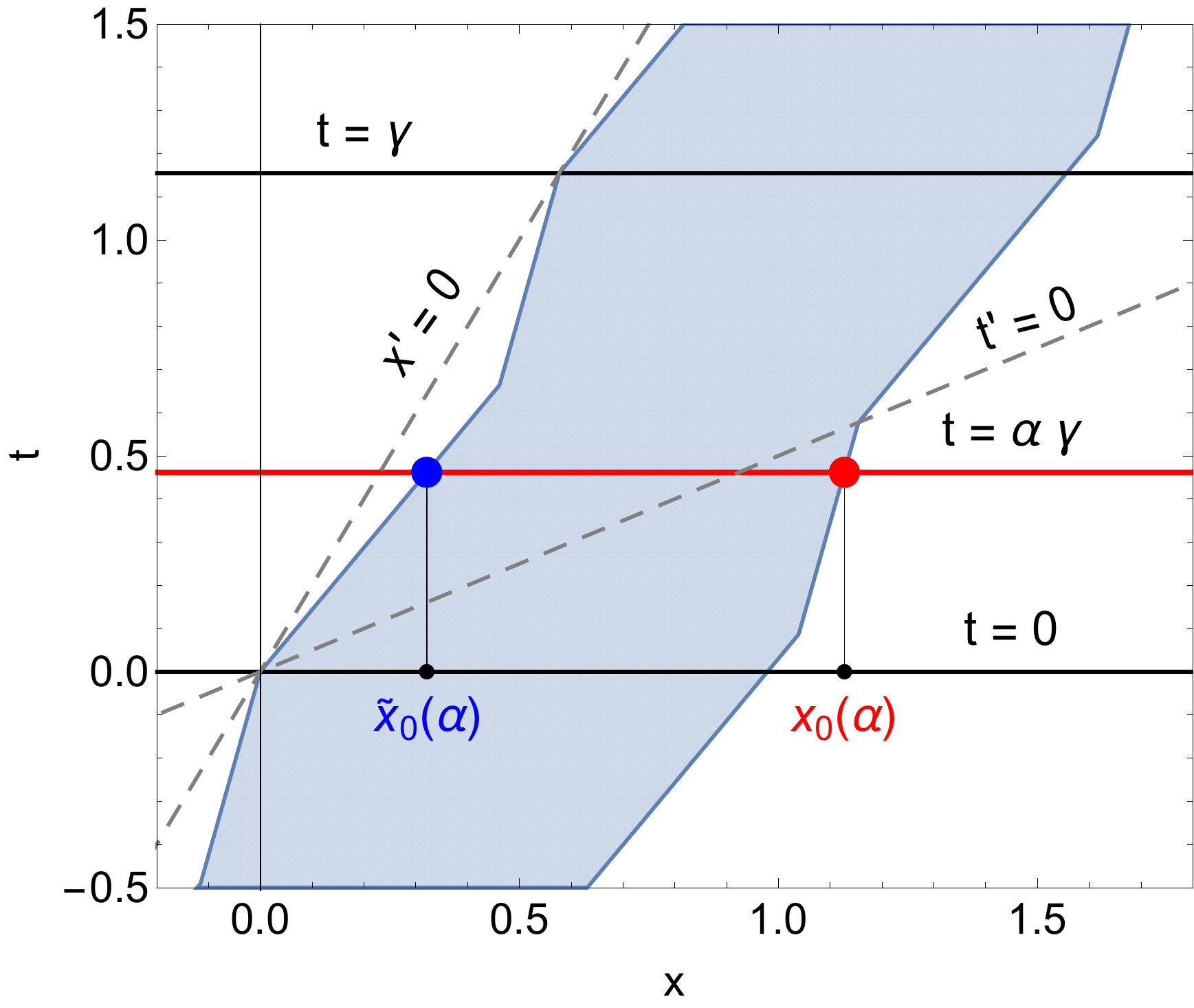}}
\caption{(a) Oscillon $\psi(\gamma\alpha,x; v,V)$ with $V=0.5$ and $v=0.3$ for sevaral values of $\alpha$. (b) A worldsheet of the oscillon and the  surface of simultaneity at $t=0.4\gamma$ in the laboratory reference frame.} \label{fazyevol}
\end{figure}

In order to set up  the phase of oscillation at $t=0$ one can make use of the translational symmetry $t\rightarrow t+t_0$ of the signum-Gordon equation. Due to the periodicity of the solution the parameter $t_0$ can be chosen  as $t_0=\alpha \gamma$. A sequence of generalized exact oscillons with different phases $\alpha$ is plotted in Fig.\ref{fazyevol}(a). The configurations $\alpha=0$ and $\alpha=1$ differ exclusively by a spatial translation which implies that they have equal phases.
In Fig.\ref{fazyevol}(b) we plot a worldsheet of the generalized exact oscillon in the laboratory reference frame in which it moves with the velocity $u=+V$. The endpoints of the oscillon are marked by $\tilde x(\alpha)$ (the left one) and $x(\alpha)$ (the right one). It is pretty clear from this diagram that the length of the oscillon 
\be
L(\alpha)=x(\alpha)-\tilde x(\alpha)\label{length}
\ee
in the laboratory reference frame is a periodic function of the time when $v\neq0$ and it is equal to $1/\gamma$ (for $\lambda=1$ ) when $v=0$.

\begin{figure}[h!]
\centering
{\includegraphics[width=0.5\textwidth,height=0.27\textwidth, angle =0]{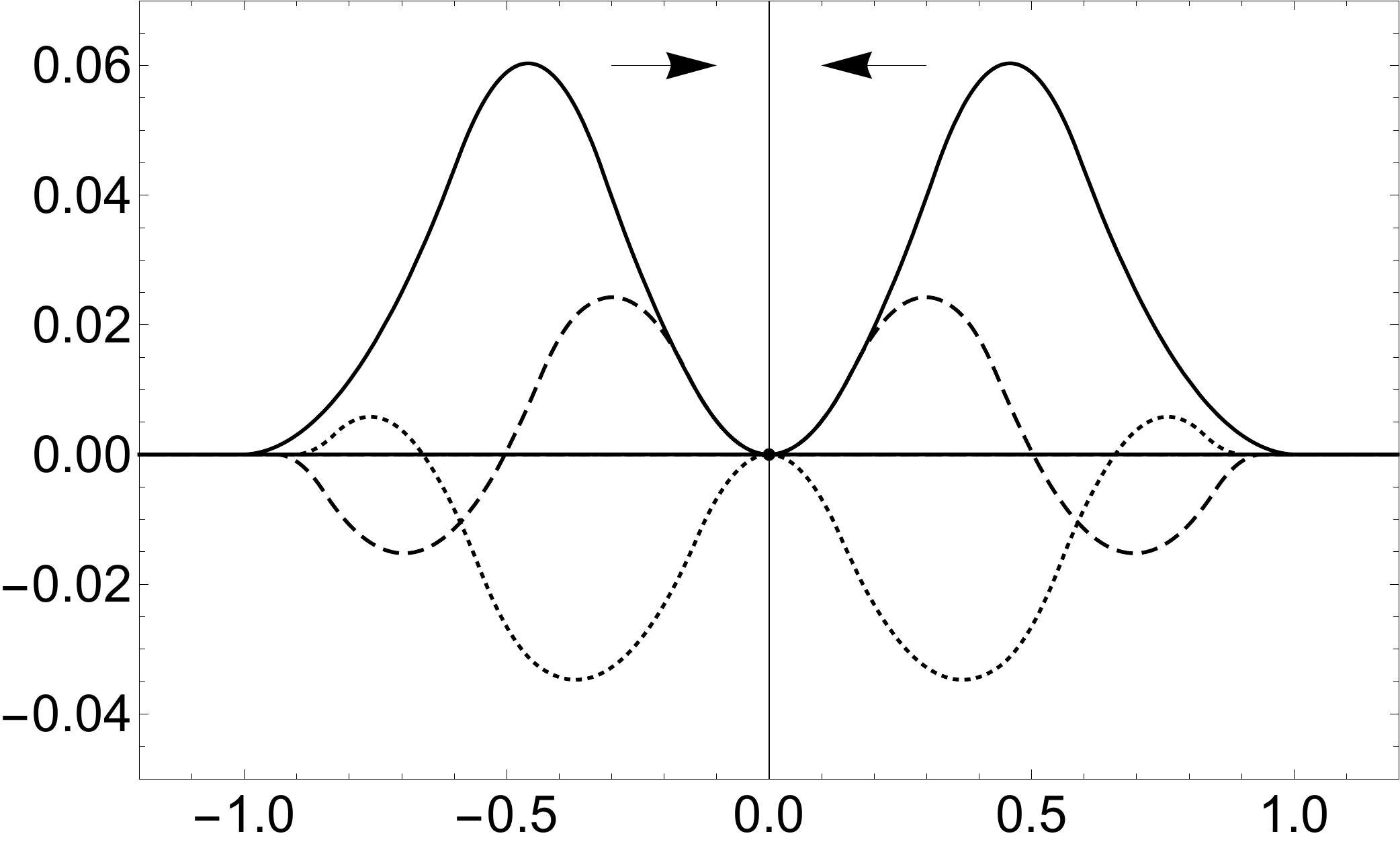}}
\caption{Symmetric initial configurations of the oscillons $\Psi^{(s)}(x)$ at $t=0$ when their supports begin to collide. The configurations correspond to $v=0.2$, $V=\frac{1}{2+v}-0.1\approx 0.3545$. The phases of oscillations are given by $\alpha=0$ (solid line), $\alpha = 0.1772$ (dashed line) and $\alpha= 0.6222$ (dotted line).}\label{3fazy}
\end{figure}

In order to get an initial configuration for the scattering process we translate the oscillon to the left by a distance  $x_0(\alpha)$. The second oscillon is obtained by the mirror reflection of the first one. Hence the initial configurations that are subject of considerations in this paper are those given by \eqref{inisymanti} with $t_0=\alpha \gamma$ and $x_0=x_0(\alpha)$. In Fig.\ref{3fazy} we plot the initial profiles of the field $\Psi^{(s)}(x)$ with different phases.

\subsubsection{Determination of the endpoints $x(\alpha)$ and $\tilde x(\alpha)$ of a oscillon}
We start with the determination of the function $x_0(\alpha)$ which describes the position of the{\it right endpoint}  of the oscillon. To obtain it is sufficient to restrict considerations to $v\ge 0$  because any non-travelling oscillon satisfies relation  $\phi(t,x;-v)=\phi(t,1-x)$. So we shall consider here  the boosts in two directions, namely $u=\pm V$, where $0\le V<1$. In all formulae containing ``$\pm$'' the upper sign corresponds to $u=+V$ and the lower one to $u=-V$.

\begin{figure}[h!]
\centering
\subfigure[$ \quad u=+V$]{\includegraphics[width=0.45\textwidth,height=0.22\textwidth, angle =0]{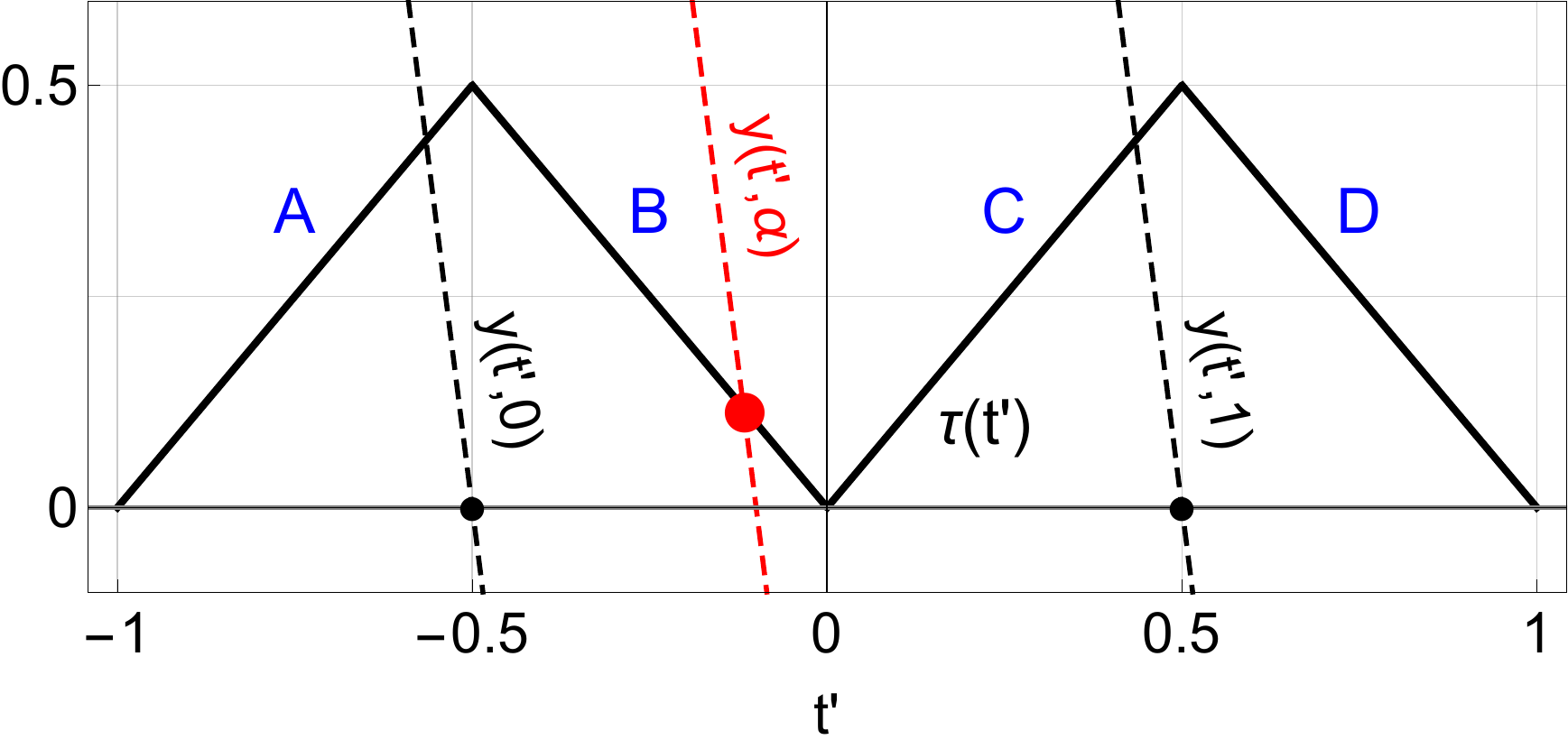}}
\subfigure[$ \quad u=-V$]{\includegraphics[width=0.45\textwidth,height=0.22\textwidth, angle =0]{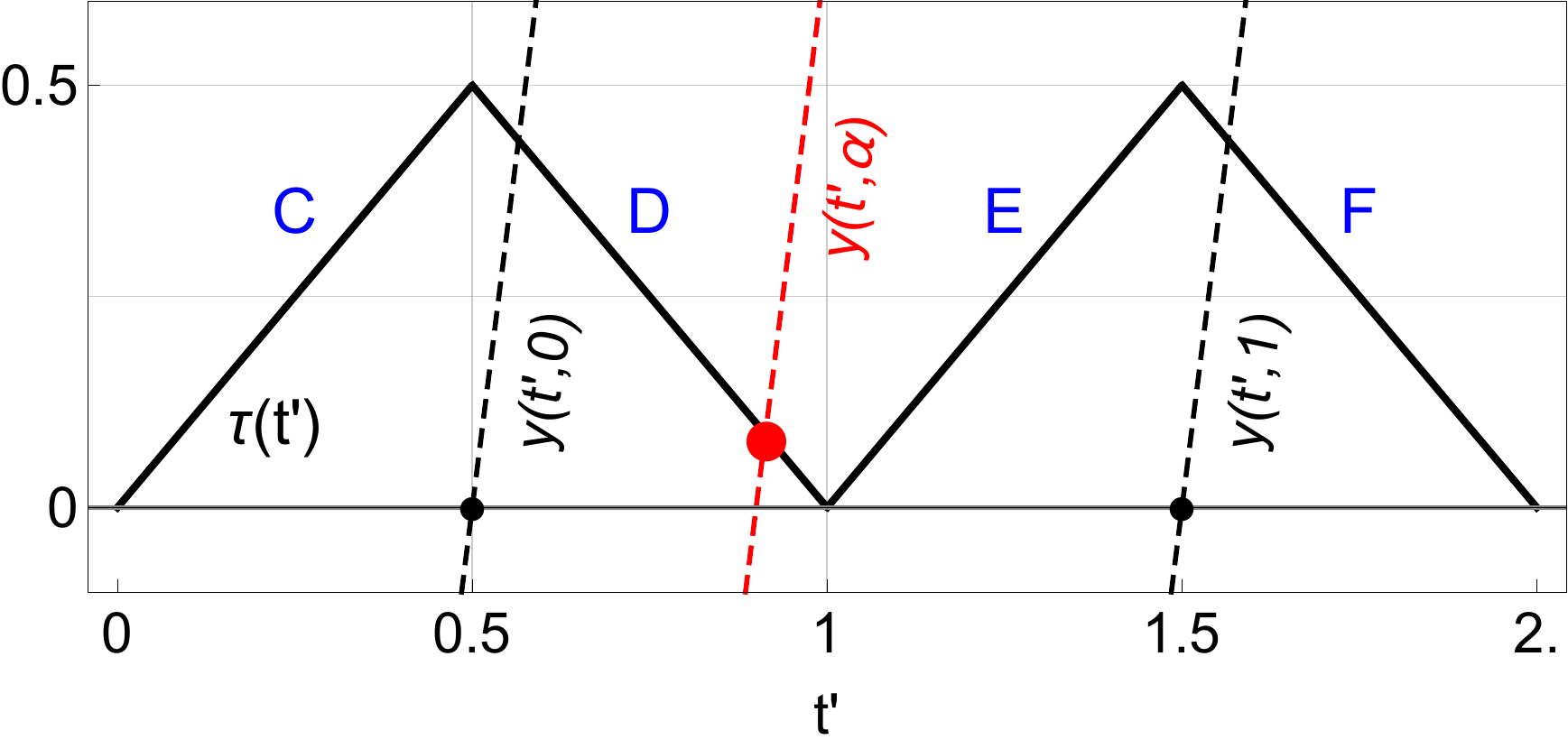}}
\caption{Functions $\tau(t')$ and $y(t',\alpha)$ for $V=0.5$ and $v=0.3$ (here  $V>V_c=\frac{10}{23}$). The middle dashed line represents solution of equation \eqref{rownaniet} with certain $\alpha$; here $\alpha=0.4$.}\label{pila}
\end{figure}

The simultaneity line $t=\alpha\gamma$ in the laboratory reference frame $S$ is described by the equation
\be
t'=\alpha\mp Vx'\label{simultaneity}
\ee
in the reference frame of the oscillon $S'$. In coordinates in $S'$ the right border of the oscillon is a worldline
given by
\be
x'=1+v\tau(t'),\label{borderosc}
\ee
where $\tau(t')$ is a function defined in \eqref{funtau}. Eliminating $x'$ from \eqref{simultaneity} and \eqref{borderosc} we find that $t'(\alpha)$ is a solution of the equation
\be
\tau(t')=y(t',\alpha)\label{rownaniet}
\ee
where
\[
y(t',\alpha)\equiv\pm\frac{1}{vV}\Big(\alpha-t'\mp V\Big)
\]
 is a straight line and $\tau(t')$ is a saw-shape function plotted in Fig.\ref{pila}. The plot shows two cases which differ by  the sign of the boost velocity $u$.

Note that due to the periodicity of the oscillon  the parameter $\alpha$ is restricted to the inteval $0\le \alpha< 1$.
This can be seen from  Fig.\ref{fazyevol}(b) (case $u=+V$), where the simultaneity lines $t=0$ and $t=\gamma$, parallel to the axis $x$, cross the left border of the oscillon at $(t',x')=(0,0)$ and $(t',x')=(1,0)$ and therefore two oscillons seen in the laboratory frame $S$ at the instants of time $t_0=0$ and $t_0=\gamma$ have the same shape. The same is true for $u=-V$ case.

Two straight lines $y(t',\alpha)$ with $\alpha=0$ and $\alpha=1$, plotted in Fig.\ref{pila}(a), cross the axis $t'$ at $t'=-V$ and $t'=1-V$. Similarly, in the case $u=-V$ plotted Fig.\ref{pila}(b), they cross the axis $t'$ at $t'=V$ for $\alpha=0$ and $t'=1+V$ for $\alpha=1$.  In order to cover the whole range of velocities $V\in[0,1)$ we shall consider the equation \eqref{rownaniet} on two intervals; namely, on the interval $-1<t'<1$ for $u=+V$ and on the interval $0<t'<2$ for $u=-V$. The saw-shape function $\tau(t')$ is given by the pieces of straight lines $\tau(t')=at'+b$, where the coefficients $(a,b)$ have different values for  different instants of time $t'$ 
\begin{eqnarray}
(a,b)=\left\{
\begin{array}{llllr}
&(1,1)\quad&{\rm for}\quad&t'\in(-1,-\frac{1}{2})&\quad(A)\nonumber\\
&(-1,0)&{\rm for}&t'\in(-\frac{1}{2},0)&(B)\nonumber\\
&(1,0)&{\rm for}&t'\in(0,\frac{1}{2})&(C)\nonumber\\
&(-1,1)&{\rm for}&t'\in(\frac{1}{2},1)&(D)\nonumber\\
&(1,-1)&{\rm for}&t'\in(1,\frac{3}{2})&(E)\nonumber\\
&(-1,2)&{\rm for}&t'\in(\frac{3}{2},2)&(F)\nonumber
\end{array}\right..
\end{eqnarray}

There are two different cases dependent on the absolute value of the boost velocity $V=|u|$. They are separated by the critical case for which the velocity has value \[V_c\equiv \frac{1}{2+v}.\]
The straight line $y(t',\alpha)$ crosses maxima of $\tau(t')$ {\it i.e.} $\tau(t'_{max})=\frac{1}{2}$ at $t'_{max}=\{-\frac{1}{2},\frac{1}{2}\}$ for $u=+V$ and at $t'_{max}=\{\frac{1}{2},1\}$  in the case $u=-V$.  It crosses the minimum $\tau(t'_{min})=0$ at $t'_{min}=0$ for $u=+V$ and at $t'_{min}=1$  for $u=-V$. The corresponding values of the parameter $\alpha$ are denoted by $\alpha^{(\pm)}_{l}$ for lower maximum, $\alpha^{(\pm)}_{u}$ for upper maximum and $\alpha^{(\pm)}_0$ for the minimum. The signs ``$\pm$'' correspond to $u=\pm V$. They have the form
\begin{eqnarray}
\begin{array}{lll}
\alpha^{(+)}_{l}=\frac{1}{2}[-1+V(2+v)],&\hskip1cm &\alpha^{(-)}_{l}=\frac{1}{2}[1-V(2+v)],\\
\alpha^{(+)}_{0}=V,& &\alpha^{(-)}_{0}=1-V,\\
\alpha^{(+)}_{u}=\frac{1}{2}[1+V(2+v)],& &\alpha^{(-)}_{u}=\frac{1}{2}[3-V(2+v)].
\end{array}
\end{eqnarray}

When the boost velocity takes the critical value $V=V_c$, the lowest value $\alpha=0$ corresponds to $\alpha_l^{(\pm)}$ and the highest value $\alpha=1$ to $\alpha_u^{(\pm)}$. Thus only two intervals $(B)$  and $(C)$ for $u=+V$ and $(D)$ and $(E)$ for $u=-V$ are relevant in this case. The interval $\binom{B}{D}$ is covered by $\alpha$ which satisfies  $0\le \alpha\le \alpha^{(\pm)}_0$ and $\binom{C}{E}$ is covered by $\alpha$ such that $\alpha^{(\pm)}_0\le \alpha<1$. The upper letter stands for the upper sign $(+)$ and the lower one for  $(-)$.

When the boost velocity is less than the critical value, $V<V_c$, then the straight line $y(t',\alpha=0)$ crosses the saw-shape chain $\tau(t')$ at a certain point given by $t'$ belonging to the interval $\binom{B}{D}$ and  $y(t',\alpha=1)$ crosses the chain $\tau(t')$ at the point with $t'$ that belongs to $\binom{D}{F}$.  Thus in this case the relevant intervals are $\binom{B}{D}$, $\binom{C}{E}$ and $\binom{D}{F}$.  The parameter $\alpha$ takes values $0\le \alpha\le \alpha_0^{(\pm)}$ in the interval $\binom{B}{D}$,  it takes values $ \alpha_0^{(\pm)}\le \alpha\le \alpha_u^{(\pm)}$ in $\binom{C}{E}$ and  values $\alpha_u^{(\pm)}\le \alpha< 1$ in $\binom{B}{F}$. Note that the intervals $\binom{B}{D}$ and $\binom{D}{F}$ are covered only partially by $\alpha$.

On the other hand, for $V>V_c$ (the case sketched in Fig.\ref{pila}) the relevant intervals are $\binom{A}{C}$, $\binom{B}{D}$ and $\binom{C}{E}$. In the interval $\binom{A}{C}$ the parameter $\alpha$ belongs to $0\le \alpha\le \alpha_l^{(\pm)}$, in $\binom{B}{D}$ it belongs to $\alpha_l^{(\pm)}\le \alpha\le \alpha_0^{(\pm)}$ and in $\binom{C}{E}$ it belongs to $\alpha_0^{(\pm)}\le \alpha< 1$. In this case the intervals $\binom{A}{C}$ and $\binom{C}{E}$ are covered only partially by $\alpha$.

The solution of the equation \eqref{rownaniet} is given by
\be
t'(\alpha)=\frac{\alpha\mp V(1+b v)}{1\pm a vV},\label{rozwiazanietlinia}
\ee
where the values of $(a,b)$ are determined by the velocity $V$ and the phase $\alpha$.  A position of the right endpoint of the oscillon, obtained from the spacetime interval, takes the value
\be
x_0(\alpha)=\sqrt{\gamma^2\alpha^2-t'(\alpha)^2+x'^2},\nonumber
\ee
where $x'$, introduced in \eqref{borderosc}, is a function of $\alpha$  given by $x'(\alpha)=1+v \Big(a t'(\alpha)+b\Big)$. After some manipulations one gets
\be
x_0(\alpha)=\frac{\gamma}{1\pm avV}\Big[(1-V^2)(1+vb)+\alpha (\pm V+v a)\Big].\label{xzero}
\ee
This results shows that $x_0$ is a linear function of $\alpha$ in the intervals in which $(a,b)$ remain constant functions of $\alpha$.

Next we determinate the function $\tilde x_0(\alpha)$ which describes the position of the {\it left endpoint}  of the oscillon. The left border of the oscillon is described by the worldline $x'=v\tau(t')$ in the rest frame of the oscillon. In order to get the function $t'(\alpha)$ one has to solve the equation
\be
\tau(t')=z(t',\alpha)\qquad \text{where} \qquad z(t',\alpha)=\pm \frac{\alpha-t'}{vV}. \nonumber
\ee
Taking $\tau(t')=a't'+b'$ we find that $t'(\alpha)$ is given by
\[
t'(\alpha)=\frac{\alpha\mp b'vV}{1\pm a'vV}.
\]
The coefficients $(a',b')$ correspond to $0\le t'\le \frac{1}{2}$ and $\frac{1}{2}\le t'<1$ and they are given by
 \begin{eqnarray}
(a',b')=\left\{
\begin{array}{llllr}
&(1,0)&{\rm for}&0\le t'\le \frac{1}{2}&(C),\nonumber\\
&(-1,1)&{\rm for}&\frac{1}{2}<t'<1&(D).\nonumber
\end{array}\right.
\end{eqnarray}
We denote the solutions of the equation $\tau(t'_{max})=\frac{1}{2}$ by $\alpha_c^{(\pm)}$. They are given by
\[
\alpha_s^{(\pm)}:=\frac{1}{2}(1\pm vV).
\]
For $0\le\alpha\le\alpha_s^{(\pm)}$ the line $z(t,\alpha)$ crosses $\tau(t')$ at $(C)$ and for $\alpha_s^{(\pm)}\le\alpha<1$ it crosses $\tau(t')$ at $(D)$.  Finally, taking similar steps for the case of the `right endpoint' we find
\be
\widetilde x_0(\alpha)=\frac{\gamma}{1\pm a'vV}\Big[(1-V^2)(vb')+\alpha (\pm V+v a')\Big]\label{xzeroleft}.
\ee

\subsubsection{Remarks on the initial configurations}
\begin{figure}[h!]
\centering
\subfigure[$ \quad \alpha=0$]{\includegraphics[width=0.45\textwidth,height=0.35\textwidth, angle =0]{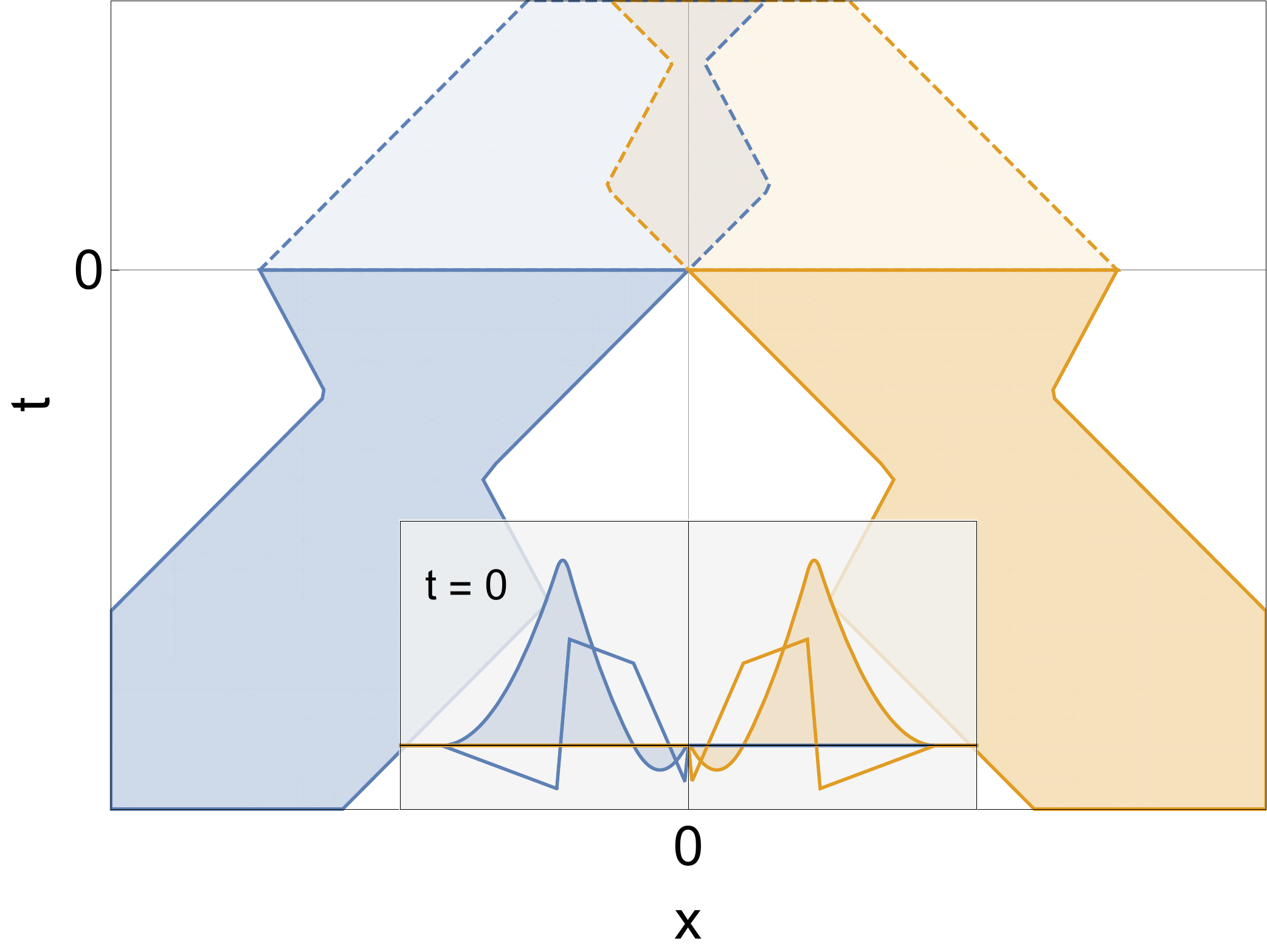}}
\subfigure[$ \quad \alpha=0.3$]{\includegraphics[width=0.45\textwidth,height=0.35\textwidth, angle =0]{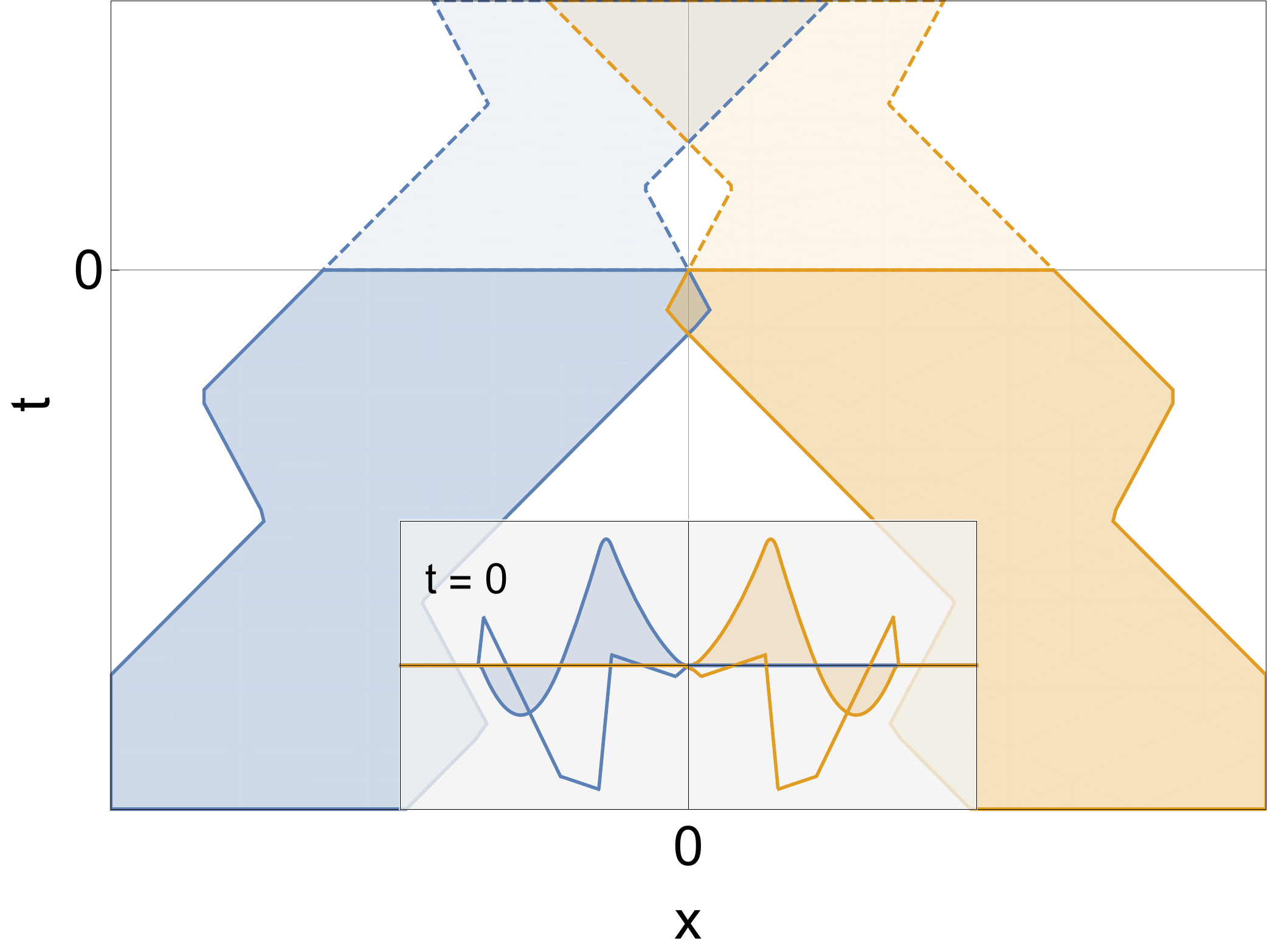}}
\caption{Symmetric configurations with $v=0.8$ and $V=0.5$ and different phases.}\label{fig:inisym}
\end{figure}

Having determined the expressions for $x_0(\alpha)$ and $\tilde x_0(\alpha)$ we can now construct arbitrary initial configurations containg generalized exact oscillons which begin to collide at $t=0$. In the case of oscillons with $|u|<v$ an additional caution is necessary. In Fig.\ref{fig:inisym} we present the worldsheets of two generalized exact oscillons 
\be
\Psi^{(s)}(t,x)=\psi(t+\alpha\gamma, x+x_0(\alpha); v, V)+\psi(t+\alpha\gamma, -x+x_0(\alpha); v, V),\nonumber
\ee
which form a symmetric configuration. The initial field configuration for the scattering process is given by $\Psi^{(s)}(x)=\Psi^{(s)}(t,x)|_{t=0}$, $\Psi_t^{(s)}(x)=\partial_t\Psi^{(s)}(t,x)|_{t=0}$ with $|u|<v$. Fig.\ref{fig:inisym}(b) shows that for some values of the phase of the colliding oscillons they would collide {\it before} the instant of time $t=0$.
This proves that the corresponding configuration $(\Psi^{(s)}(x), \Psi^{(s)}_t(x))$ cannot be obtained by approaching two travelling oscillons. A requirement that worldsheets of colliding oscillons have no intersections for $t<0$ excludes such field configurations.

Although we have paid the main attention only to the symmetric (antisymmetric) initial configurations it is quite clear that nonsymmetric configurations with vanishing total momentum can be obtained by taking colliding oscillon with different phases.
\begin{figure}[h!]
\centering
\subfigure[$ \quad \alpha_1=0,\,\, \alpha_2=0.3$]{\includegraphics[width=0.45\textwidth,height=0.35\textwidth, angle =0]{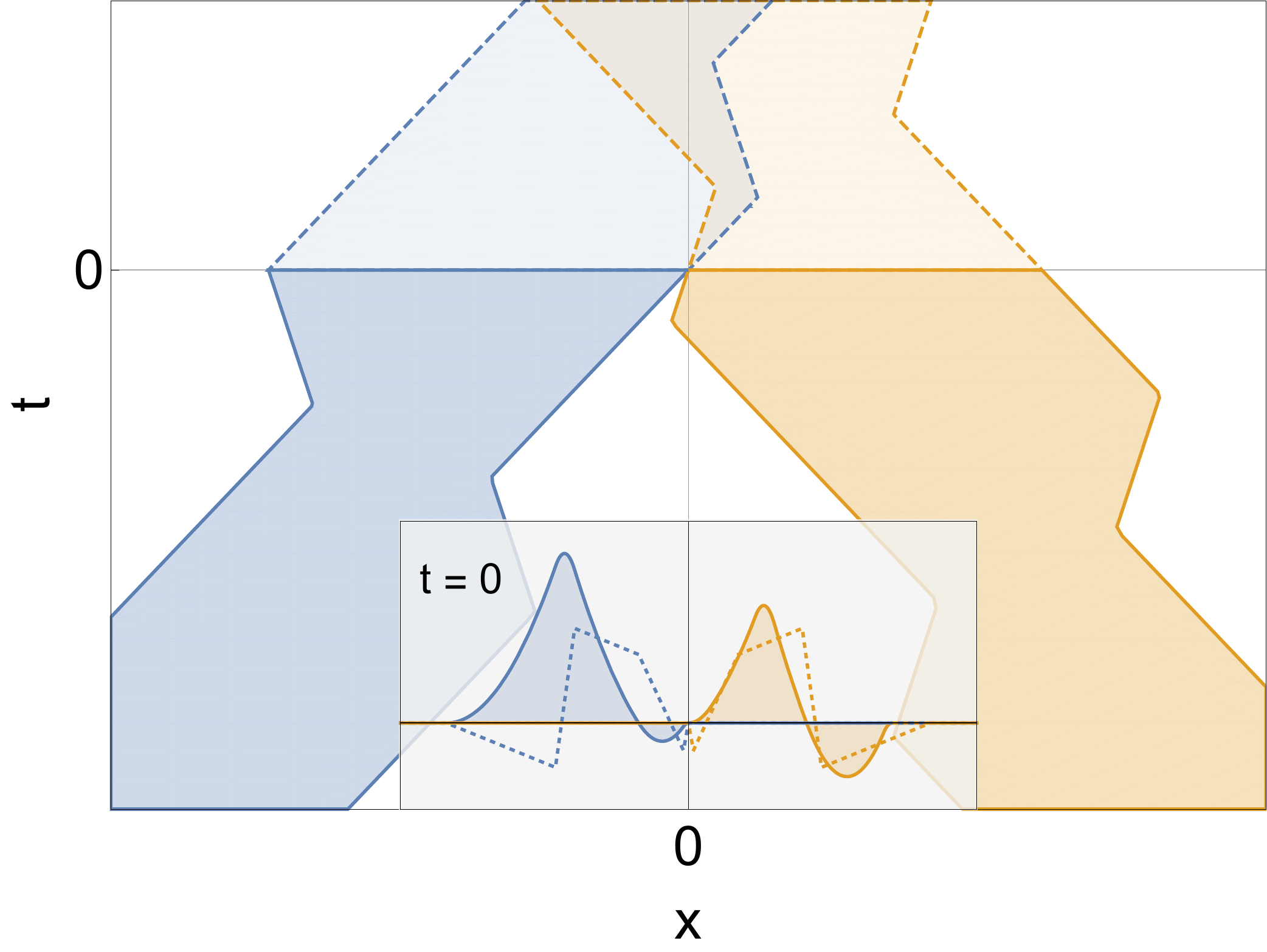}}
\subfigure[$ \quad \alpha_1=0.65,\,\, \alpha_2=0.35$]{\includegraphics[width=0.45\textwidth,height=0.35\textwidth, angle =0]{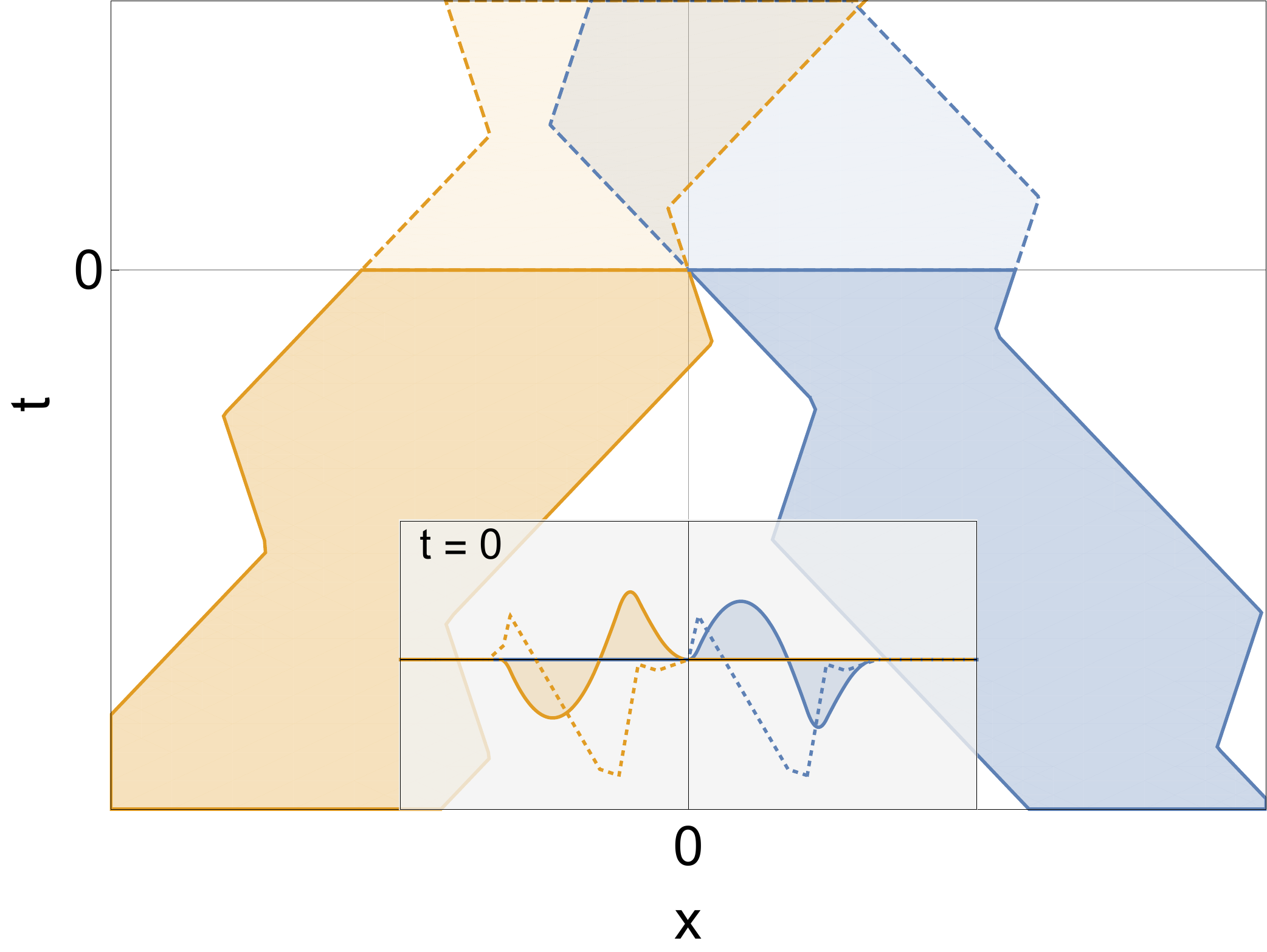}}
\caption{Generic configurations with $v=0.7$ and $V=0.5$. (a) shift by the right endpoint, (b) shift by the left endpoint.}\label{fig:genericconf}
\end{figure}
An example of two oscillons with phases $\alpha_L$ and $\alpha_R$ is given by
\be
\Psi(t,x)=\psi(t+\alpha_L\gamma, x+x_0(\alpha_L); v, V)+\psi(t+\alpha_R\gamma, -x+x_0(\alpha_R); v, V)\label{2phases}
\ee
and  shown in Fig.\ref{fig:genericconf}(a). This field configuration is obtained by shifting one of the oscillons by distance $x_0(\alpha_L)$ (placing it to the left of $x=0$) and the second one by $x_0(\alpha_R)$ and reflection $x\rightarrow-x$  (placing it to the right of $x=0$). Another possibility is shown in Fig. \ref{fig:genericconf}(b). The first oscillon, with $u=-V$, is shifted by $\tilde x_0(\alpha_R)$ (which places it to the right of $x=0$) and the second oscillon, with $u=+V$, is shifted by $x_0(\alpha_L)$ (which places it to  to the left of $x=0$). In this case
\be
\Psi(t,x)=\psi(t+\alpha_R\gamma, x+\tilde x_0(\alpha_R); v, -V)+\psi(t+\alpha_L\gamma, x+x_0(\alpha_L); v, V).\nonumber
\ee
It is quite clear from  diagrams in Fig.\ref{fig:genericconf} that there exist a set of  phases for which the worldsheets of oscillons overlap for $t<0$.

\subsection{Antisymmetric configurations}
 \label{sec:antisym-config}
In this section we present some numerical results for the scattering of two oscillons which form an antisymmetric initial configuration of the signum-Gordon field. We have chosen antisymmetric initial data to be discussed here first because the result of the scattering of oscillons in such a case is not as complex as that  for  symmetric configurations.  The fact that initial configuration is antisymmetric implies that 
\be
\Psi^{(a)}(t,x;v,V)|_{x=0}=0\label{bcant}
\ee
for any instant of time $t$. This condition expresses the vanishing of  the net force exercised on a degree of freedom at $x=0$  ({\it e.g.} ball, pendulum in a discretized mechanical realization of the model). In other words, the only effect of interaction between left and right oscillon is a fixing of the value of scalar field $\chi=0$ at $x=0$. Thus the evolution of the system in regions $x<0$ and $x>0$  effectively splits into two independent problems containing an initial oscillon and the boundary condition \eqref{bcant}.

\begin{figure}[]
\centering
\subfigure[$\quad V=0.5$]{\includegraphics[]{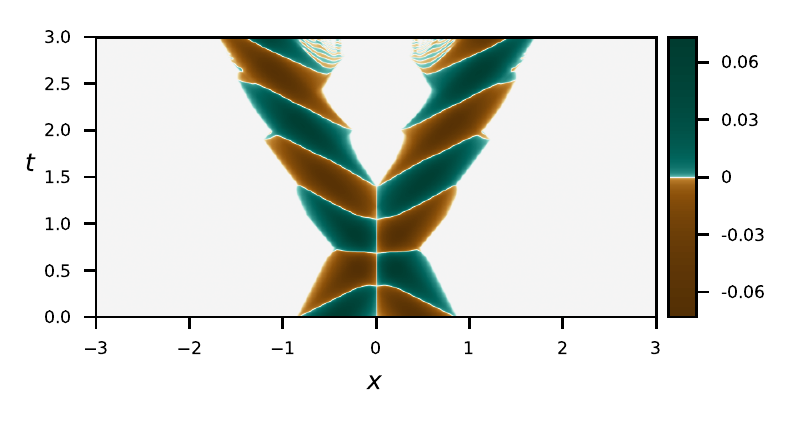}}
\subfigure[$\quad V=0.6$]{\includegraphics[]{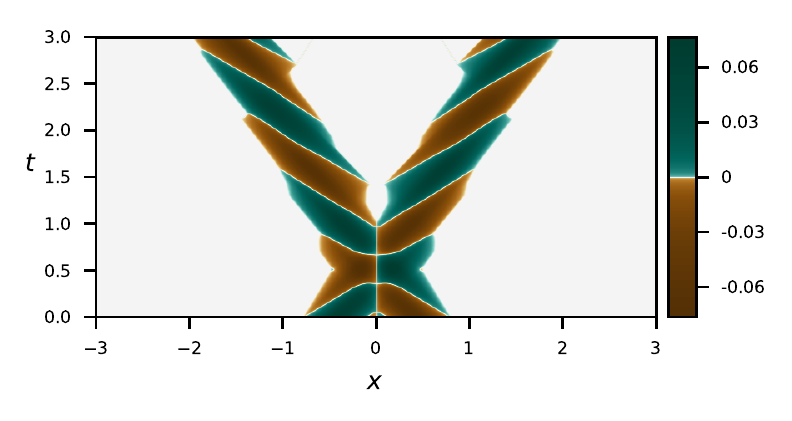}}
\subfigure[$\quad V=0.7$]{\includegraphics[]{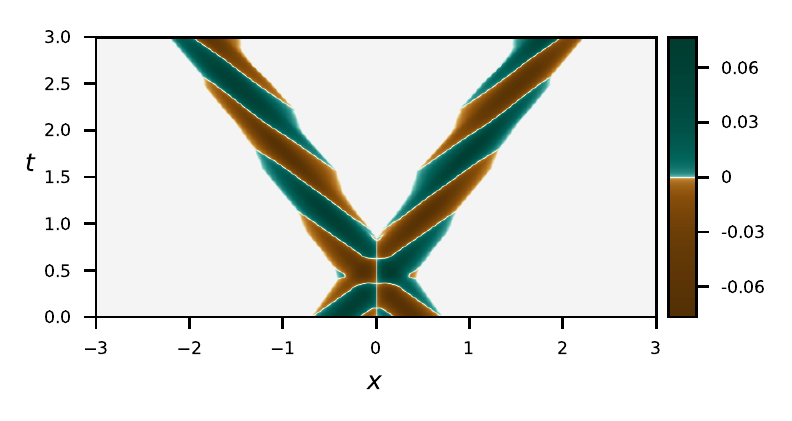}}
\subfigure[$\quad V=0.8$]{\includegraphics[]{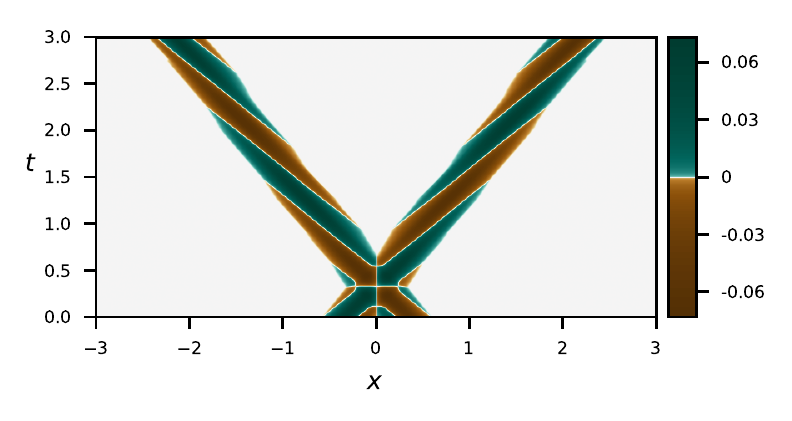}}
\subfigure[$\quad V=0.95$]{\includegraphics[]{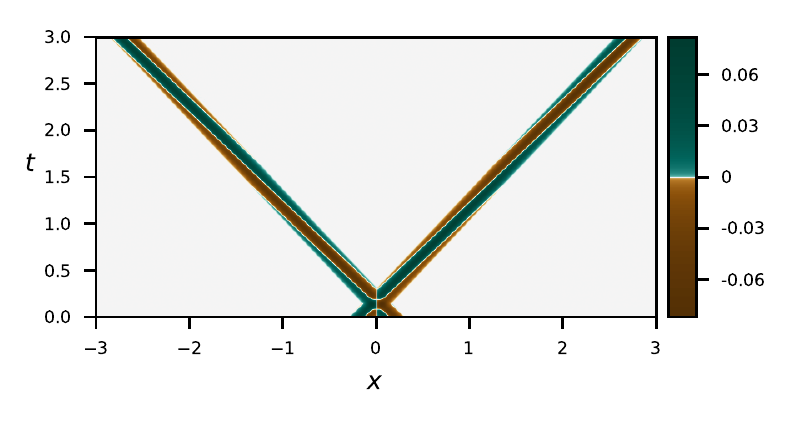}}
\subfigure[$\quad V=0.97$]{\includegraphics[]{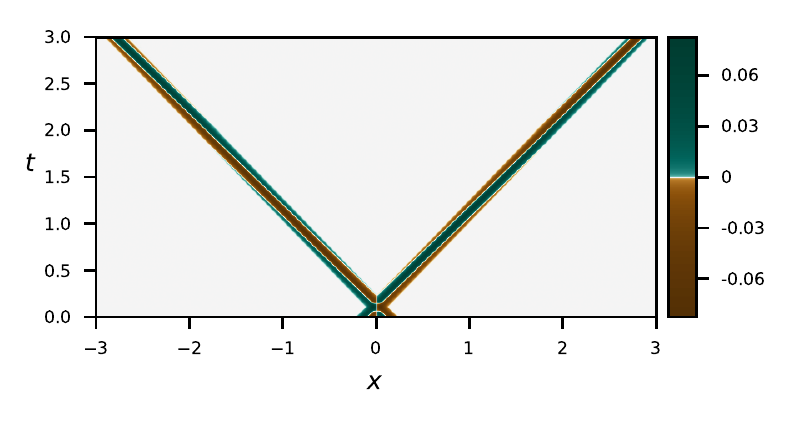}}\caption{The dependence of the scattering of oscillons (antisymmetric configuration) on their initial speed $V$. The initial configuration contains oscillons with phase $\alpha=0$ and no zig-zag motion of their border.}
\label{fig:scatvelanti}
\end{figure}

In our numerical study we have evolved an antisymmetric configuration without  assuming the condition $\chi=0$ at $x=0$. Looking at results presented in Fig.\ref{fig:scatvelanti} we clearly see that the condition \eqref{bcant} is satisfied. This effect manifests itself in the presence of vertical white segments at diagrams which are located at $x=0$. Another important observation is the absence of radiation in the central region of the diagrams independently on the value of the initial speed of oscillons. The sources of radiation  generated in this process are irregular borders of two outgoing oscillons, see Fig.\ref{fig:scatvelanti}(a).

One can  note that irregularities of the border are more likely to appear for small velocities $V$ of colliding oscillons than for the larger ones.
Moreover, in spite of being irregular the outgoing oscillons radiate significantly less than it would be expected for strongly perturbed oscillons. In fact, the outgoing oscillons with irregular borders belong to a more general class of oscillons.  We discuss this subject in more deatil in Sec.\ref{nonuniform}.

\begin{figure}[h!]
\centering
\subfigure[$\quad \alpha=0.1$]{\includegraphics[]{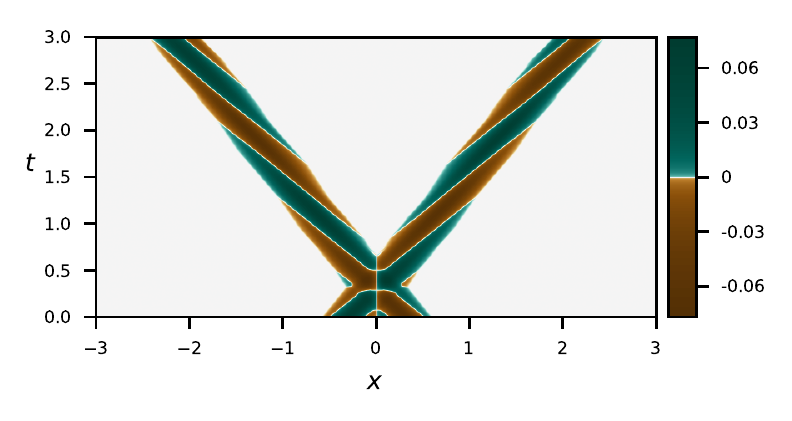}}
\subfigure[$\quad \alpha=0.5$]{\includegraphics[]{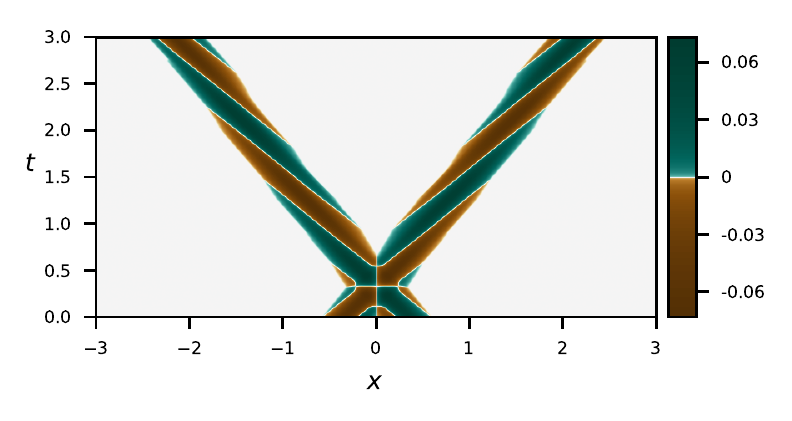}}
\subfigure[$\quad \alpha=0.75$]{\includegraphics[]{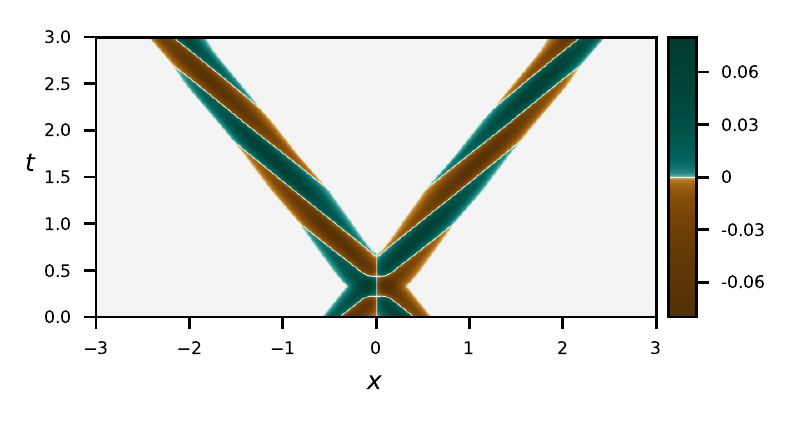}}
\subfigure[$\quad \alpha=0.9$]{\includegraphics[]{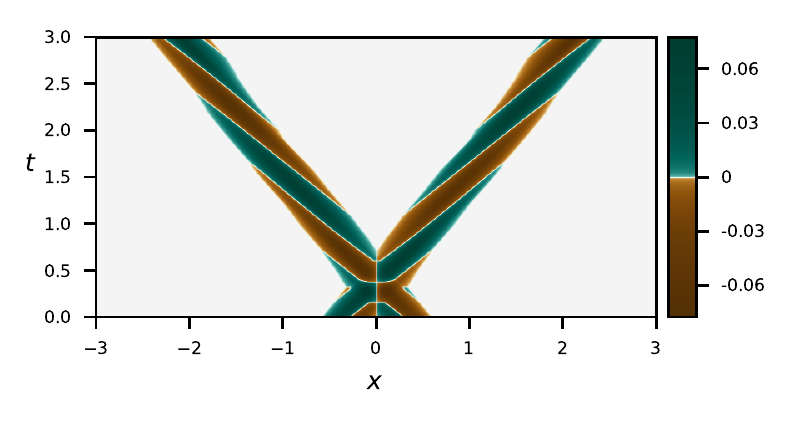}}
\caption{The scattering of oscillons (antisymmetric configuration) as a function of their phase $\alpha$. The initial configuration contains oscillons with speed $V=0.8$ and no zig-zag motion of their border.}
\label{fig:scatphaseanti}
\end{figure}

Another interesting question is the dependence of the scattering process on the phase $\alpha$ that characterizes the initial configuration. In Fig.\ref{fig:scatphaseanti}
we present the results of the scattering at $V=0.8$ for four values of the phase. The figures show that the evolution of the initial configurations does not depend much on the value of the parameter $\alpha$.

Another parameter that the incoming oscillons depend on is the speed $v$ of the oscillon border in its rest frame.
\begin{figure}[h!]
\centering
\subfigure[$\quad v=0.47$]{\includegraphics[]{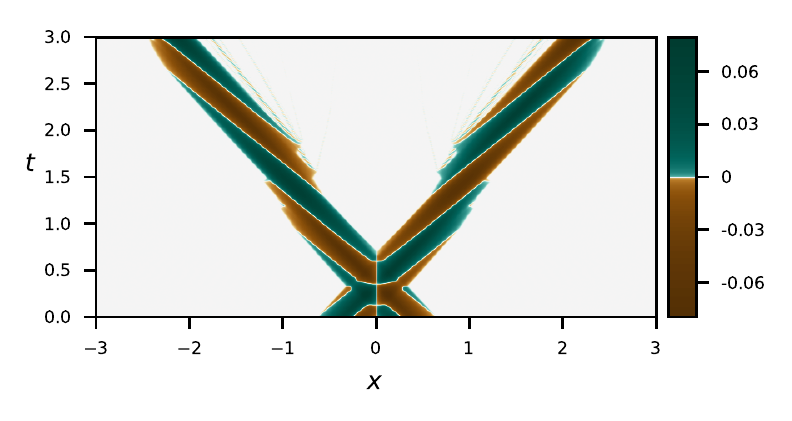}}
\subfigure[$\quad v=0.82$]{\includegraphics[]{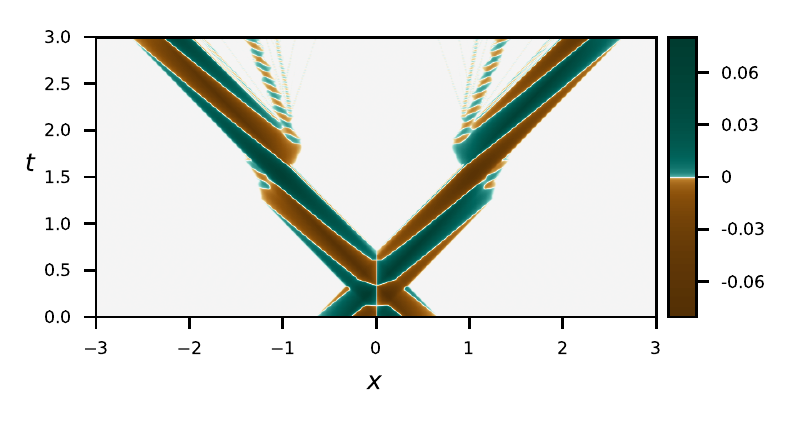}}
\caption{The scattering of oscillons (antisymmetric configuration) in dependence on parameter $v$. The initial configuration contains oscillons with speed $V=0.8$ and $\alpha=0$.}
\label{fig:scatvanti}
\end{figure}
Our numerical studies have shows that irregularities of borders of two outgoing oscillons  increase with increasing of the value of parameter $v$ which determines the swaying motion of the incoming oscillons. in Fig.\ref{fig:scatvanti} we present the results of the scattering of an anti-symmetric configuration of oscillons with speed $V=0.8$, phase $\alpha=0$ and speeds of the border $v=0.47$ and $v=0.82$. The figures show that as the borders of outgoing oscillons became more and more irregular the number and size of the oscillons emitted from these borders increases.

It is quite notable that the radiation generated in the process of the scattering of anti-symmetric configurations is emitted only from the border irregularities of the outgoing oscillons.  Such oscillons can be interpreted as (strongly) perturbed exact oscillons of the signum-Gordon model. The surplus of their energy is converted into small oscillons which are sent away from the irregularities.

\subsection{Symmetric configurations}
\label{sec:symmetric-configurations}

\subsubsection{Overview of the numerical results}
In this section we present a general overview of the numerical results obtained in the scattering of solitons described by symmetric initial configurations parametrized by the boost velocity $V$ and initial phase $\alpha$. For better transparency we present first only the cases with $v=0$. They are sufficient to provide us a basic feeling about the  typical results of the scattering. In section \ref{subs} we discuss further examples of oscillons with non-zero velocity of the border  $v$.

\begin{figure}[h!]
\centering
{\includegraphics[width=0.45\textwidth,height=0.3\textwidth, angle =0]{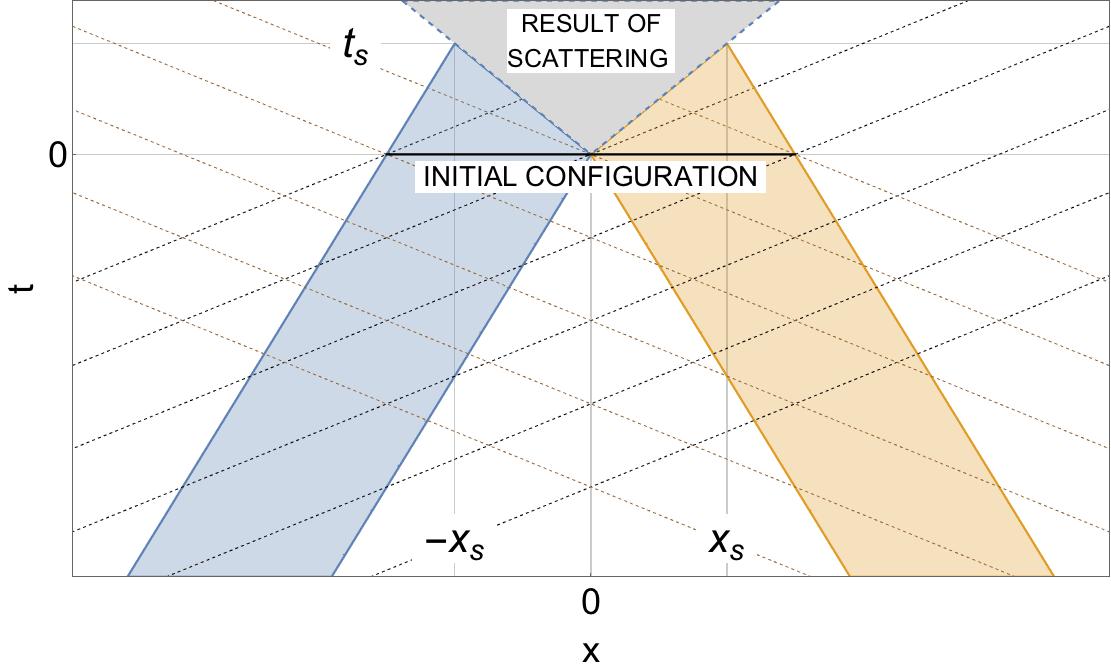}}
\caption{Worldsheets of two incoming oscillons for $v=0$. The oscillons  are unperturbed inside two triangular regions above the line $t=0$. The zeros of the oscillon  correspond to the intersections of its worldsheet with a family of dashed lines.}
\label{fig:wstegi1}
\end{figure}

A typical situation with $v=0$ is shown in Fig.\ref{fig:wstegi1}. Two incoming exact oscillons form the initial configuration at $t=0$. This configuration is adopted as an input for our numerical simulations.

\begin{figure}[h!]
\centering
\subfigure[$\quad V=0.5$]{\includegraphics[]{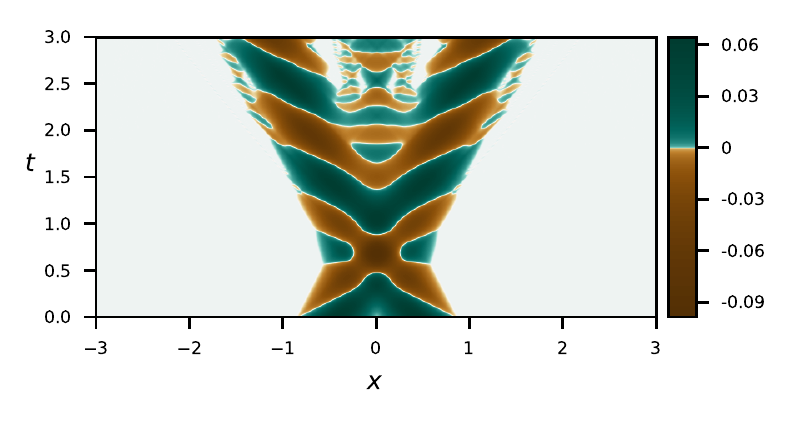}}
\subfigure[$\quad V=0.6$]{\includegraphics[]{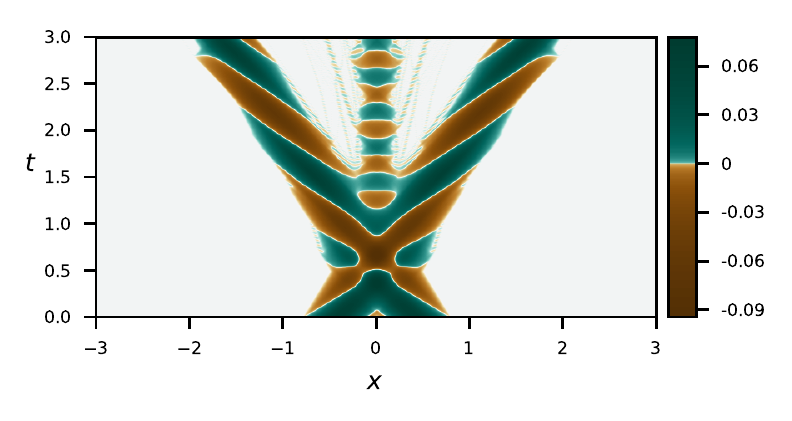}}
\subfigure[$\quad V=0.7$]{\includegraphics[]{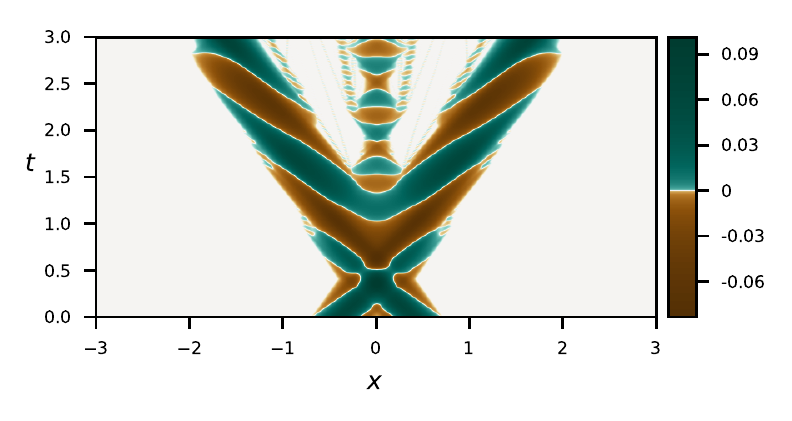}}
\subfigure[$\quad V=0.8$]{\includegraphics[]{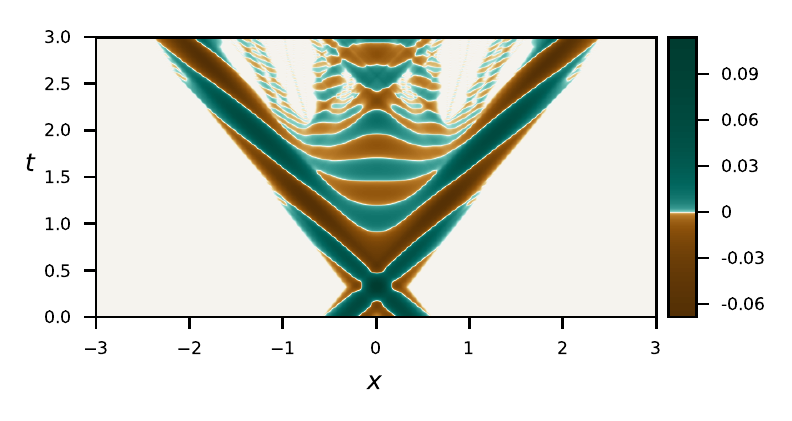}}
\subfigure[$\quad V=0.95$]{\includegraphics[]{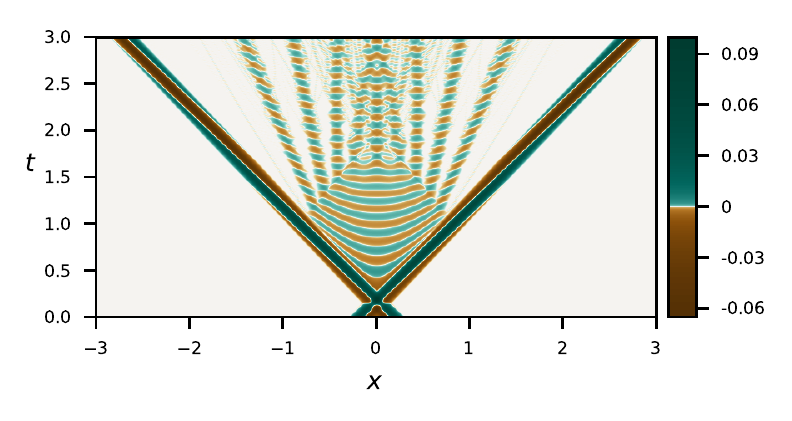}}
\subfigure[$\quad V=0.97$]{\includegraphics[]{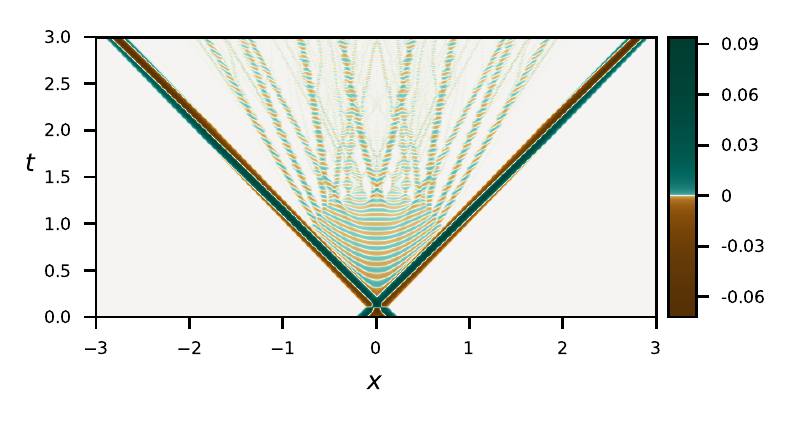}}
\caption{The scatterings of oscillons (symmetric configuration) as a function of their initial speed $V$. The initial configurations contain oscillons with phase $\alpha=0$ and no zig-zag motion of their border $v=0$.}
\label{fig:scatvel}
\end{figure}

We have performed many numerical simulations for distinct velocities and phases of scattered oscillons.  In Fig.\ref{fig:scatvel} we plot the scattering process of two  exact oscillons with $v=0$ and $\alpha=0$ for six values of boost velocity $V$.
All initial configurations contain two exact oscillons whose supports  touch each other at $t=0$. Numerics shows that the result of collision strongly depends on the velocity $V$ of colliding oscillons. All six diagrams contain two main oscillon-like objects (they perhaps can be identified with perturbed oscillons) that emerge shortly after collision.   They move with the velocity which is almost equal to velocity of colliding oscillons. The emerging oscillon-like objects obtained in the process of the scattering are significantly less regular when initial velocities of oscillons are small. For higher velocities the main outgoing oscillons are quite regular. Such field configuration are very close to the exact (generalized) oscillons. In the rest of the paper we call them quasi-oscillons.  A fundamental difference between diagrams is visible in their central region where radiation appears in the form of jets.  Looking more carefully at the radiation we see that it contains certain structures  that strongly resemble oscillons. They look a bit like perturbed oscillons. Note also that the emerging oscillons obtained for small velocities $V$ emit oscillon-like objects directly from their irregular border.  A presence of radiation in the scattering process reflects the non-integrable character of the signum-Gordon model.

\begin{figure}[h!]
\centering
\subfigure[$\quad \alpha=0$]{\includegraphics[]{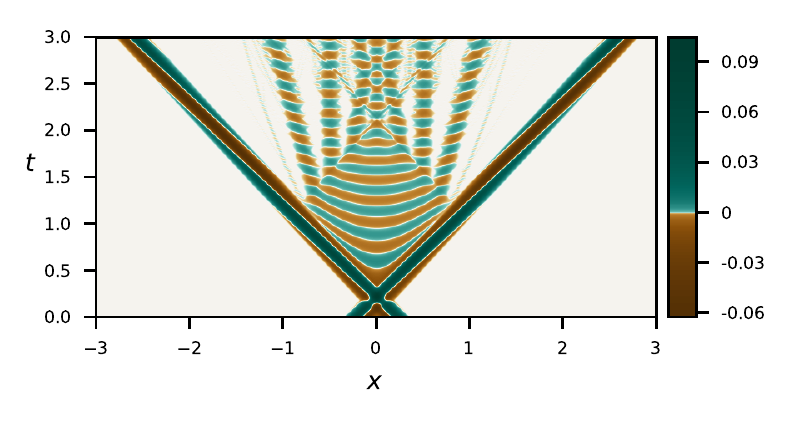}}
\subfigure[$\quad \alpha=0.25$]{\includegraphics[]{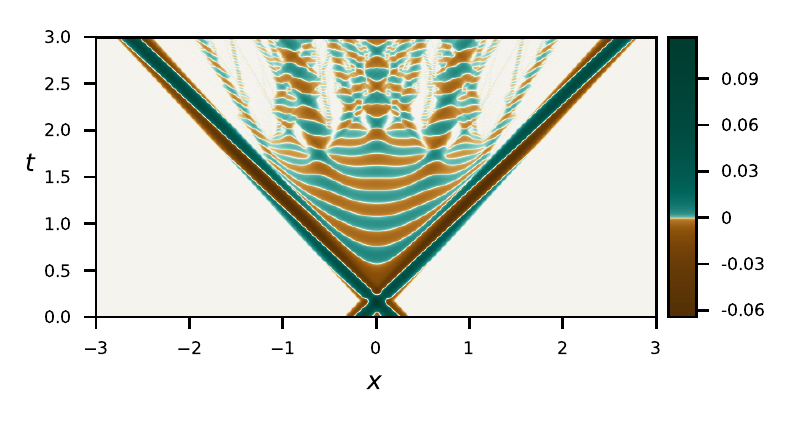}}
\subfigure[$\quad \alpha=0.84$]{\includegraphics[]{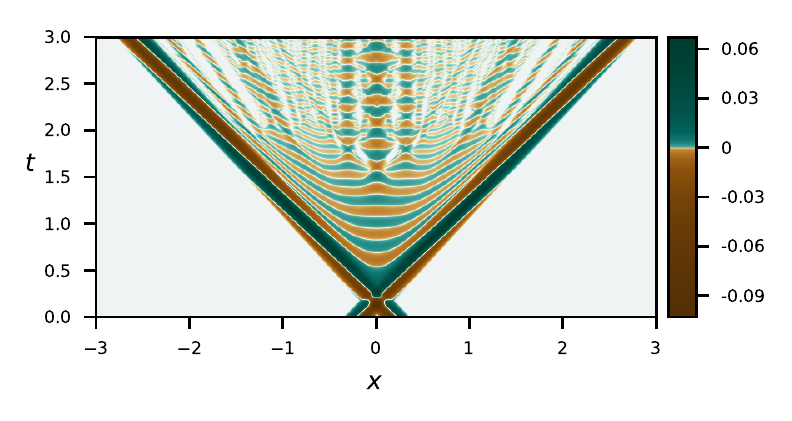}}
\subfigure[$\quad \alpha=0.93$]{\includegraphics[]{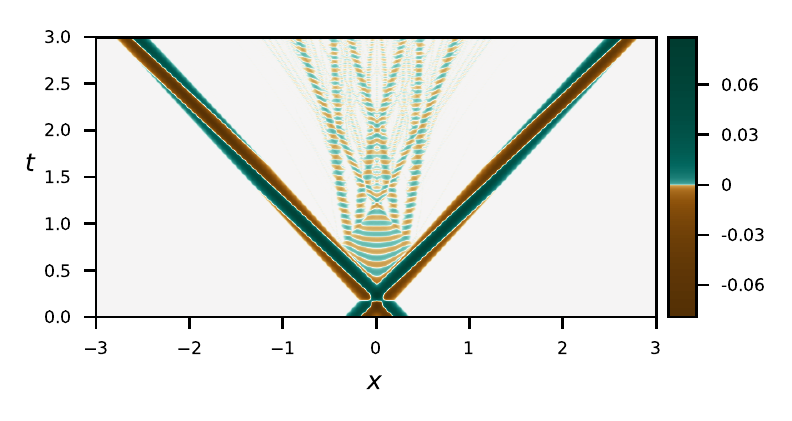}}
\caption{The scattering process for initial symmetric configurations containing oscillons with speed $V=0.93$ and no zig-zag motion of their border and different initial phases $\alpha$. }
\label{fig:scatphase}
\end{figure}

In Fig.\ref{fig:scatphase} we present the numerical results of scattering of two oscillons with $V=0.93$, $v=0$ for four initial phases $\alpha=0$, $\alpha=0.25$ , $\alpha=0.84$ and $\alpha=0.93$. The figure demonstrates that the initial phase of the colliding oscillons is indeed a relevant scattering parameter. The form of the jets in each  subfigure of Fig.\ref{fig:scatphase}  is clearly different. This difference clearly shows that the result of the scattering process  is very sensitive to the value of the phase $\alpha$.

Looking at figures Fig.\ref{fig:scatvel} and Fig.\ref{fig:scatphase} we see that in the first stage of the scattering the interacting oscillons exist on support that shrinks from its initial size $2L$ to a certain minimal size $2L_{\rm min}$ where $L=\sqrt{1-V^2}$ is the size of the oscillon in the laboratory reference frame. This process takes some time $t_s$. For $t>t_s$ we observe emerging of two main oscillons and the appearance of waves of energy identified with  radiation. In order to evaluate the time $t_s$ we assume that the left (right) border of the oscillon, which initially moves with velocity $u=V$ ($u=-V$), moves freely with the velocity that the oscillon has in the laboratory reference frame until it hits the future light cone of the event $(0,0)$; see Fig.\ref{fig:wstegi1}. This assumption allows to determine the event $P_s$ with the coordinates $(t_s, x_s)$ in the laboratory reference frame. 
\begin{figure}[h!]
\centering
\subfigure[]{\includegraphics[width=0.4\textwidth,height=0.25\textwidth, angle =0]{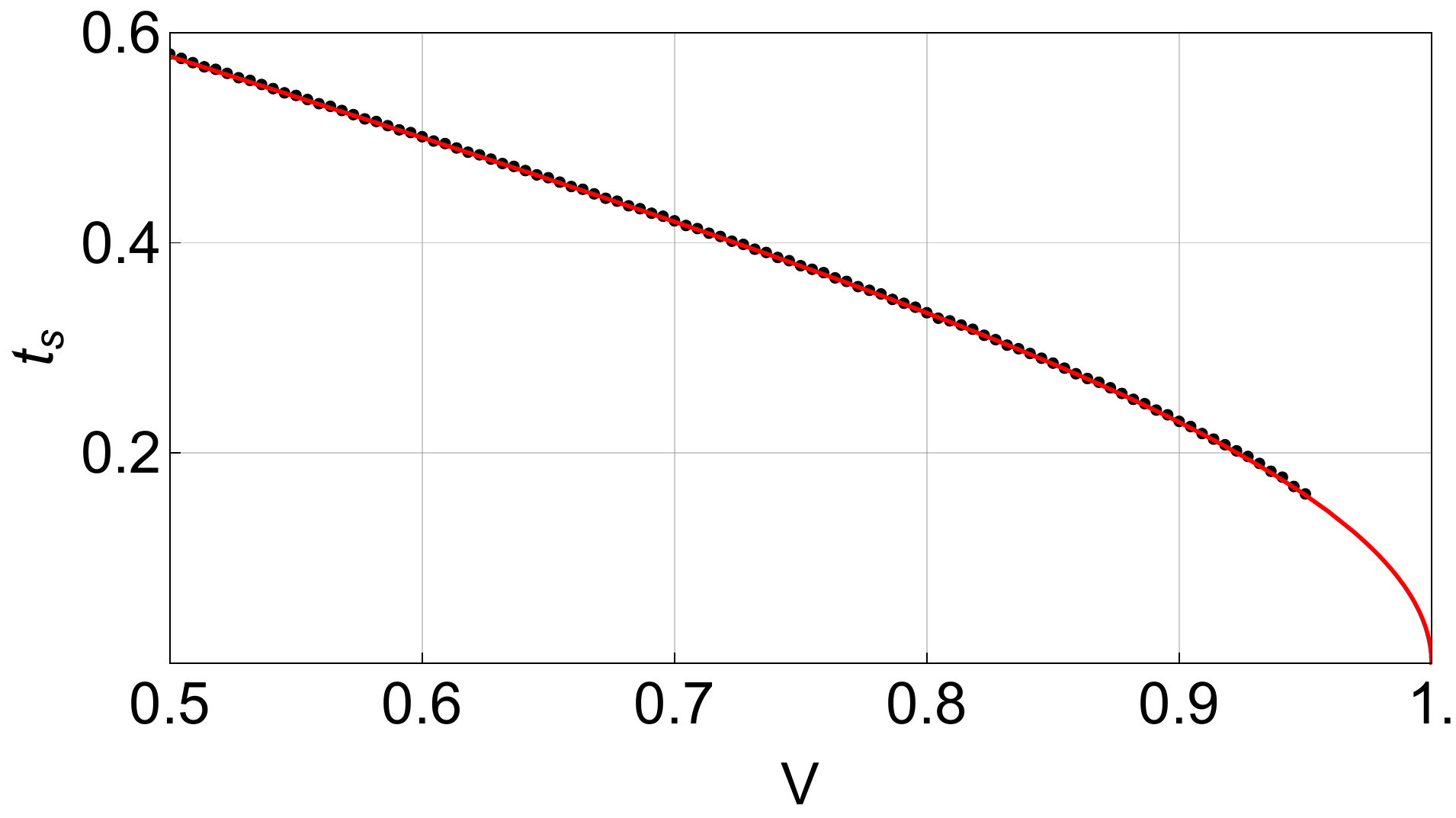}}
\hskip 0.7cm
\subfigure[]{\includegraphics[width=0.4\textwidth,height=0.25\textwidth, angle =0]{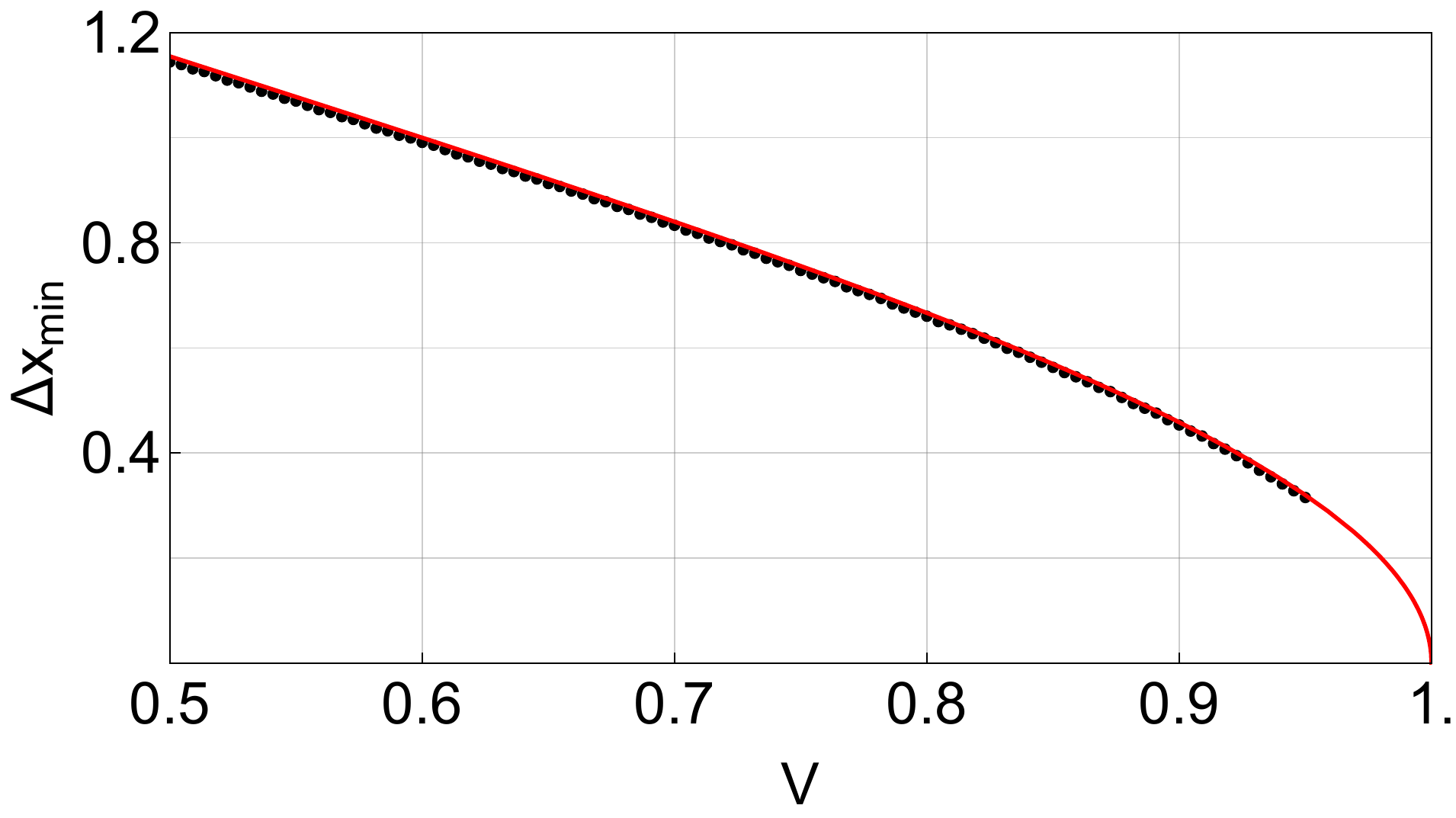}}
\caption{The cases of different velocities $V$. (a) Characteristic time $t_s$ and (b) the minimum size $\Delta x_{\rm min}$. Dots represent numerical data whereas solid curves stand for expressions $t_s=\sqrt{\frac{1-V}{1+V}}$ and $\Delta x_{\rm min}=2t_s$.}
\label{fig:timesize}
\end{figure}
Thus the left border of the left oscillon is given by $x=-\sqrt{1-V^2}+Vt$ and it intersects the light cone line $x=-t$ at
\be
t_s=\sqrt{\frac{1-V}{1+V}}=-x_s.\label{shrink}
\ee
The formula \eqref{shrink} is valid exclusively for $v=0$.  Due to the symmetry of the initial configuration we see that the minimal support size of the scattering oscillons can be estimated by the expression $\Delta x_{\rm min}=2\sqrt{\frac{1-V}{1+V}}$. In Fig.\ref{fig:timesize} we show numerical data (dots) and analytical curves (solid lines) representing the characteristic time of the scattering $t_s$ and the minimum size of the oscillon configuration $\Delta x_{\rm min}$ at $t_s$.

%
%

\subsubsection{High velocities - formation of shock waves}

Next we discuss in more detail some of our numerical results. First we note that for small velocities the numerical solution is very irregular. Looking at figures Fig.\ref{fig:scatvel}(a)-(c) we clearly see a formation of a strongly perturbed oscillon centred at $x=0$.  This perturbed oscillon is certainly unstable and it radiates out smaller oscillon-like objects. The situation changes for scatterings at high velocities as then the numerical solution has a more regular pattern.
In particular, when velocity of incident exact oscillons is close to unity  we observe another interesting solution.
An example of such a  solution is presented in Fig.\ref{fig:scatvel}(e). The solution presented there was obtained for the scattering of two exact oscillons with initial phases $\alpha=0$, speeds $V=0.93$ and no swaying motion {\it i.e.} for $v=0$. A very characteristic feature of this numerical solution is a presence of regular waves that are localized in a diamond-like shape region on the Minkowski diagram. Such waves emerge shortly after the collision  and eventually decay into a sequence of oscillon-like structures. It turns out that the nature of these waves can be understood in terms of the so-called {\it shock waves} that are exact solutions of the signum-Gordon model  reported in \cite{scalling}.

A shock wave is a particular solution of the signum-Gordon model with two discontinuities that propagate with the speed of light. Here, however, we do not observe such wavefronts  due to the presence of two oscillons that move with subluminal velocities. The collapse of the wave suggests that our numerical solution is only an approximation to the exact shock wave which would exist for arbitrary times $t>0$.

We can check the hypothesis as to the nature of our wave solution comparing its zeros to the zeros of the analytical shock wave. According to \cite{scalling}
a shock wave solution belongs to the class of the signum-Gordon solutions described by $\phi(t,x)=\theta(-z)W(z)$, where $z=\frac{1}{4}(x^2-t^2)$. The function $W(z)$ obeys the ordinary equation $zW''(z)+W'(z)={\rm sgn}(W(z))$ and it consists of infinitely many partial solutions $W_k(z)$, $k\in{\mathbb Z}$ matched at points $z=-a_k$. Each partial solution satisfies the equation $zW_k''(z)+W_k'(z)=(-1)^k$ and the conditions $W_k(-a_k)=0=W_{k+1}(-a_k)$ and $W'_k(-a_k)=W'_{k+1}(-a_k)$. So, it can be written in the form
\[
W_k(z)=(-1)^k\left(z+a_k+b_k\ln\frac{|z|}{a_k}\right),
\]
where
\be
\frac{b_{k+1}}{a_k}=2-\frac{b_k}{a_k},\qquad {\rm and}\qquad\frac{a_{k+1}}{a_k}=1+\frac{b_{k+1}}{a_k}\ln \frac{a_{k+1}}{a_k}.\label{rekurencja}
\ee
\begin{figure}[h!]
\centering
\subfigure[]{\includegraphics[width=0.45\textwidth,height=0.23\textwidth, angle =0]{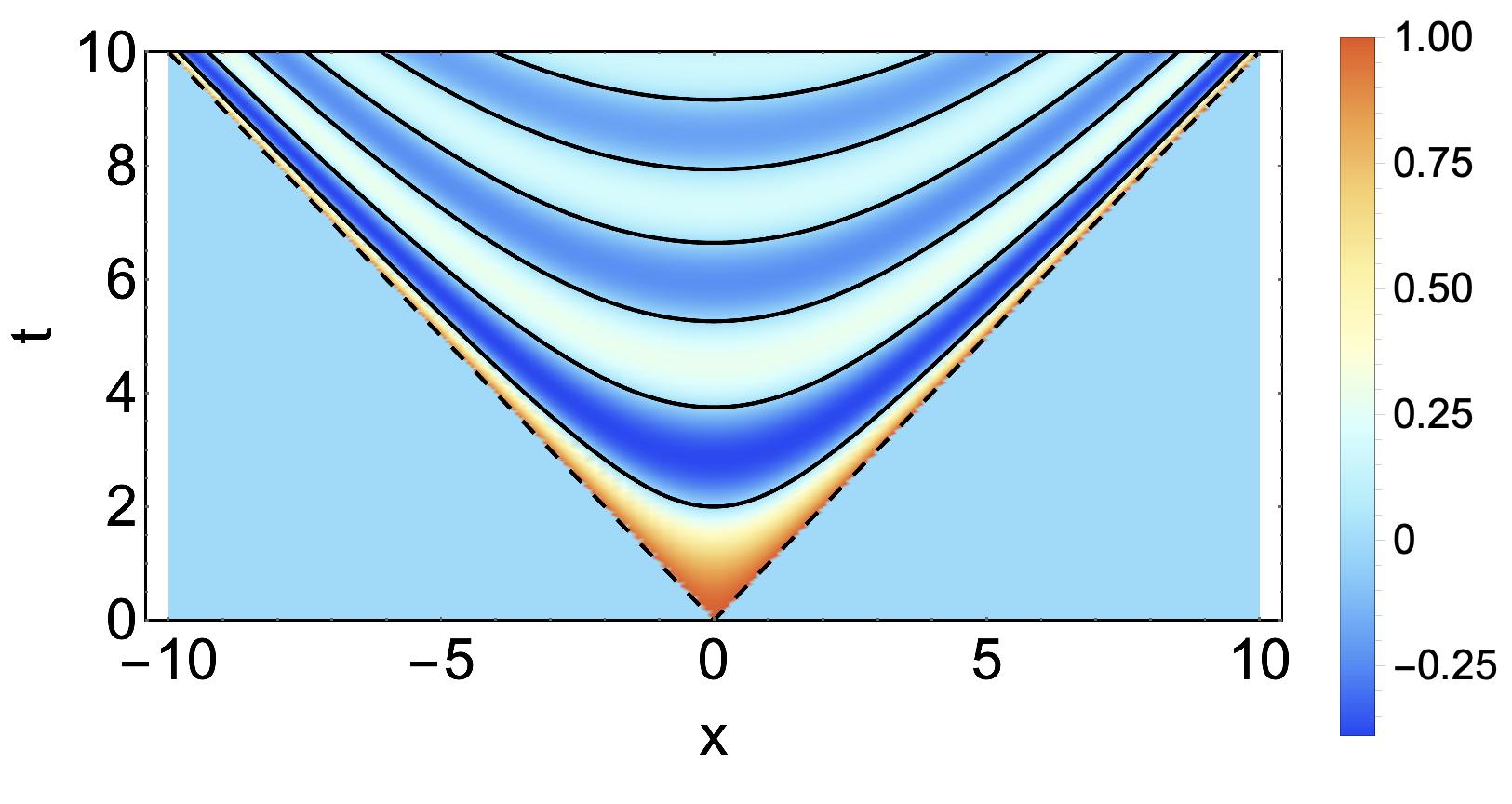}}
\hskip0.5cm
\subfigure[]{\includegraphics[width=0.45\textwidth,height=0.23\textwidth, angle =0]{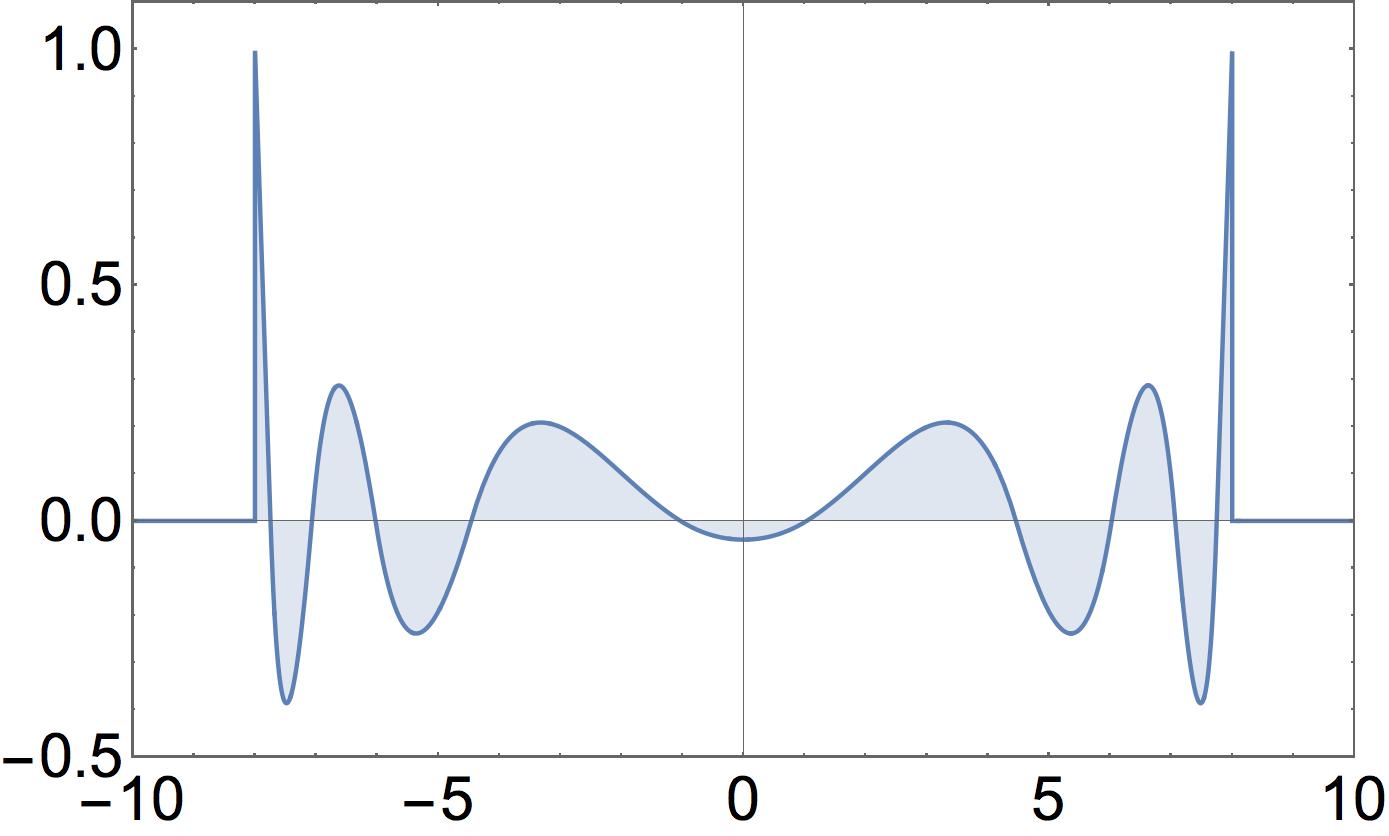}}
\caption{The exact shock wave solution for $a_0=1$. (a) The solution on the space-time diagram  and (b) the snapshot of the solution for $t=8.0$.}
\label{fig:shock}
\end{figure}

Note that we must have $b_0=0$ in order to avoid the singularity of the logarithm at $z=0$. The first zero $a_0$ is a free parameter which determines values of all other constants via recurrence relations \eqref{rekurencja}. In particular, one gets  $b_1=2a_0$. Denoting $\alpha_{k+1}:=\frac{1}{2}\frac{b_{k+1}}{a_k}$ and $y_{k+1}:=\frac{a_{k+1}}{a_k}$ one gets relations \eqref{rekurencja} in the form
\be
\alpha_{k+1}=1-\frac{\alpha_k}{y_k},\qquad{\rm and}\qquad y_{k+1}=1+2\alpha_{k+1}\ln y_{k+1}.\label{rekurencja2}
\ee
Note that $\alpha_1=1$.
\begin{figure}[h!]
\centering
{\includegraphics[width=0.75\textwidth,height=0.38\textwidth]{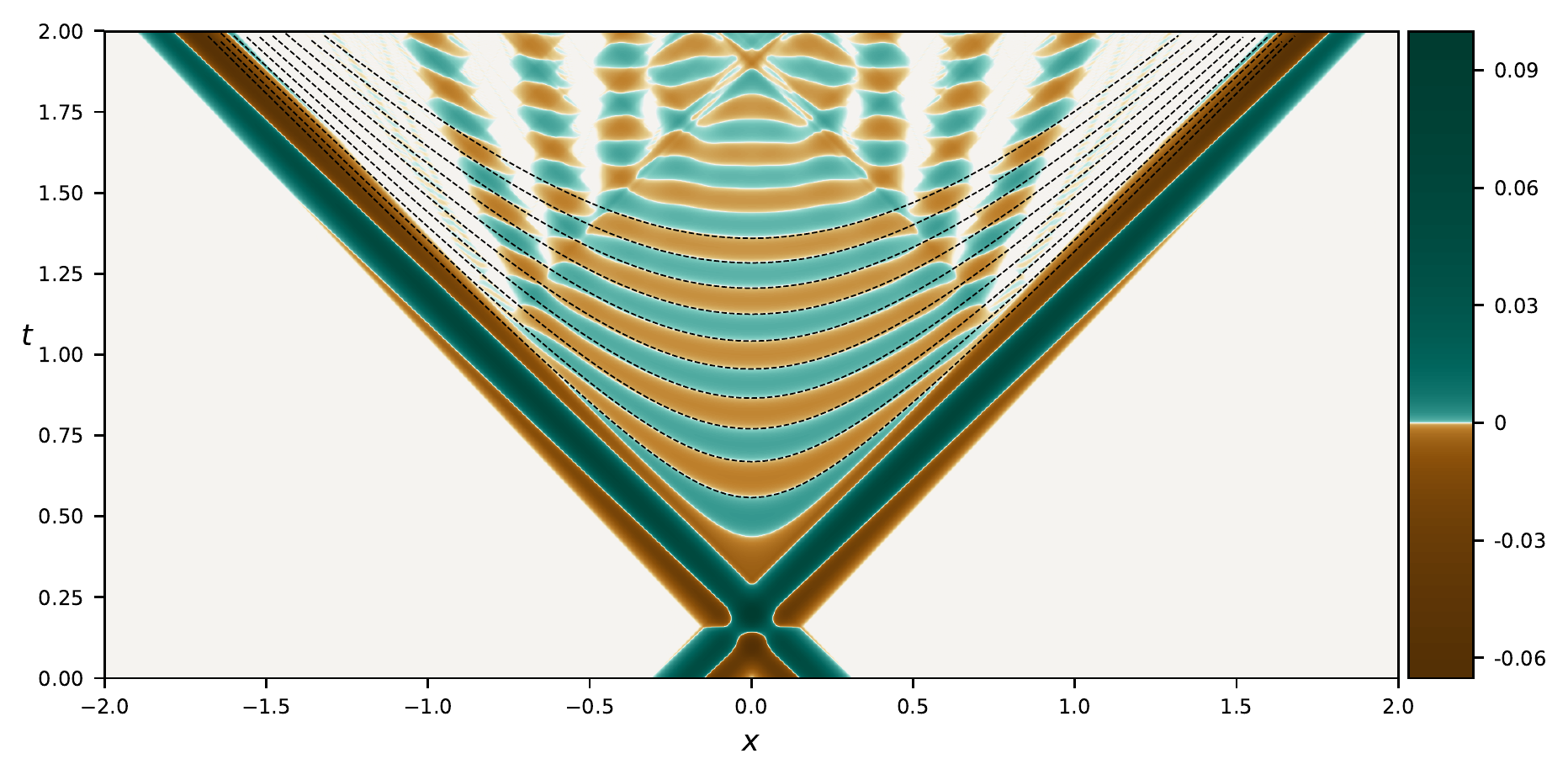}}
\caption{Scattering of two exact oscillons for symmetric field configuration. The incident oscillons move with the speed {$V=0.95$} in the laboratory reference frame. The diamond-shaped structure in the middle resembles the shock wave solution. The fit of ten first zeros was obtained for $a_0=0.00541$ and $T_0=0.282$.}\label{fig-sym-V=0.95}
\end{figure}
Furthermore, it follows from \eqref{rekurencja2} that $\alpha_{k+1}$ is determined by $a_k$ and $b_k$.  Solving numerically the second equation of \eqref{rekurencja2} one gets $y_{k+1}$ and so $a_{k+1}$ and $b_{k+1}$ can be determined too. Thus we note that the zeros of the field $\phi(t,x)$ are localized on the hyperbolas
\begin{equation}
	x_k(t)=\pm\sqrt{t^2-4a_k},
\end{equation}
at the spacetime diagram. A few first such zeros are sketched  in Fig.\ref{fig:shock}(a).

Having determined the analytical form of the positions of zeros we can now fit the zeros to the numerical data presented in Fig.\ref{fig-sym-V=0.95}.
First we note that the  formation of the waves in the process of the scattering of two oscillons  does not start at $t=0$ but at some further instant of time $T_0$. For this reason the fitting requires, besides $a_0$, another parameter: a time translation $T_0$.  Thus $x_k(t)=\pm\sqrt{(t-T_0)^2-4a_k}$. The trajectories of the zeros of the exact shock wave solution reproduce very well the trajectories of our numerical solution. This supports our hypothesis about the shock wave character of the regular oscillations localized in the diamond-shape region of Fig.\ref{fig-sym-V=0.95}. Clearly, the numerical shock-like wave solutions are similar to the exact ones only in a limited region of the Minkowski diagram. Outside of this region they break and produce oscillon-like structures.

\subsubsection{Vanishing of the radiation} \label{sec:sym-rad-vanish}
\begin{figure}[h!]
\centering
\subfigure[$\quad \alpha=0.414$]{\includegraphics[]{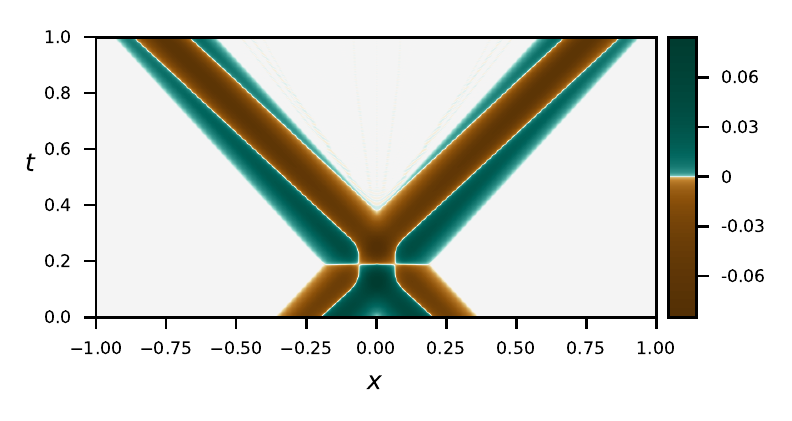}}
\subfigure[$\quad \alpha=0.914$]{\includegraphics[]{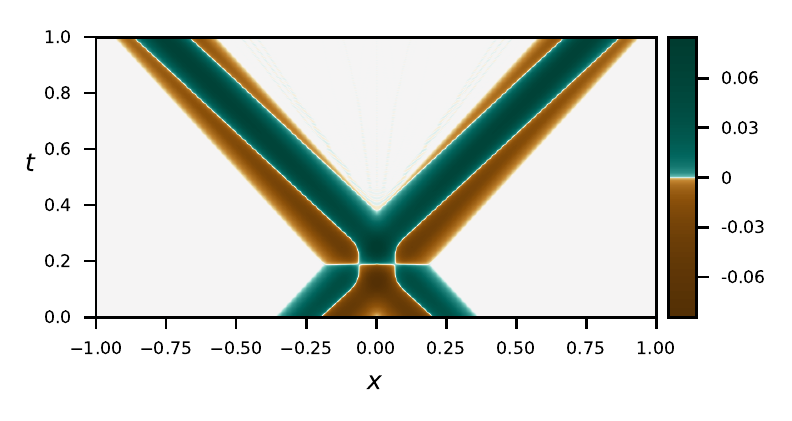}}
\caption{The scatterings of the symmetric initial configuration with $V=0.93$, $v=0$. The scattered oscillons have phase (a) $\alpha=0.414$ and (b) $\alpha=0.914$.}
\label{fig:shock2}
\end{figure}

In our numerical studies we have also spotted a very interesting case. Namely, some symmetric initial configurations  evolve in such a way that the resulting field contains only the main quasi-oscillons {\it i.e.} the amount of radiation generated in this process is virtually insignificant. This phenomenon was observed in the high speed range ($V$ approximately above $0.7$ ) {\it i.e.} when two main outgoing quasi-oscillons had very regular form.  We have also found that, for a given velocity $V$,  there are two phases $\alpha$ for which the radiation is absent. Fig.\ref{fig:shock2} shows two examples of the scattering processes  containing initial configurations with identical velocities $V=0.93$ and different phases $\alpha$. We have chosen $v=0$ in both cases. The phases of configurations shown in Fig.\ref{fig:shock2}(a) and Fig.\ref{fig:shock2}(b)  differ by a factor $\Delta\alpha=0.5$.
\begin{figure}[h!]
\centering
\subfigure[$\quad\alpha=0.414$]{\includegraphics[width=0.42\textwidth,height=0.37\textwidth, angle =0]{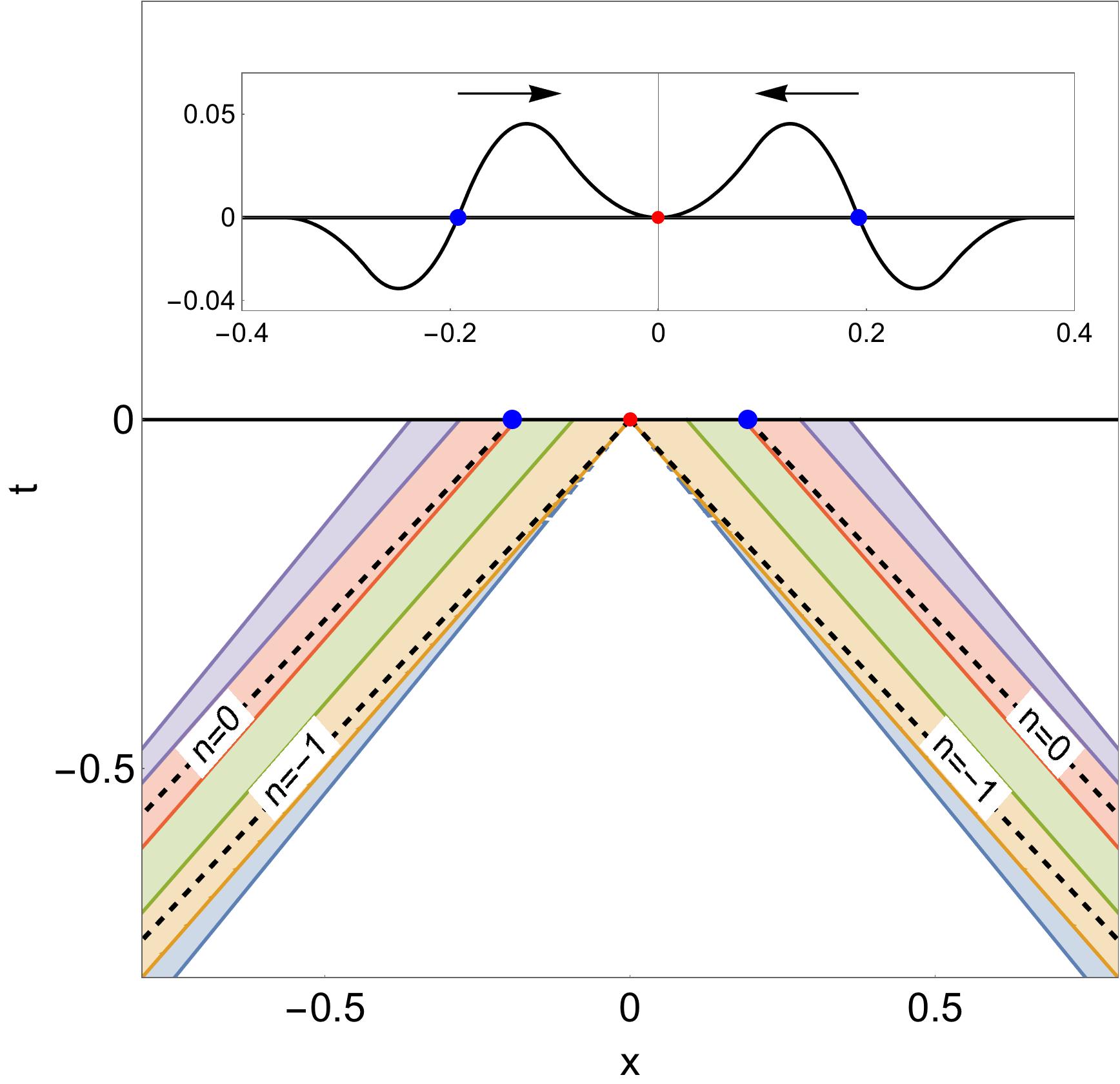}}
\hskip0.5cm
\subfigure[$\quad\alpha=0.914$]{\includegraphics[width=0.42\textwidth,height=0.37\textwidth, angle =0]{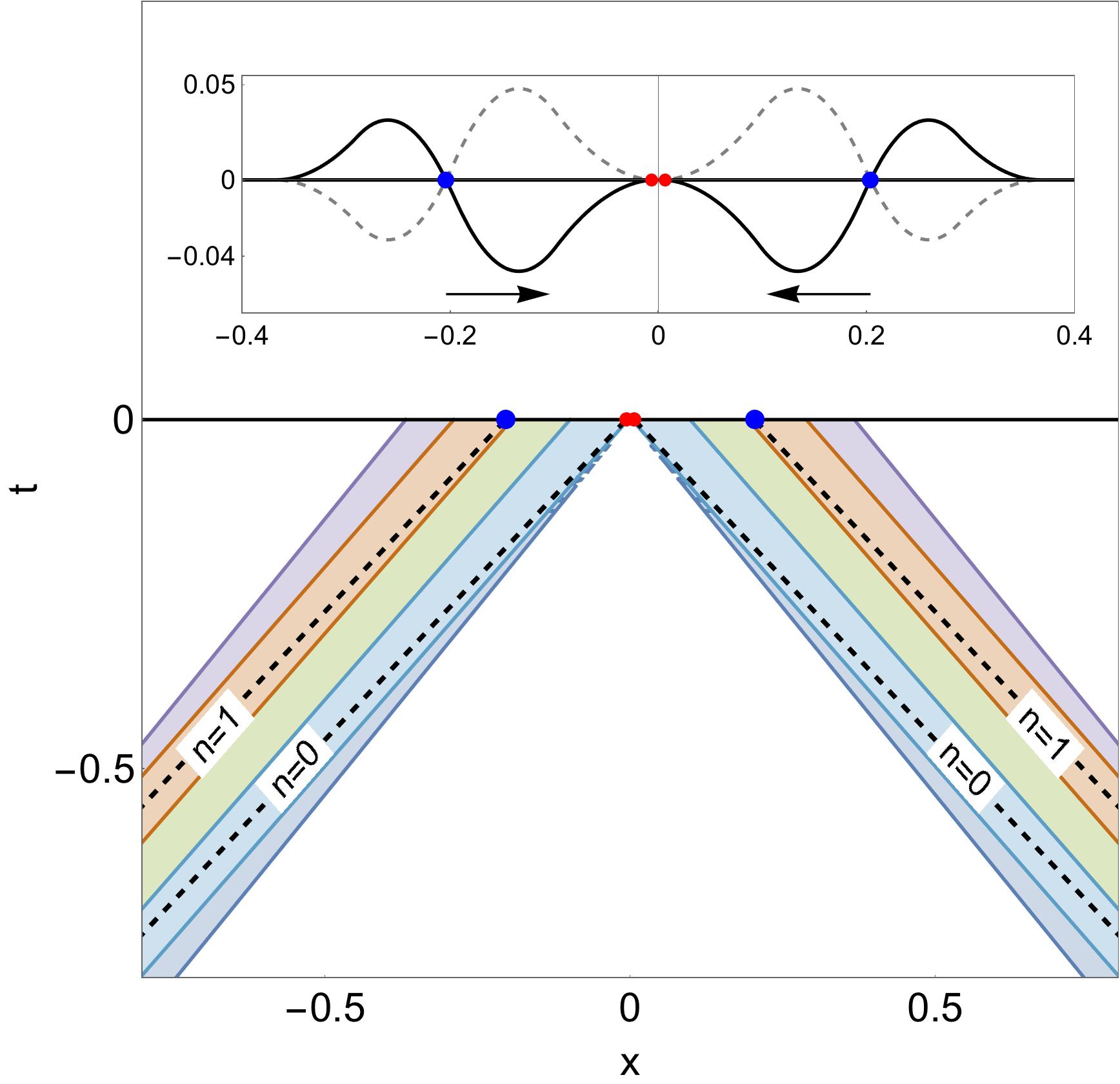}}
\caption{Initial profile of the signum-Gordon field and worldsheets of the incoming oscillons  for $V=0.93$, $v=0$ and (a) $\alpha=0.414$, (b) $\alpha=0.914$. The zeros are marked by dashed lines. Theirs positions at $t=0$ are given by (a) $B^{(-1)}(\alpha)=-0.006$, $B^{(0)}(\alpha)=-0.204$ and  (b) $B^{(0)}(\alpha)=-0.006$, $B^{(1)}(\alpha)=-0.204$.}
\label{fig:sup}
\end{figure}

The observed difference of phases can be understood by comparing the zeros of the incoming oscillons in figures (a) and (b). The initial configurations at $t=0$ and the worldsheets of the oscillons for $t<0$ are plotted in Fig.\ref{fig:sup}.
We have also marked there the zeros of the oscillons. They correspond to the segments of straight lines
\be
x^{(n)}(t)=\frac{t}{V} +B^{(n)}(\alpha),\qquad\text{where} \qquad B^{(n)}(\alpha) \equiv \frac{\gamma}{2V}\left(2\alpha -\frac{n}{\gamma^2}\right)-x_0(\alpha)\label{xn}.
\ee
Here $x_0(\alpha)$ is given by \eqref{xzero} and $n=0,\pm1,\pm2,\ldots$ stands for the numbering of the infinite sequence of zeros. Since $x_0(\alpha)$ is a linear function of the phase $\alpha$ then the coefficients $B^{(n)}(\alpha)$ are also linear functions of $\alpha$.
Expression \eqref{xn} shows that all the zeros of the oscillon travel with the same speed equal to $1/V$.  Moreover, this expression also shows that  two consecutive zeros numbered by $n$ and $n+1$ lie on the same straight line if the phases of these oscillons differ by a certain value $\Delta\alpha$ determined from the equation $B^{(n+1)}(\alpha +\Delta\alpha)-B^{(n)}(\alpha )=0$ with $v=0$.  This equation has a solution $\Delta\alpha=\frac{1}{2}$.
In particular, this shows that  the incoming zeros $x^{(0)}(t)$, shown in Fig.\ref{fig:sup}(a), and  $x^{(1)}(t)$, shown in Fig.\ref{fig:sup}(b), lie on the same straight line.  Moreover, it can also be checked that both initial configurations characterized by phases $\alpha$ and $\alpha+\frac{1}{2}$ differ only by the sign {\it i.e.} $\psi\rightarrow-\psi$ when $\alpha\rightarrow\alpha+\frac{1}{2}$, see the insertion plots in Fig.\ref{fig:sup}(b).

\begin{figure}[h!]
\centering
{\includegraphics[width=0.7\textwidth,height=0.4\textwidth, angle =0]{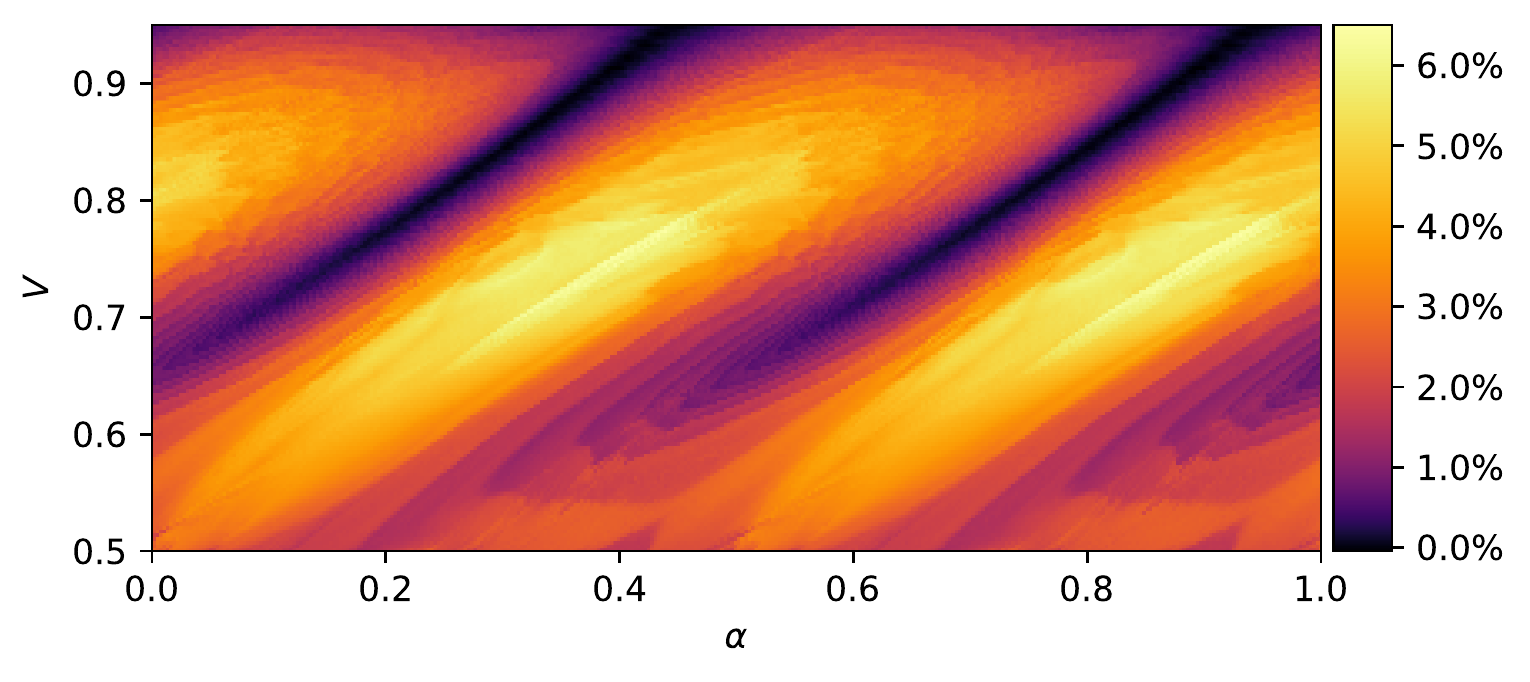}}
\caption{Fraction of the total energy of initial configuration carried out by radiation for $v=0$.}
\label{fig:symphaseshift}
\end{figure}

The case presented in Fig.\ref{fig:shock2} is not unique. Our numerical results suggest that in the regime of high velocities $V$ one can fine-tune the initial parameters $V$ and $\alpha$ (and also $v$, see the next section) in such a  way that the outgoing oscillons are not accompanied by almost any radiation. The absence of radiation demonstrates that the outgoing quasi-oscillon has virtually the same energy as the incoming exact oscillon. In this case the numerical  quasi-oscillon is very similar to the exact one.  In the most common situations there is certain difference of energies and this difference is explained by the release of a very large number of small oscillon-like structures. In Fig.\ref{fig:symphaseshift} we plot the energy radiated out by the system as a fraction of its initial energy. Two dark regions in the upper part of the diagram correspond to the choices of the initial parameters $(\alpha,V)$ such that  this fraction is very close to zero.
Looking at this picture we see that the initial configurations that minimize the radiation correspond to parameters $\alpha$ and $V$ that lie, approximately, on straight lines. For $V<0.7$ the outgoing oscillons are quite irregular; see for instance Fig.\ref{fig:scatvel}(a). Usually,  some radiation can be seen in the vicinity of outgoing irregular perturbed oscillons and this radiation is emitted from these irregularities. In consequence, the determination of the energy of outgoing oscillons is less reliable for small velocities.

\subsubsection{Oscillons with $v\neq 0$}\label{subs}

\begin{figure}[h!]
\centering
{\includegraphics[width=0.6\textwidth,height=0.4\textwidth, angle =0]{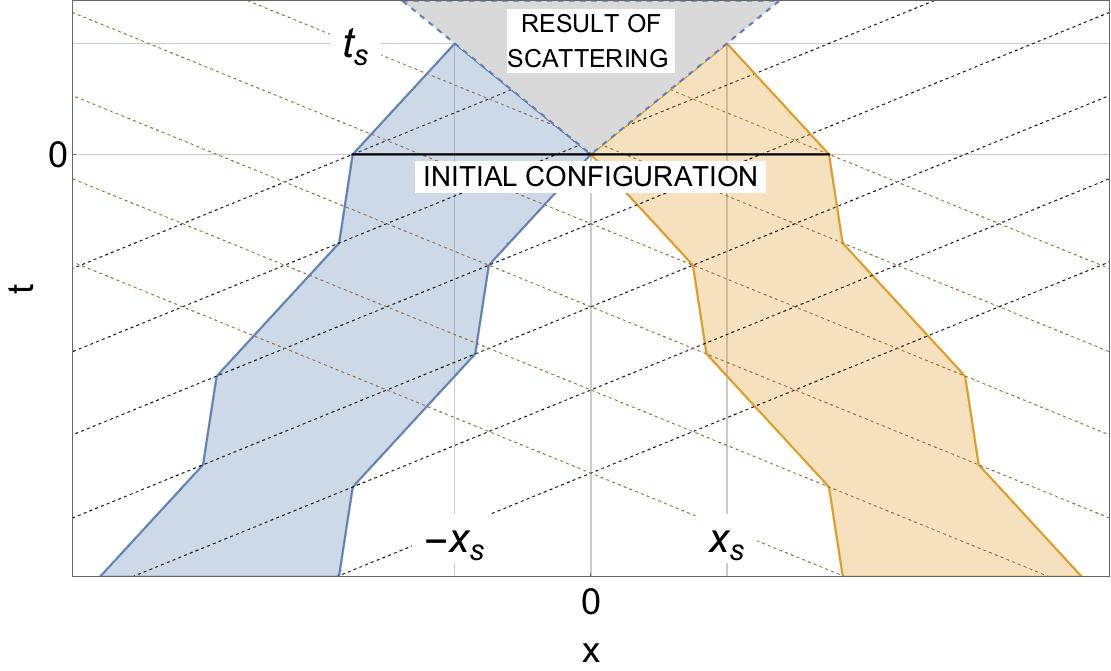}}
\caption{Worldsheets of two incoming generalized oscillons with $v\neq0$. The oscillons  are unperturbed inside two triangular regions above the line $t=0$. The zeros of each oscillon correspond to the intersections of its worldsheet with a family of dashed lines.}
\label{fig:wstegi2}
\end{figure}

In this section we present some of our results on the scattering of generalized exact oscillons {\it i.e.} oscillons which depend on $v$ -- the parameter controlling the swaying motion of the oscillon endpoints. In Fig.\ref{fig:wstegi2} we show the wordsheets of two such oscillons. 
The configuration at $t=0$ was taken as the initial data for our numerical simulation. In similarity to the case $v=0$ we can find the expressions for the characteristic time of the scattering $t_s$ and the minimum support size  $\Delta x_{\rm min}=2|x_s|$ by solving the equations $x=-L(\alpha)+w t$ and $x=-t$. We have found 
\be
t_s=\frac{L(\alpha)}{1+w}=-x_s,\qquad {\rm where}\qquad w\equiv\frac{V+v}{1+Vv},
\ee
 where $L(\alpha)\equiv x_0(\alpha)-\widetilde x_0(\alpha)$ is the size of the oscillon at $t=0$. $x_0(\alpha)$ is given by \eqref{xzero} and $\widetilde x_0(\alpha)$ by \eqref{xzeroleft}.

\begin{figure}[h!]
\centering
\subfigure[$\quad\alpha=0.414$]{\includegraphics[width=0.47\textwidth,height=0.25\textwidth]{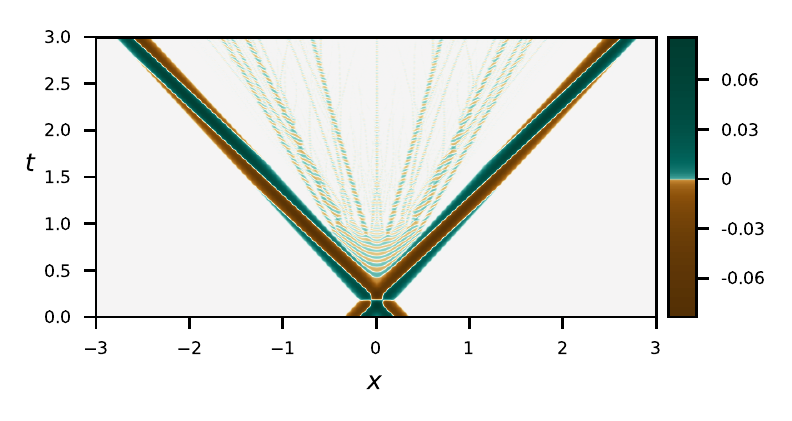}}
\subfigure[$\quad\alpha=0.420$]{\includegraphics[width=0.47\textwidth,height=0.25\textwidth]{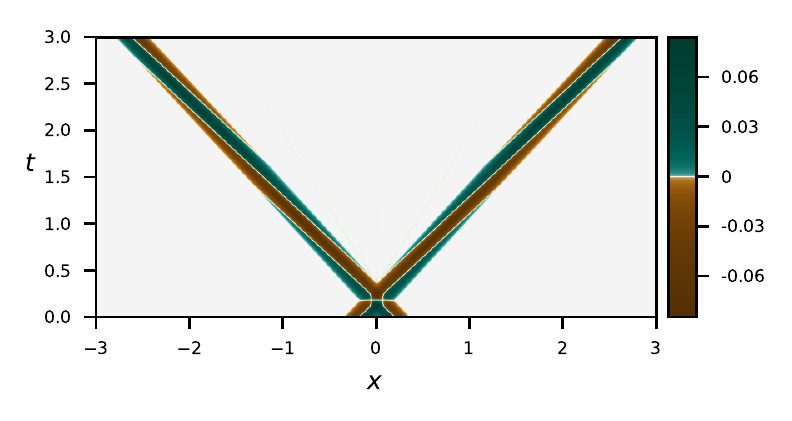}}
\subfigure[$\quad\alpha=0.914$]{\includegraphics[width=0.47\textwidth,height=0.25\textwidth]{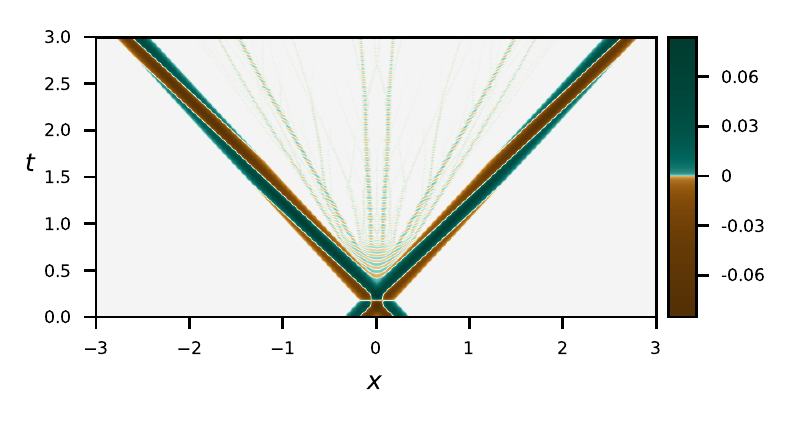}}
\subfigure[$\quad\alpha=0.918$]{\includegraphics[width=0.47\textwidth,height=0.25\textwidth]{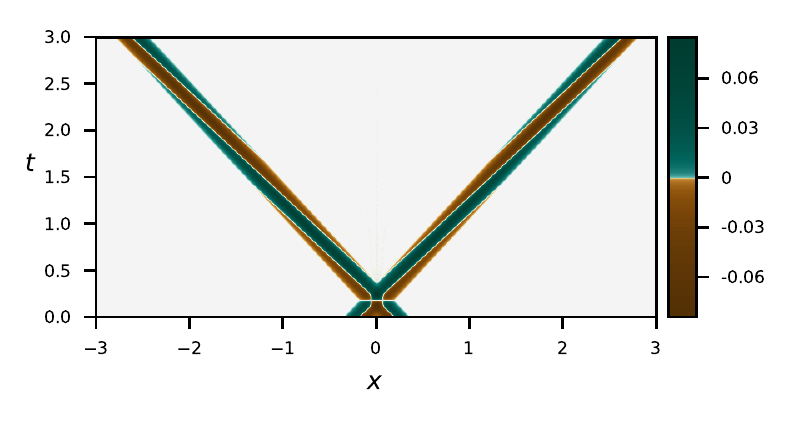}}
\caption{The scattering process for the initial symmetric configuration containing oscillons with speed $V=0.93$, velocity of the border $v=0.02$ for different phases $\alpha$.}
\label{fig:v002}
\end{figure}

An interesting question now arises of how the replacement of $v=0$ by $v\neq 0$ modifies the results of the scattering processes. For instance, we can take a $v\neq 0$ generalization of the symmetric initial configuration with $V=0.93$ and $\alpha=0.414$.
In Fig.\ref{fig:v002}(a) we present the result of the scattering for $v=0.02$. We see that even such small value of the parameter $v$ is sufficient for the appearance  of shock waves which further transform into a cascade of oscillons. This demonstrates  that the scattering process is quite sensitive to the value of the parameter $v$. In order to minimize this radiation one can adjust properly the parameter $\alpha$. We have found that in this specific case the radiation vanishes for $\alpha=0.420$ and $\alpha=0.918$, see Fig.\ref{fig:v002}(b) and (d). For higher values of $v$ there was much more radiation emitted during the process of scattering.

\begin{figure}[h!]
\centering
\subfigure[$\quad\alpha=0.414$]{\includegraphics[width=0.47\textwidth,height=0.24\textwidth]{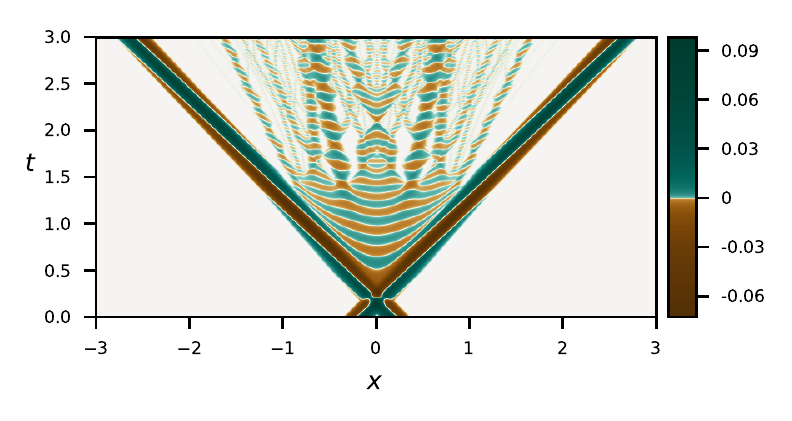}}
\subfigure[$\quad\alpha=0.469$]{\includegraphics[width=0.47\textwidth,height=0.24\textwidth]{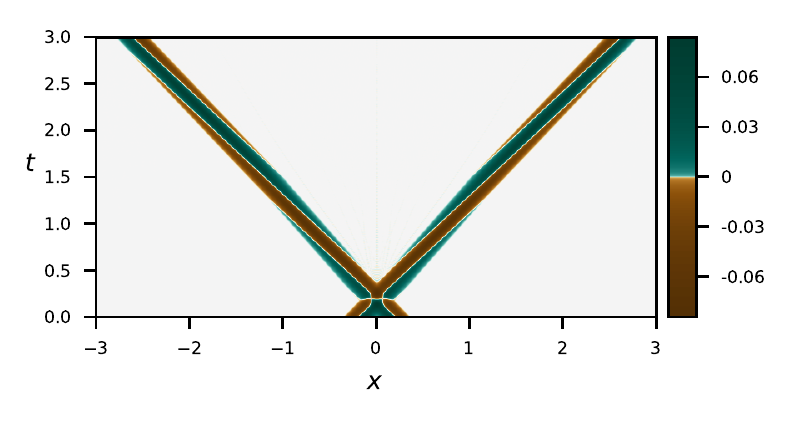}}
\subfigure[$\quad\alpha=0.914$]{\includegraphics[width=0.47\textwidth,height=0.24\textwidth]{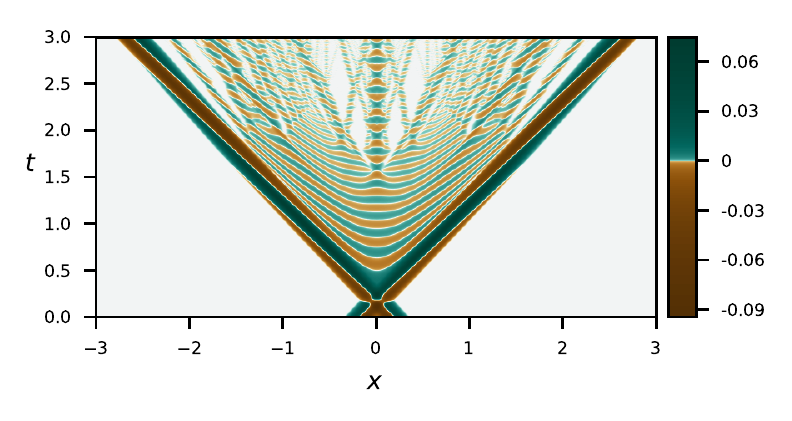}}
\subfigure[$\quad\alpha=0.949$]{\includegraphics[width=0.47\textwidth,height=0.24\textwidth]{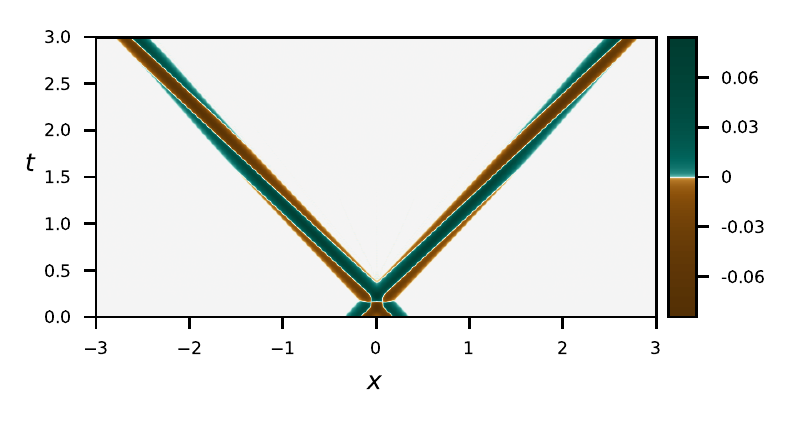}}
\caption{The scattering process for initial symmetric configuration containing oscillons with speed $V=0.93$, velocity of the border $v=0.2$ and different phases $\alpha$.}
\label{fig:v02}
\end{figure}

\begin{figure}[h!]
\centering
\subfigure[$\quad\alpha=0.414$]{\includegraphics[width=0.47\textwidth,height=0.24\textwidth]{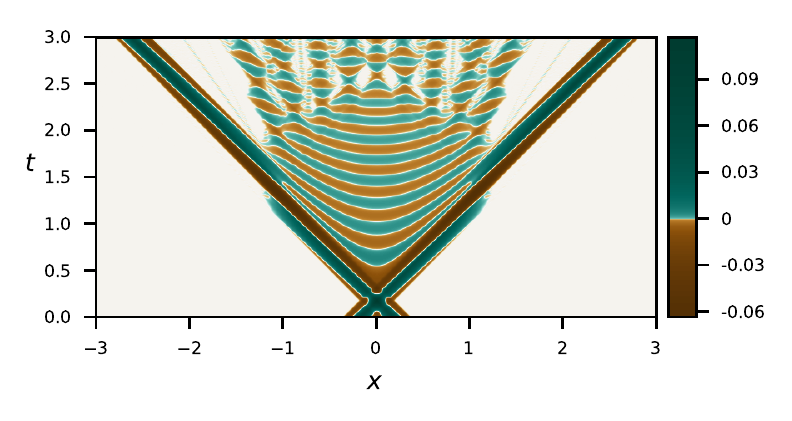}}
\subfigure[$\quad\alpha=0.594$]{\includegraphics[width=0.47\textwidth,height=0.24\textwidth]{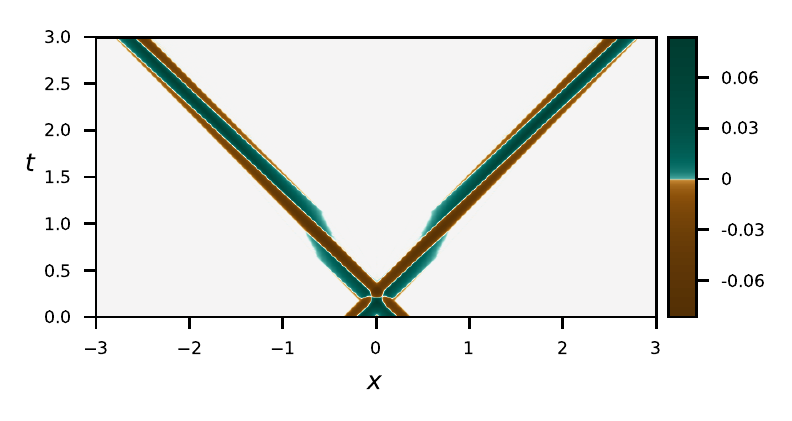}}
\subfigure[$\quad\alpha=0.914$]{\includegraphics[width=0.47\textwidth,height=0.24\textwidth]{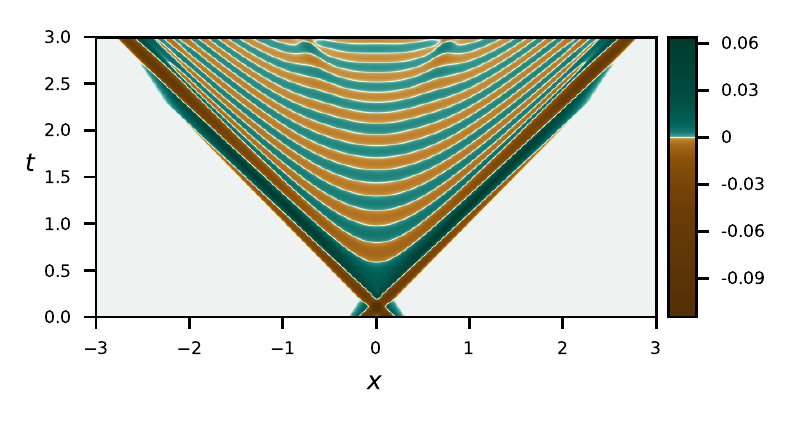}}
\subfigure[$\quad\alpha=0.047$]{\includegraphics[width=0.47\textwidth,height=0.24\textwidth]{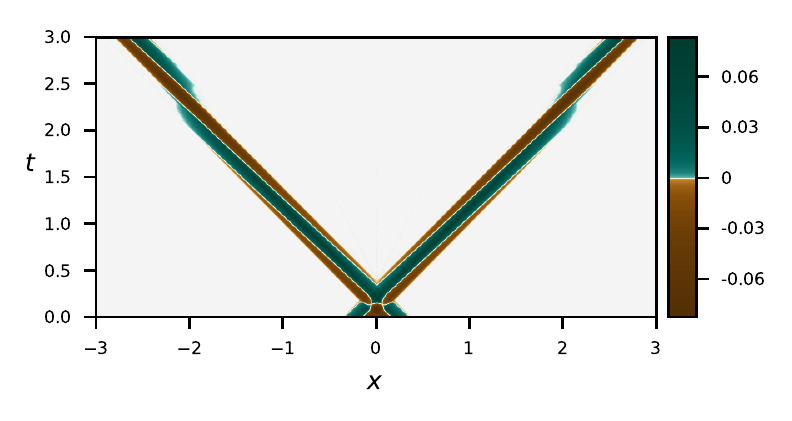}}
\caption{The scattering processes for the initial symmetric configuration containing oscillons with speed $V=0.93$, velocity of the border $v=0.7$ for different phases $\alpha$.}
\label{fig:v07}
\end{figure}

In Fig.\ref{fig:v02} and  Fig.\ref{fig:v07} we present the cases of $v=0.2$ and $v=0.7$. In both cases we have found two values of the phase $\alpha$ that minimize the radiation.

\begin{figure}[h!]
\centering
\subfigure[$\quad v=0.02$]{\includegraphics[width=0.3\textwidth,height=0.18\textwidth, angle =0]{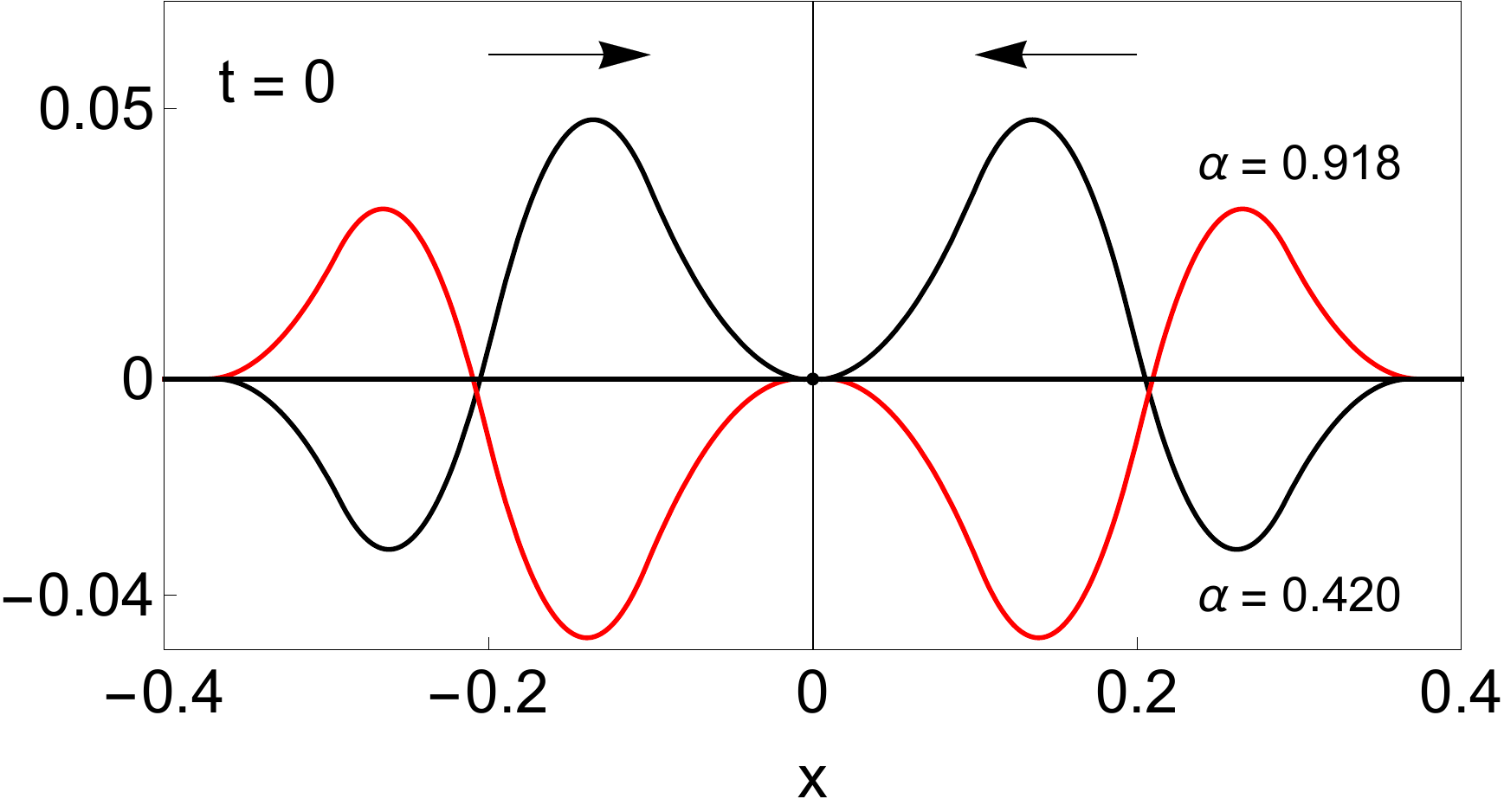}}
\hskip0.5cm
\subfigure[$\quad v=0.2$]{\includegraphics[width=0.3\textwidth,height=0.18\textwidth, angle =0]{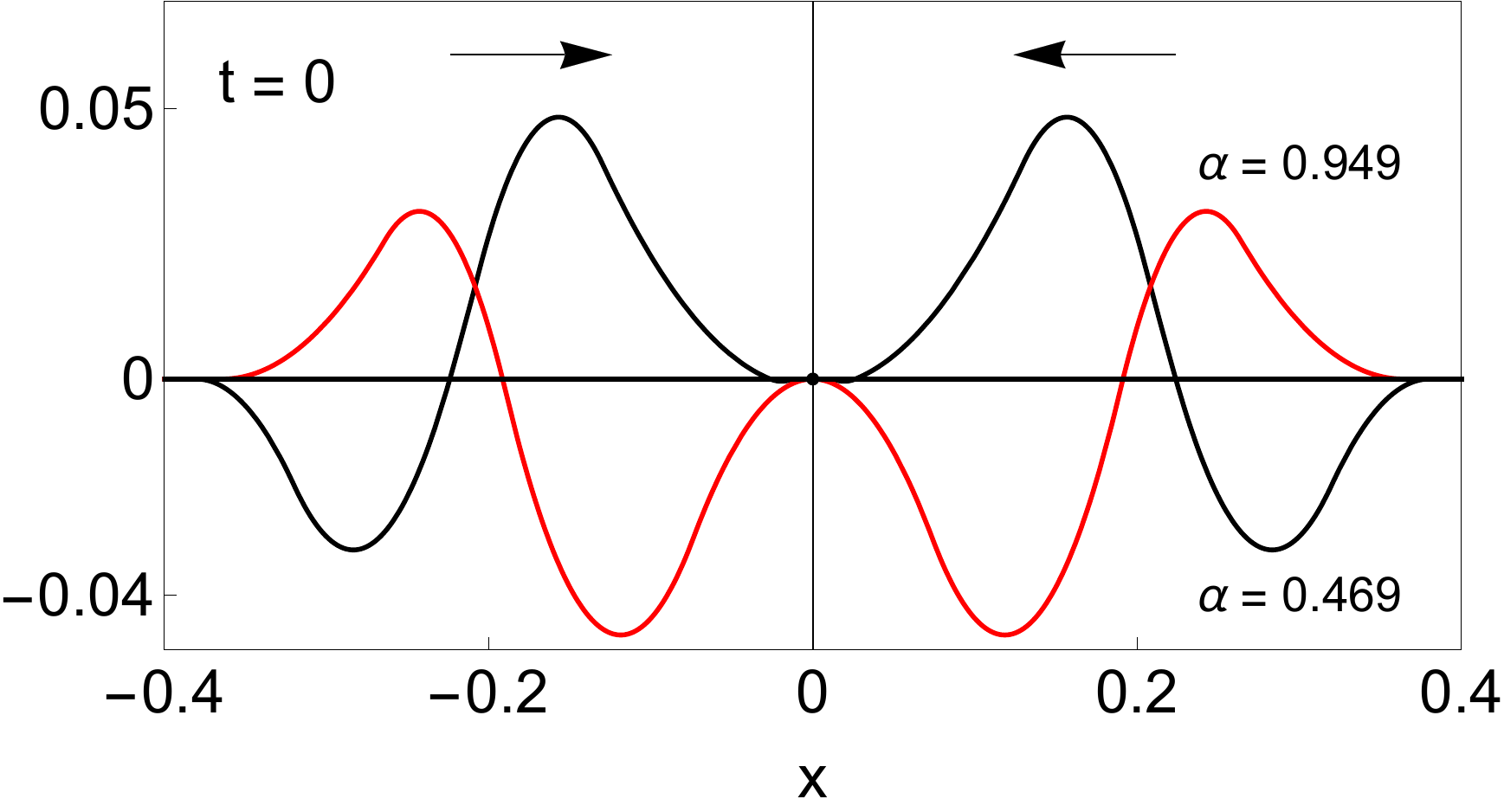}}
\hskip0.5cm
\subfigure[$\quad v=0.7$]{\includegraphics[width=0.3\textwidth,height=0.18\textwidth, angle =0]{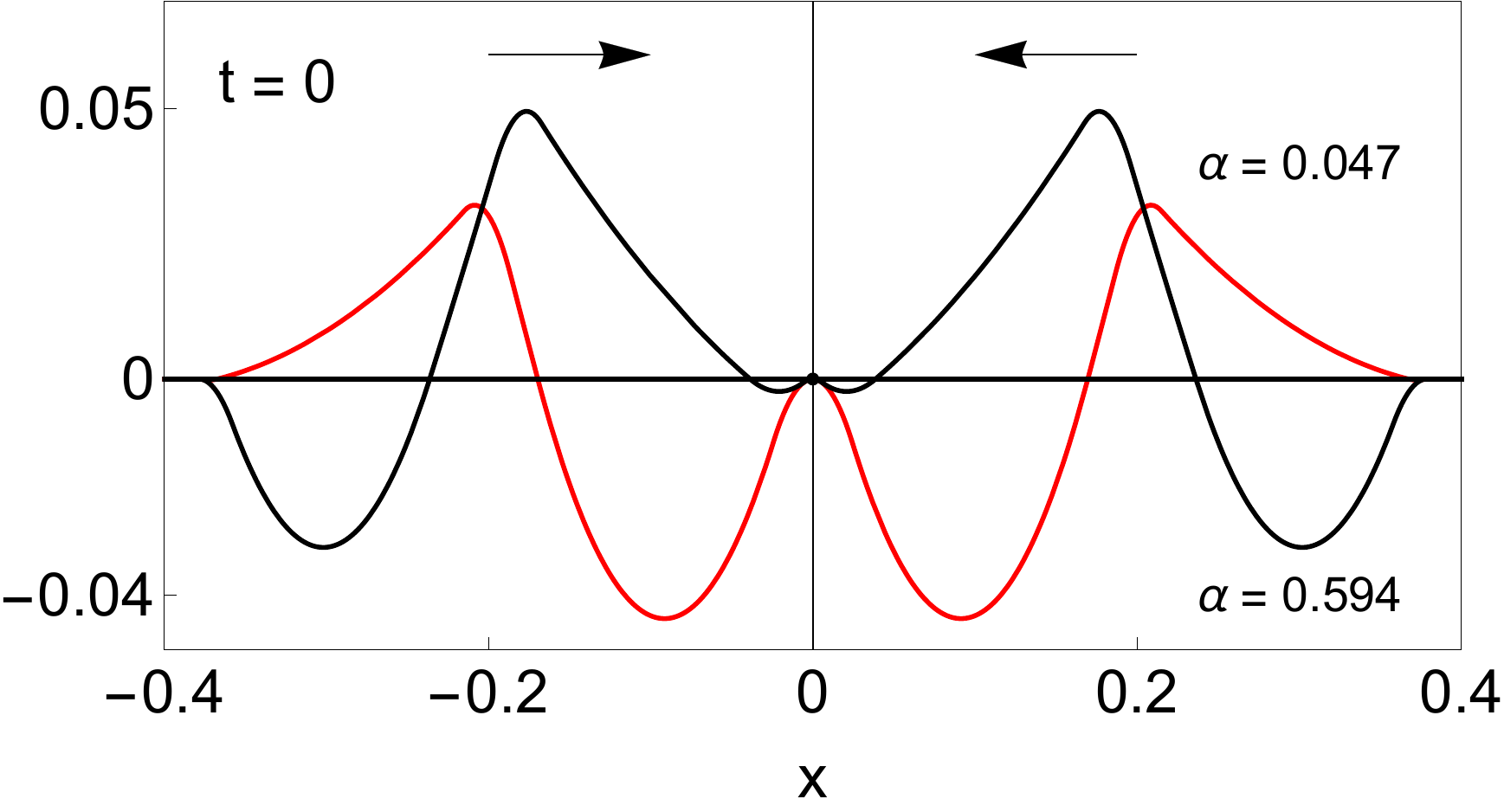}}
\caption{Initial configurations that minimize the radiation for $V=0.93$.}
\label{fig:inivanish}
\end{figure}

Looking carefully at the initial configurations we see that there is a significant difference between the case $v=0$ and $v\neq 0$. In Fig.\ref{fig:inivanish} we present the initial field configurations that minimize the radiation. Comparing configuration $\Psi^{(s)}(x; v, V,\alpha+\Delta\alpha)$, where $\Delta\alpha$ is taken from the numerical simulations, with $\Psi^{(s)}(x; v, V,\alpha)$ we see that the configuration with $\alpha+\Delta\alpha$ is not equal to the negative of the configuration with $\alpha$. The difference between $\Psi^{(s)}(x; v, V,\alpha+\Delta\alpha)$ and $-\Psi^{(s)}(x; v,V,\alpha)$ increases with the increase of $v$. On the other hand, $\Psi^{(s)}\rightarrow-\Psi^{(s)}$ is a symmetry of the signum-Gordon equation so if $\Psi^{(s)}(x; v, V, \alpha)$ minimizes the radiation then $-\Psi^{(s)}(x; v, V, \alpha)$ minimizes the radiation too. Thus, for any set of parameters $(v,V)$ there are four different initial configurations $\pm \Psi^{(s)}(x; v,V,\alpha)$ and $\pm \Psi^{(s)}(x; v, V,\alpha+\Delta\alpha)$ that minimize the radiation. They reduce to two for $v=0$.

\begin{figure}[h!]
\centering
{\includegraphics[width=0.6\textwidth,height=0.3\textwidth, angle =0]{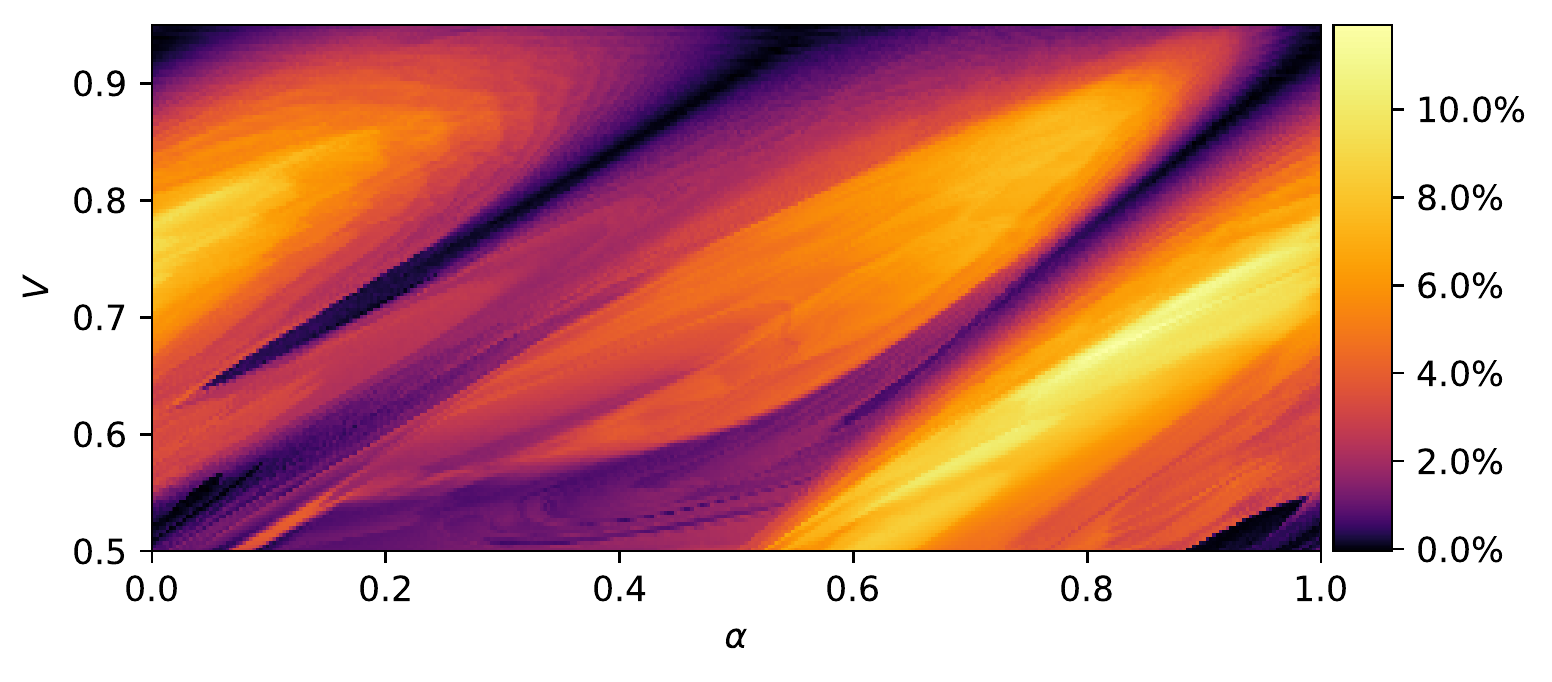}}
\caption{Fraction of the total energy of the initial configuration carried out by the radiation for $v=0.45$.}
\label{fig:symphaseshift2}
\end{figure}

In Fig.\ref{fig:symphaseshift2} we have plotted the fraction of the initial energy carried by the radiation. The figure was produced for $v=0.45$. The dark regions, corresponding to very low values of the radiated energy,  are less regular when compared with  Fig.\ref{fig:symphaseshift} for the case of vanishing $v$.


\subsection{Non-symmetric configurations}
\begin{figure}[h!]
\centering
\subfigure[$\quad\alpha_L=0,\quad\alpha_R=0.4$]{\includegraphics[width=0.47\textwidth,height=0.24\textwidth]{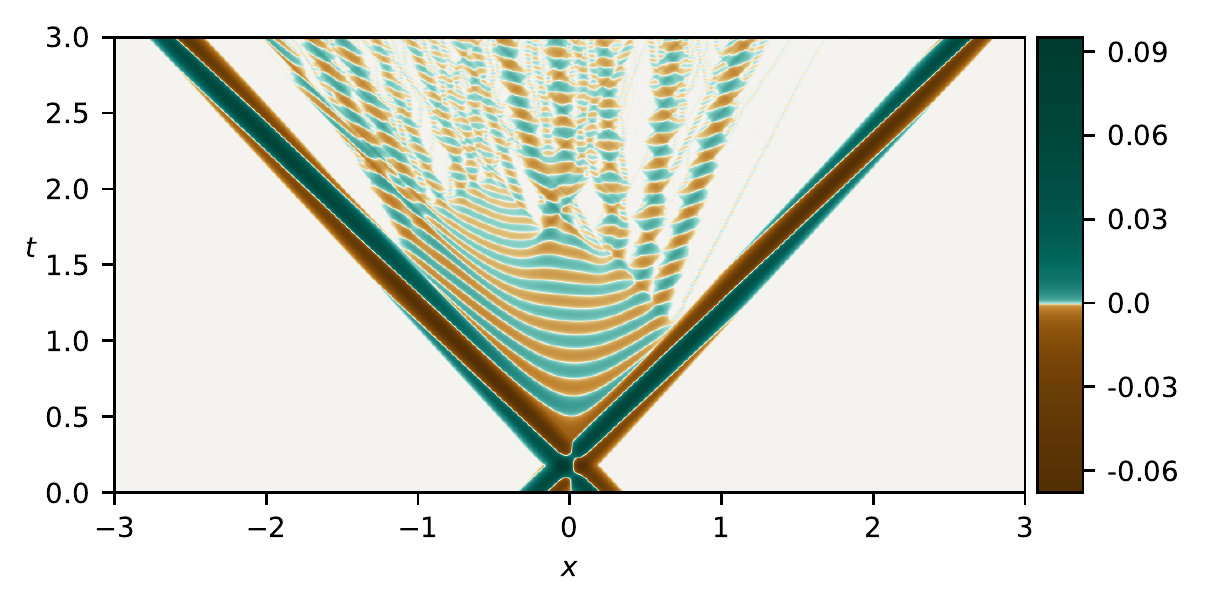}}
\subfigure[$\quad\alpha_L=0,\quad\alpha_R=0.816$]{\includegraphics[width=0.47\textwidth,height=0.24\textwidth]{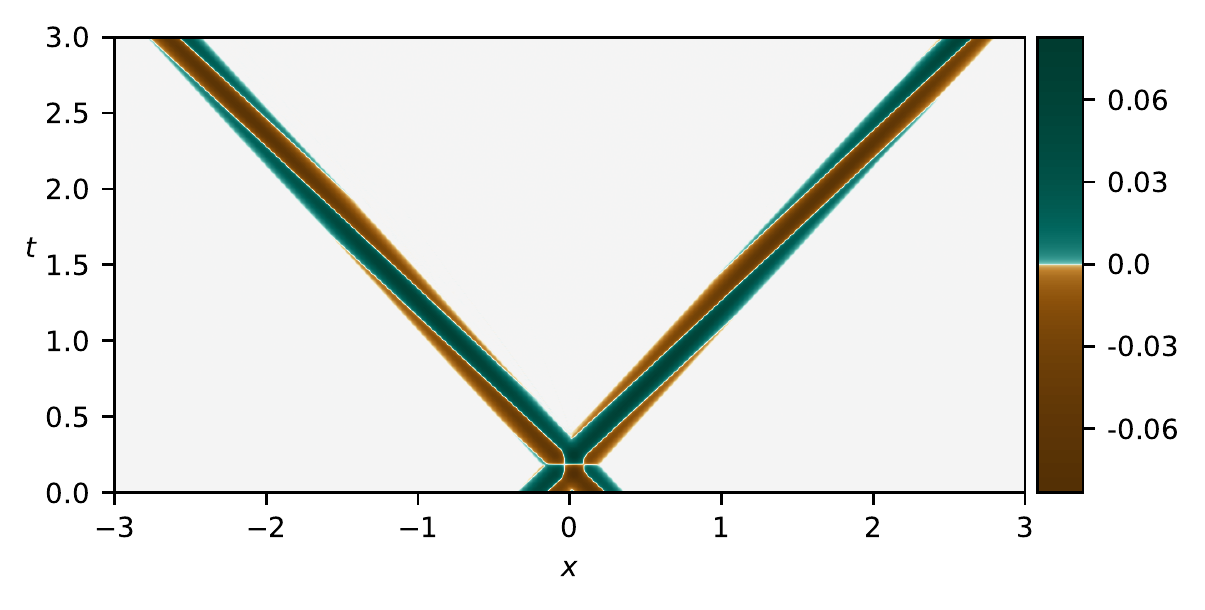}}
\subfigure[$\quad\alpha_L=0,\quad\alpha_R=0.876$]{\includegraphics[width=0.47\textwidth,height=0.24\textwidth]{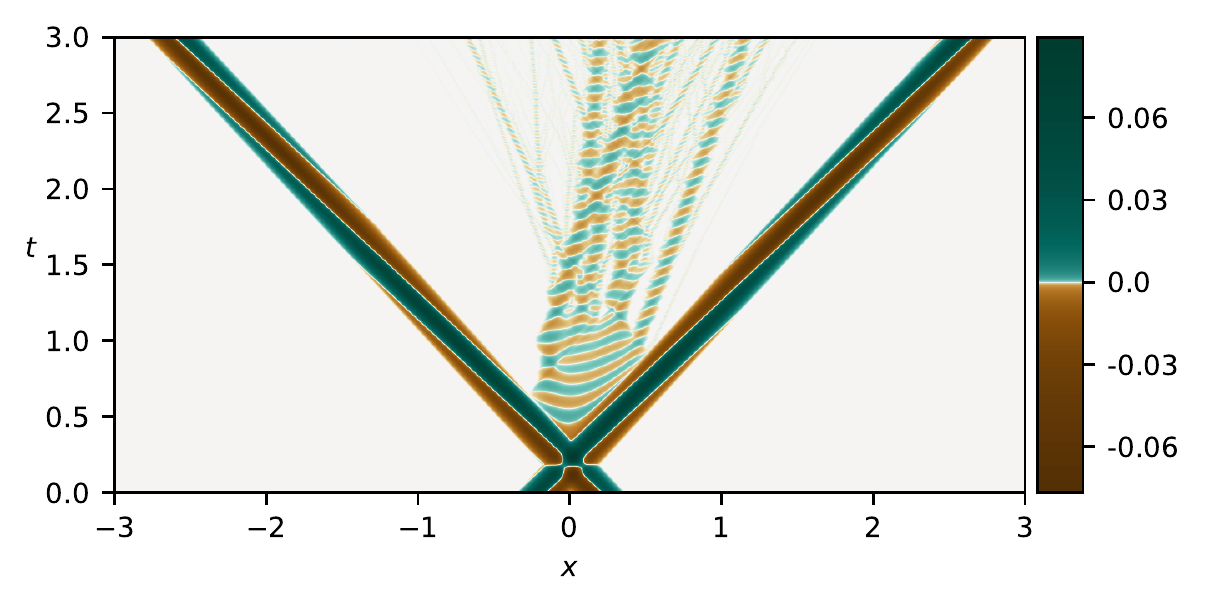}}
\subfigure[$\quad\alpha_L=0.7,\quad\alpha_R=0.097$]{\includegraphics[width=0.47\textwidth,height=0.24\textwidth]{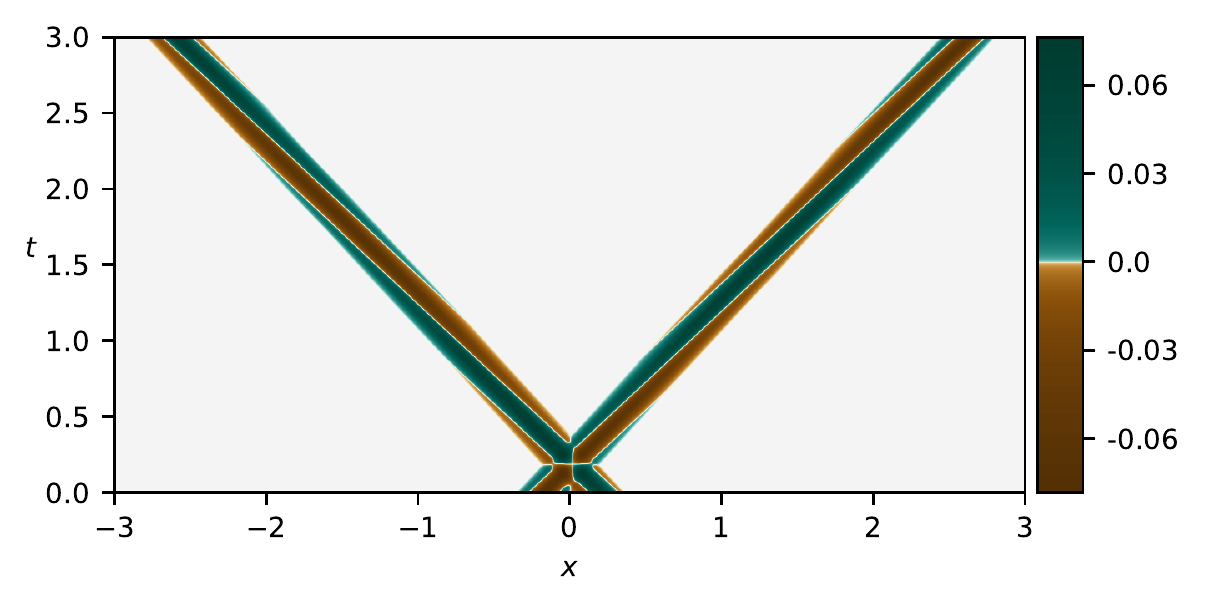}}
\caption{The scattering process for initial configuration containing oscillons with speed $V=0.93$, velocity of the border $v=0$ and different phases $\alpha_L$ and $\alpha_R$.}
\label{fig:a-sym}
\end{figure}

The simplest non-symmetric configurations are described by oscillons that differ only by their phases. In order to probe this class of scatterings, we use initial conditions given by (\ref{2phases}) taken at time $t=0$. In Figs.\ref{fig:a-sym}(a)-(d) we plot four cases of the time evolution for these processes. Interestingly, there are all non-symmetric configurations for which no radiation is present. Figs.\ref{fig:a-sym}(b) and \ref{fig:a-sym}(d) show two of these cases.

\begin{figure}[h!]
\centering
{\includegraphics[width=0.6\textwidth]{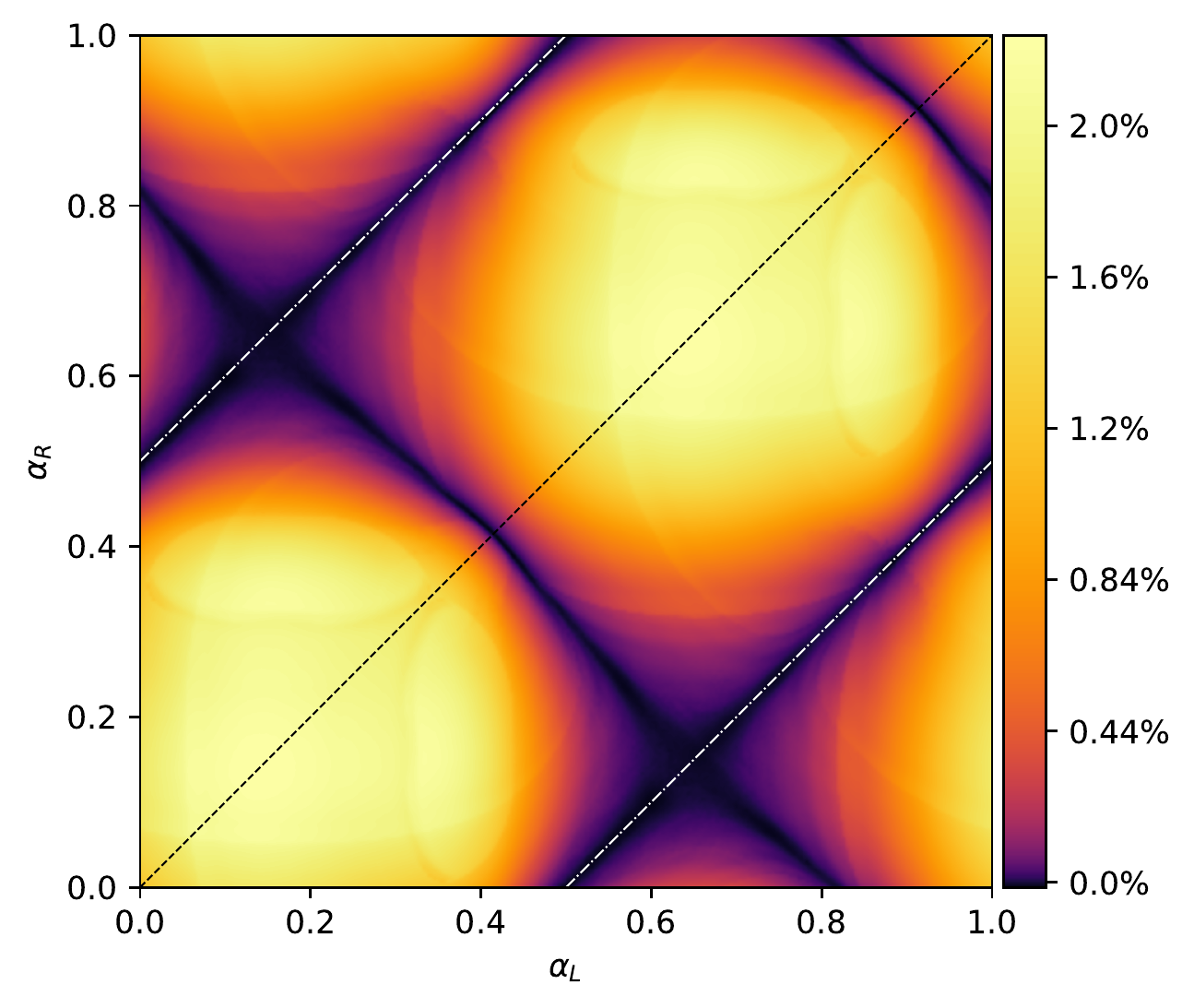}}
\caption{Fraction of the total energy of initial configuration carried out by the radiation as a function of phases $\alpha_L$ and $\alpha_R$ of incoming exact oscillons  for $V=0.93$. The black dashed line corresponds to the set of all such symmetric configurations and the white dot-dashed lines correspond to the set of all such anti-symmetric  configurations.}
\label{fig:a-sym-phase-map}
\end{figure}

In order to have a clearer picture of the amount of input energy converted into the radiation, we present a density plot much like the ones presented in Figs. \ref{fig:symphaseshift} and \ref{fig:symphaseshift2}, except that in this case we have fixed the values $V=0.93$ and $v=0$, and we vary the parameters $\alpha_L$ and $\alpha_R$ (now independent of each other). This plot is shown in Fig. \ref{fig:a-sym-phase-map}.

The case $\alpha_L=\alpha_R$ corresponds tothe  symmetric configurations of the initial conditions, and is marked in the plot as a black dashed line. This line passes through two minima in the emitted radiation, as is expected according from the results discussed in Sec. \ref{sec:symmetric-configurations}. Along this line, the plot has a coincident set of values with the plot in Fig. \ref{fig:symphaseshift} along the line given by $V=0.93$. Also, as mentioned in Sec. \ref{sec:sym-rad-vanish}, for null swaying speed of the oscillon endpoints ($v=0$), a shift of $\frac12$ in the phase of a given oscillon produces the same oscillon with a sign change in the value of its field and its time derivative. For this reason, the relation $\alpha_R=\alpha_L\pm\frac12$ corresponds to the anti-symmetric initial field configurations. This relation is marked in the figure as the two dot-dashed white lines. In agreement with the results from Sec. \ref{sec:antisym-config}, these lines lie on top of the dark regions that correspond to no-radiation zones.

The plot can be seen as periodic both in $\alpha_L$  and in $\alpha_R$ (although the phase is originally defined as a value between $0$ and $1$, the field configuration is precisely the same for both these values), leaving it with a toroidal topology. So, the two black strips marked by the white dot-dashed lines form a belt around the torus and become a single continuous region.

We note, still, a second strip of no-radiation forming a $90$ degree angle with these lines. One could expect the eventual vanishing of radiation in non-symmetric initial scattering configurations \textit{e.g.} Figs. \ref{fig:a-sym}(b) and \ref{fig:a-sym}(d). Yet, the regularity of this region (it is a straight line) is quite remarkable and requires further considerations.

\begin{figure}[h!]
\centering
{\includegraphics[width=0.65\textwidth]{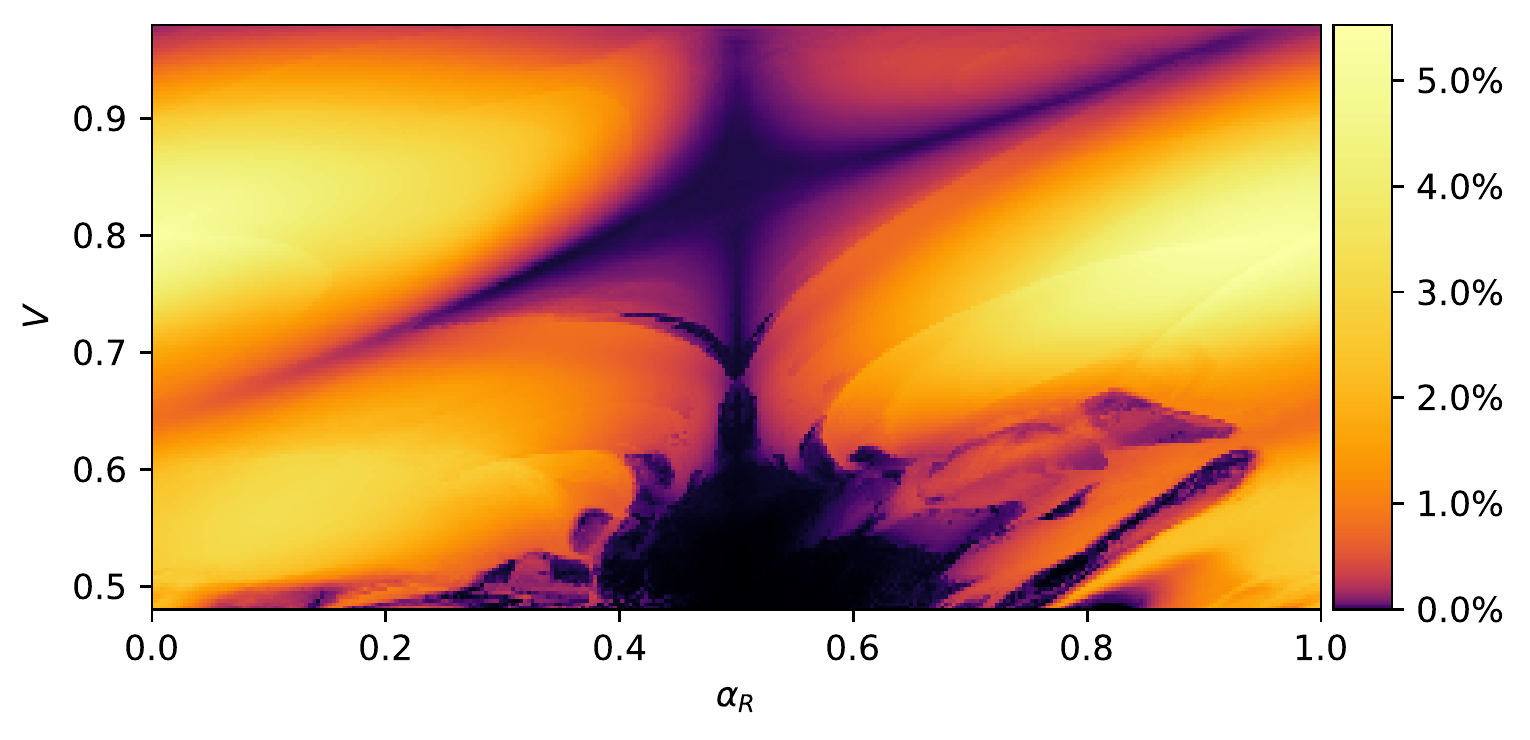}}
\caption{Fraction of the total energy of the initial configuration carried out by the radiation as a function of the input speed $V$ and phase of right input oscillon $\alpha_R$ while holding fixed the right input oscillon's phase $\alpha_R=0$.}
\label{fig:a-sym-V-aR-map}
\end{figure}

We present a similar plot with $\alpha_L$ fixed in which we varied $\alpha_R$ and $V$. This plot is presented  in Fig. \ref{fig:a-sym-V-aR-map} and, along the line given by $V=0.93$, its values coincide with those of Fig. \ref{fig:a-sym-phase-map} for $\alpha_L=0$.


We note that the strip along $\alpha_R=\frac12$ corresponds to the anti-symmetric configurations and so it shows little or no radiation around it. Below the value $V\simeq0.7$ there appears a large irregular region devoid of radiation. From the standpoint of numerical stability of our methods for the measurement of radiation in regions of low $V$, as mentioned in Sec. \ref{sec:fractal}, this region could well be a numerical artefact. In a brief investigation of this hypothesis, though, we have found this void to be an accurate description and the region, indeed, represents a zone of no-radiation interactions for very low energy input (low boosts). So the value $V\simeq0.7$ is critical in the sense that, below it, the entire region for this particular set of values in the parameter space, generates initial conditions to the scattering that produces no radiation, and the region, itself, does not seem to have a very well defined shape.

\begin{figure}[]
\centering
\subfigure[]{\includegraphics[width=0.49\textwidth]{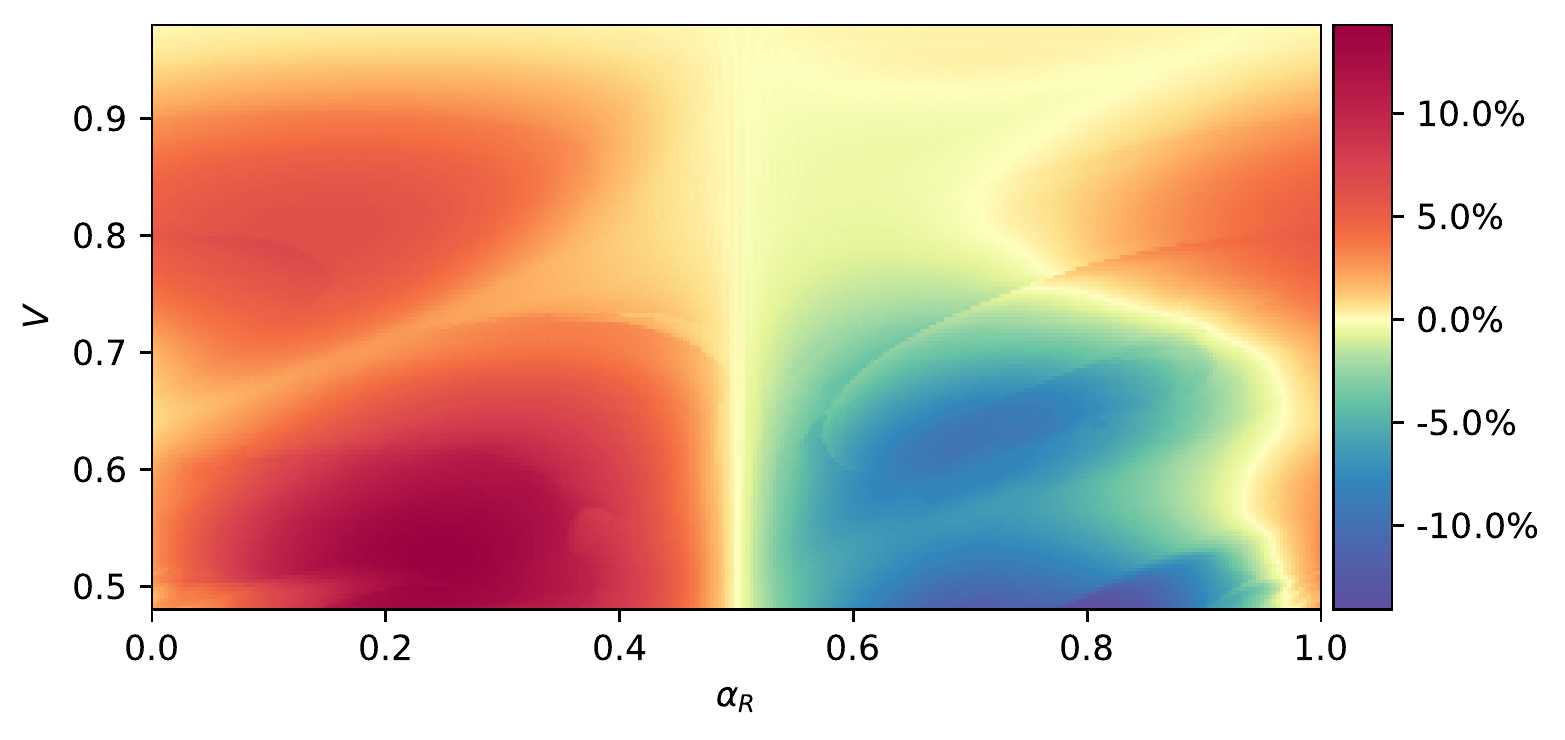}}
\subfigure[]{\includegraphics[width=0.49\textwidth]{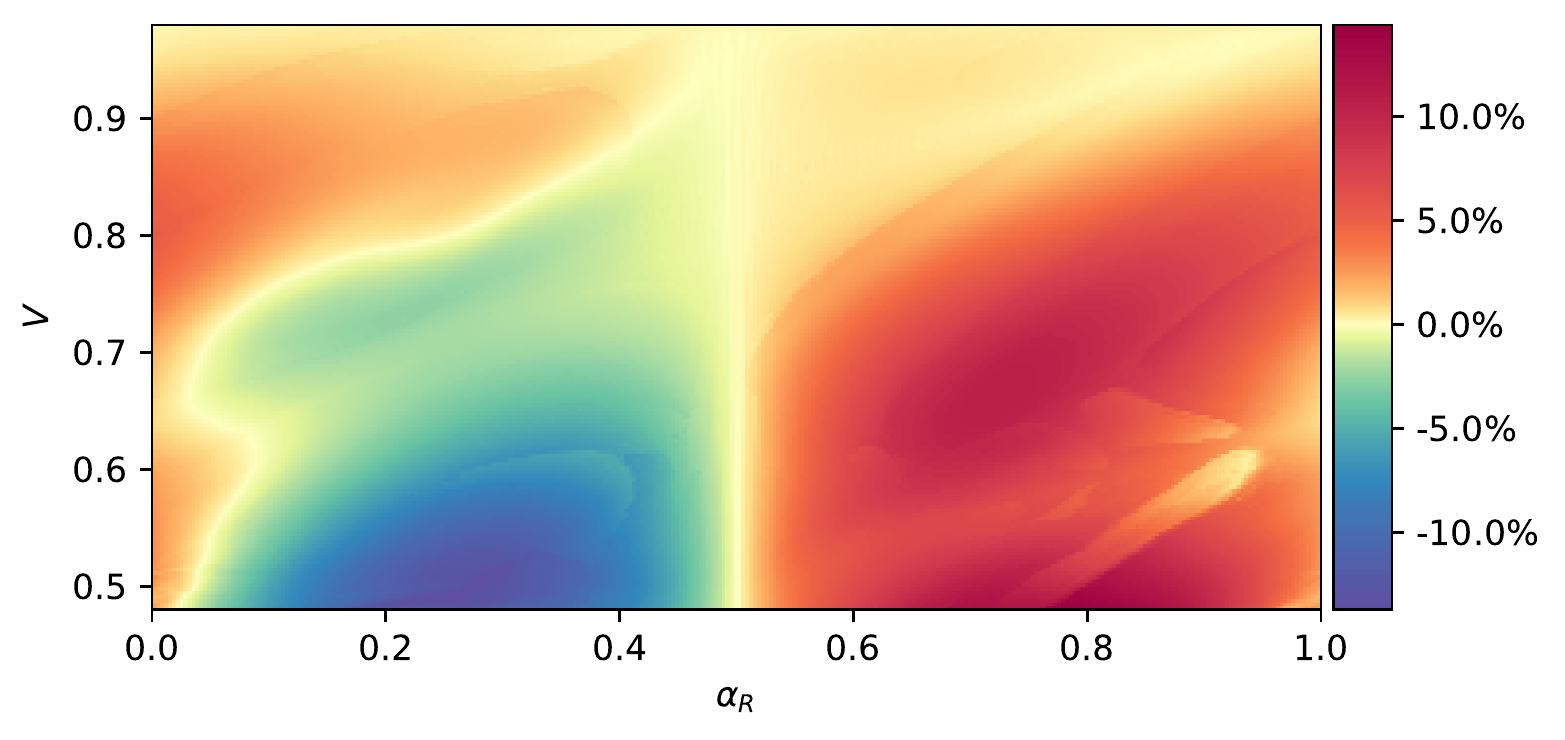}}
\caption{Balance of energy after the interaction (in percentage of incident energy) for each oscillon, where (a) is for the left outgoing main oscillon and (b) the right outgoing main oscillon.}
\label{fig:rad-balance}
\end{figure}

One possible explanation of this void is that there is, indeed, some generation of the radiation in the scattering process but this radiation happens to be absorbed by the outgoing oscillons, which in turn makes them become more perturbed. This hypothesis can be checked by considering the energy of each individual outgoing oscillon in this region. In Figs.\ref{fig:rad-balance}(a) and (b) we plot the energy balance of each outgoing oscillon compared to its initial energy. This balance is calculated by
\[
\begin{split}
	\Delta E_L &\equiv 1-\frac{2E_L}{E}, \\
	\Delta E_R &\equiv 1-\frac{2E_R}{E},
\end{split}
\]
where $E_L$ is the energy of the left outgoing oscillon (which entered the interaction region from the right), $E_R$ is the energy of the right outgoing oscillon and $E$ the total input energy. Note that the plot in Fig. \ref{fig:a-sym-V-aR-map} shows the value of the amount of energy lost to the radiation, which is given by
\[
\begin{split}
	\frac{E_{rad}}{E}	&=1-\frac{E_L+E_R}{E} \\
			&=\frac12(\Delta E_L + \Delta E_R),
\end{split}
\]
so that the total radiation is just the sum of the energy balances of individual scattered oscillons.

\begin{figure}[h!]
\centering
{\includegraphics[width=0.65\textwidth]{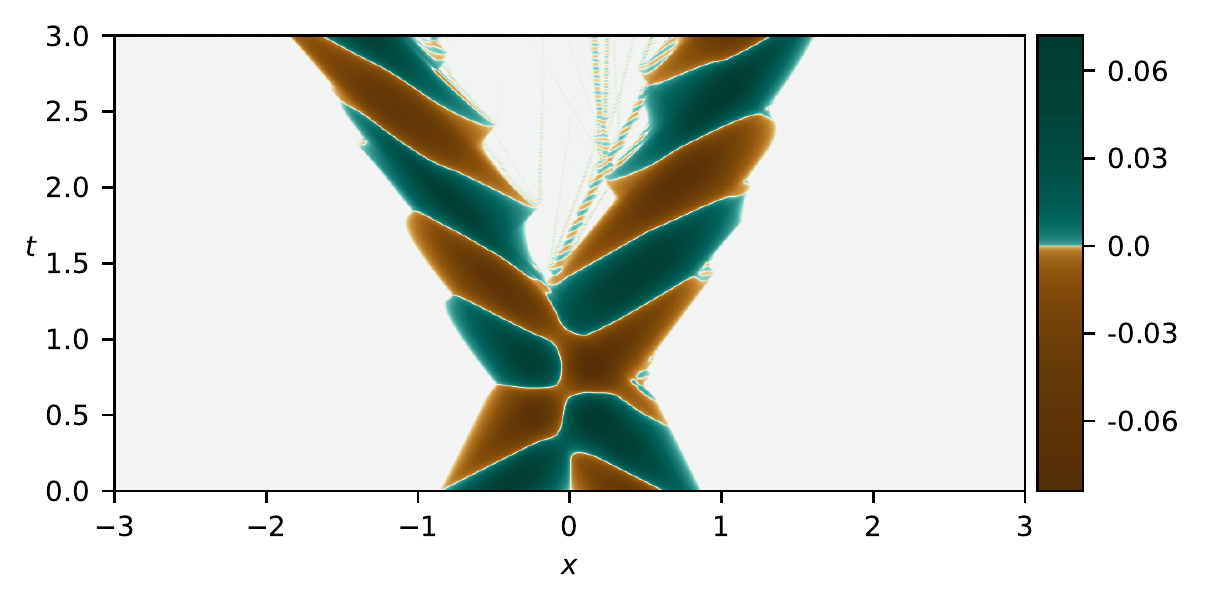}}
\caption{Scattering process for $V=0.5$ and $\alpha_R=0.642$, with parameters $\alpha_L,v_L,v_R$ all set to zero.}
\label{fig:gen-V05-a20642}
\end{figure}
From our figures (Figs. \ref{fig:rad-balance}(a) and (b)) we  see that each oscillon looses/gains a considerable amount of energy in the process. This amount, in some casees, is close to up to $15\%$ of each oscillon's incident energy, yet the total radiation generated is no larger than about $5.5\%$ of the total incident energy. In Fig. \ref{fig:gen-V05-a20642} we present the case for $V=0.5$, $\alpha_L=0$, $\alpha_R=0.642$, $v_L=0=v_R$, which corresponds to a configuration that produces no radiation (as it is located within the void region) and, at the same time, has a relatively large energy transfer between the two interacting oscillons (the energy balance plots of these regions show a large energy change of both oscillons). Note that the left input oscillon (outgoing towards the right) is larger than the one outgoing to the left. Most such configurations seem to reproduce this behaviour, and the channel through which this energy is transferred, in all cases, seems to be related to the scale of the outgoing oscillons.

\subsection{Oscillons with accelerating borders}\label{nonuniform}

Looking at Fig.\ref{fig:scatvel} and Fig.\ref{fig:scatvelanti} we note that the outgoing oscillons are significantly different from the exact generalized oscillon with uniformly moving endpoints. The main difference is in the form of worldlines describing the borders of the oscillons which take the form of continuous curves rather than zig-zag piecewise straight lines. Moreover, the curves are surprisingly regular in shape and this suggests that the outgoing objects are not just simple perturbations of the generalized exact oscillons. Recently, in \cite{swierczynski} further generalization of the signum-Gordon oscillins has been proposed. This generalization leads to the emergence of oscillons with borders that are described by arbitrary time-like curves. Here we present a construction of oscillons with curvilinear borders and produce some plots of such oscillons after applying to them Lorentz boosts.

\subsubsection{General properties}
For an oscillon with period $ T $ the complete solution can be constructed from the restriction of the solution $ \varphi(t,x) $ to the interval $ 0 \leq t \leq T/2 $. Similarly to the oscillons already known, we can get periodic and localized solutions imposing the conditions
\begin{align}
	\varphi(0, x) &= \varphi\qty(\frac{T}{2}, x) = 0, \label{eq:initial-field}
\end{align}
\begin{align}
	 \partial_t \varphi(0, x)&=
	\begin{cases}
		0          & \text{if }\quad x < 0\\
		\oldchi(x) & \text{if }\quad 0 \leq x \leq T,\\
		0          & \text{if }\quad x > T
	\end{cases}
	\label{eq:initial-dv}
\end{align}
where $ \oldchi(x) $ is a continuous function such that $ \oldchi(x) \leq 0 $ for all $ 0 \leq x \leq T $ and $ \oldchi(0) = \oldchi(T) = 0 $. As a consequence, we assume that the solution $\varphi(t,x)$ is negative for $0 \leq t \leq T/2$.

The oscillon solution is localized in the sense that it is nonzero only in the region between two time-like curves $\gamma_L$ and $\gamma_R$ -- the borders of the oscillon. The right border is a displacement by $T$ of the left curve, so that the size of the oscillon remains constant. For $ t \in [0, T/2] $ the borders move to the right by $\Delta$ and, for $ t \in [T/2, T] $, the borders move in the opposite direction -- returning to the original position.

The conditions \eqref{eq:initial-field} and \eqref{eq:initial-dv} are satisfied if we take the following ansatz for the non-zero part of the solution:
\begin{equation}
	\varphi(t, x) =
	\begin{cases}
		F(x + t) - F(x - t + T) + \frac{t^2}{2} - \frac{T^2}{8} & \text{if }\quad -t < x < t \\
		F(x + t) - F(x - t) + \frac{t^2}{2}                     & \text{if }\quad t < x < T - t \\
		F(x + t - T) - F(x - t) + \frac{t^2}{2} - \frac{T^2}{8} & \text{if }\quad T - t < x < T + t, \\
	\end{cases}\label{cases}
\end{equation}
where $F(x)$ satisfies:
\begin{align}
	F(T) &= F(0) - \frac{T^2}{8}, \\
	\dv{F(x)}{x} &= \frac{\oldchi(x)}{2}.\label{cond2}
\end{align}

For less general oscillons, we can demand that the solutions approach zero smoothly at the left border $\gamma_L$ {\it i.e.}  the partial solution \eqref{cases} for $-t < x < t$ gives 
\begin{equation}
\partial_x\varphi(t,x)|_{\gamma_L}=0,\label{condleft}
\end{equation}
  where points belonging to the curve $\gamma_L$ have coordinates related by $x=x(t)$. Expressing condition \eqref{condleft}  in light-cone coordinates
\begin{align}
	y_\pm & = x \pm t
\end{align}
we get
\begin{equation}
\left(1+\frac{dy_+}{dy_-}\right)\left(1-2\frac{dy_+}{dy_-}\right)\partial_+\varphi(y_+,y_-)\Big|_{\gamma_L}=0,\label{condleft2}
\end{equation}
where we have used the fact that $(y_+,y_-)$ are related and $\frac{dy_+}{dy_-}=-\frac{1}{2}\frac{dx}{dt}$. Since $\gamma_L$ is a time-like curve  $|\frac{dy_+}{dy_-}|<\frac{1}{2}$ and so condition $\eqref{condleft2}$ is equivalent to  $\partial_+\varphi(y_+,y_-)|_{\gamma_L}=0$. This leads to $F'(y_+)=-\frac{1}{4}(y_+-y_-)$ for the first expression in \eqref{cases}. Taking into account expression \eqref{cond2} we finally get
\begin{equation}
	\oldchi(y_+) = - \frac{1}{2} \qty(y_+ - y_-),
	\label{eq:dv-left}
\end{equation}
where $y_-=g(y_+)$ is a function of $y_+$ representing the left border of the oscillon (worldline $\gamma_L$).
Similarly, demanding that $\partial_x\varphi(t,x)|_{\gamma_R}=0$ at the right border of the oscillon $\gamma_R$ we get a condition which, when written in terms of $y_-$, takes the form $\partial_-\varphi(y_+,y_-)|_{\gamma_R}=0$. Then, from the last expression in \eqref{cases} we get
\begin{equation}
	\oldchi(y_-) = - \frac{1}{2} \qty(y_+ - y_-).
	\label{eq:dv-right}
\end{equation}
Here $y_+=h(y_-)$ is a function of $y_-$ representing the right border of the oscillon (worldline $\gamma_R$).
Note that for points on $\gamma_L$ we have $0 \leq y_+ \leq T/2 + \Delta$ and for points on $\gamma_R$ we have $T/2 + \Delta \leq y_- \leq T$. Thus, \eqref{eq:dv-left} determines $\oldchi(x)$ for $0 \leq x \leq T/2 + \Delta$ and \eqref{eq:dv-right} for $T/2 + \Delta \leq x \leq T$.
\begin{figure}[h!]
\centering
\includegraphics[width=0.6\textwidth,height=0.3\textwidth, angle =0]{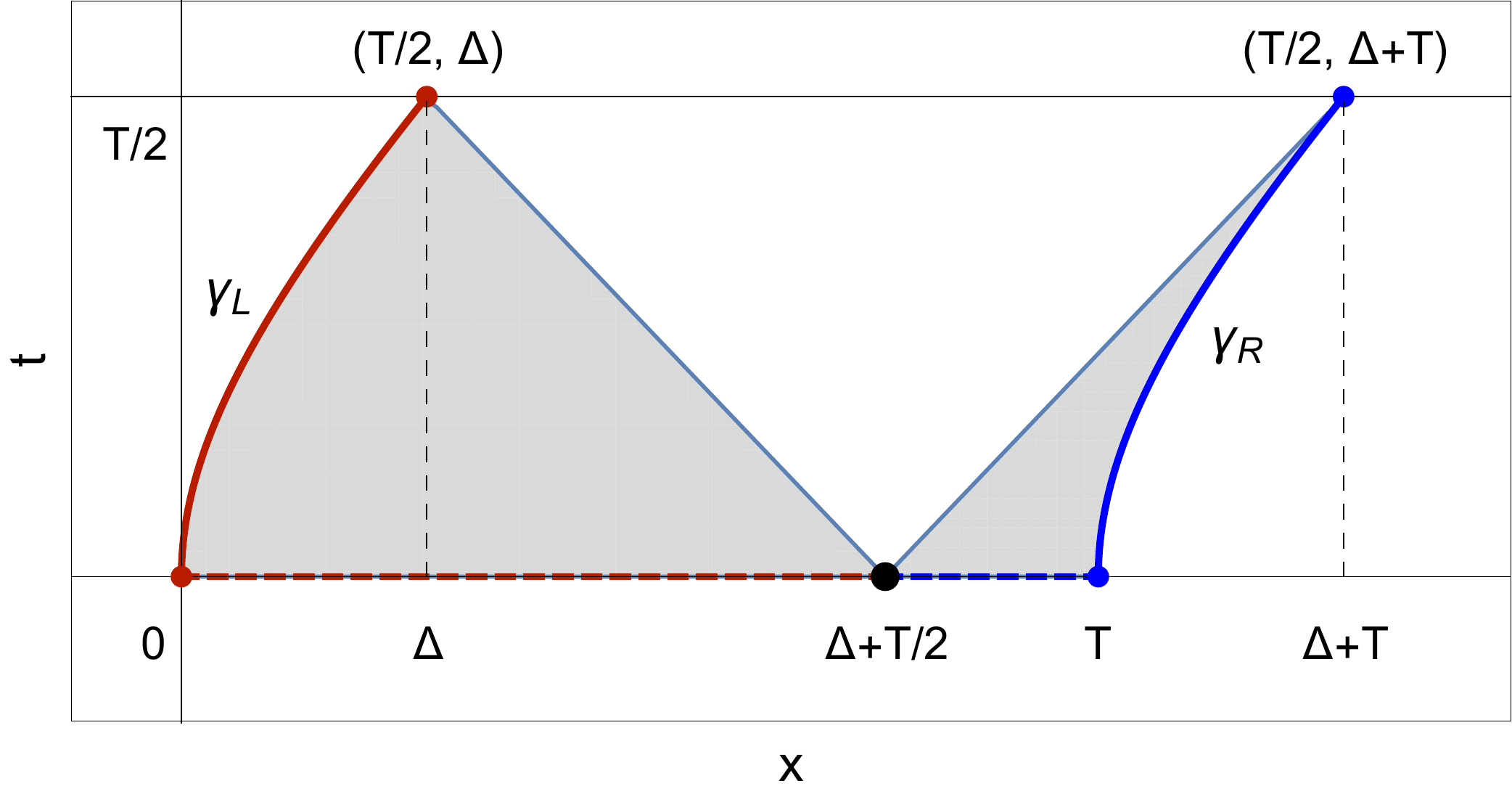}
\caption{Borders $\gamma_L$ and $\gamma_R$ of the oscillon and the supports $[0,\Delta+T/2]$ and $[\Delta+T/2, T]$ of $f(x)$. }
\label{fig:diag}
\end{figure}

Using this formalism, all oscillon solutions limited by a pair of identical time-like curves can be constructed by plugging the {\it a priori} given trajectories of the border in \eqref{eq:dv-left} and \eqref{eq:dv-right}. The  trajectories should have a form that one can to describe them explicitly as  $y_-=g(y_+)$ for $\gamma_L$  and $y_+=h(y_-)$ for $\gamma_R$. The only remaining problem is to integrate the resulting expressions to get $ F(x) $ -- and consequently $ \varphi(t, x) $.

\subsubsection{Example}
The first interesting example we have found is an oscillon with borders having a constant acceleration $a$ in the instantaneous rest frame of the border. In what follows we will use units in which $c=1$. In the reference frame of the oscillon, in which the border has acceleration $\gamma^{-3}a$, the trajectory describing the motion of such borders takes the form
\begin{equation}
	x(t) = x_0 + \frac{1}{a} \qty[ \sqrt{1 + \qty(at + \gamma_0 v_0)^2} - \gamma_0],\label{worldlinex}
\end{equation}
where $v_0$ is the velocity of the border at $t=0$, $\gamma_0 = (1 - v_0^2)^{-1/2}$ and $x_0 = 0$ ($x_0 = T$) for $\gamma_L$ ($\gamma_R$). Note that if the oscillon has no extra motion then its rest frame is just the laboratory reference frame. Using the light-cone coordinates $y_{\pm}=x\pm t$ we  put the expression \eqref{worldlinex} into the form $y_-=g(y_+)$, where $x_0=0$, and $y_+=h(y_-)$ and where $x_0=T$.
Then plugging these expressions into \eqref{eq:dv-left} and \eqref{eq:dv-right} we get our explicit expression for $\oldchi(x)$:
\begin{equation}
	\oldchi(x) =
	\begin{cases}
		- \frac{x}{2} \qty(1 + \frac{B}{x + A}) & \text{if } 0 \leq x \leq \frac{T}{2} + \Delta \\
		\frac{x - T}{2} \qty(1 + \frac{A}{x - T + B}) & \text{if } \frac{T}{2} + \Delta \leq x \leq T,
	\end{cases}
\end{equation}
where
\begin{align}
	A &= \frac{\gamma_0 (1 + v_0)}{a} \label{eq:A},\\
	B &= \frac{\gamma_0 (1 - v_0)}{a} \label{eq:B}.
\end{align}

Once we know $\oldchi(x)$, it is possible to integrate it and get partial solutions $\varphi_k (t, x)$ with $k \in \qty{C, L_1, L_2, L_3, R_1, R_2, R_3}$, each one valid in a specific subset of the region between $ \gamma_L $ and $\gamma_R$. Such solutions are given by:
\begin{align}
	\begin{split}
		\varphi_C(t, x; A, B) =& - \frac{AB}{4} + \frac{1}{4} (x + t + B - T) (x - t + A) + t \qty(t - \frac{T}{2}) \\&- \frac{AB}{4} \ln \abs{\frac{1}{AB} (x + t + B - T) (x - t + A)},
	\end{split}\\
	\begin{split}
		\varphi_{L_1}(t, x; A, B) =& \frac{t}{2} (t - x - B) + \frac{AB}{4} \ln\abs{\frac{x + t + A}{x - t + A}},
	\end{split}\\
	\begin{split}
		\varphi_{L_2}(t, x; A, B) =& \frac{AB}{4} - \frac{1}{4} (x + t + A)(x - t + B) \\&+ \frac{AB}{4} \ln\abs{\frac{1}{AB} (x + t + A)(x - t + B)},
	\end{split}\\
	\begin{split}
		\varphi_{L_3}(t, x; A, B) =& \frac{1}{2} \qty(t - \frac{T}{2})(x + t + A) - \frac{AB}{4}\ln\abs{\frac{x + t + B - T}{x - t + B}}
	\end{split}
\end{align}
where $AB=a^{-2}$.
Similarly to the oscillons previously presented and discussed we can relate the solutions $\varphi_{R_i}(t,x)$ to the solutions $\varphi_{L_i}(t,x)$ through the transformations:
\begin{align}
	x   &\rightarrow T - x, \\
	v_0 &\rightarrow -v_0, \\
	a   &\rightarrow -a.
\end{align}
Note that the last two transformations are equivalent to the transformations:
\begin{align}
	A &\rightarrow -B, \\
	B &\rightarrow -A.
\end{align}
Thus we have
\begin{equation}
	\varphi_{R_i}(t,x; A, B) = \varphi_{L_i} (t, T - x; -B, -A).
\end{equation}
\begin{figure}[h!]
\centering
\subfigure[$\quad V=0,\,\, a=1$]{\includegraphics[width=0.45\textwidth,height=0.25\textwidth, angle =0]{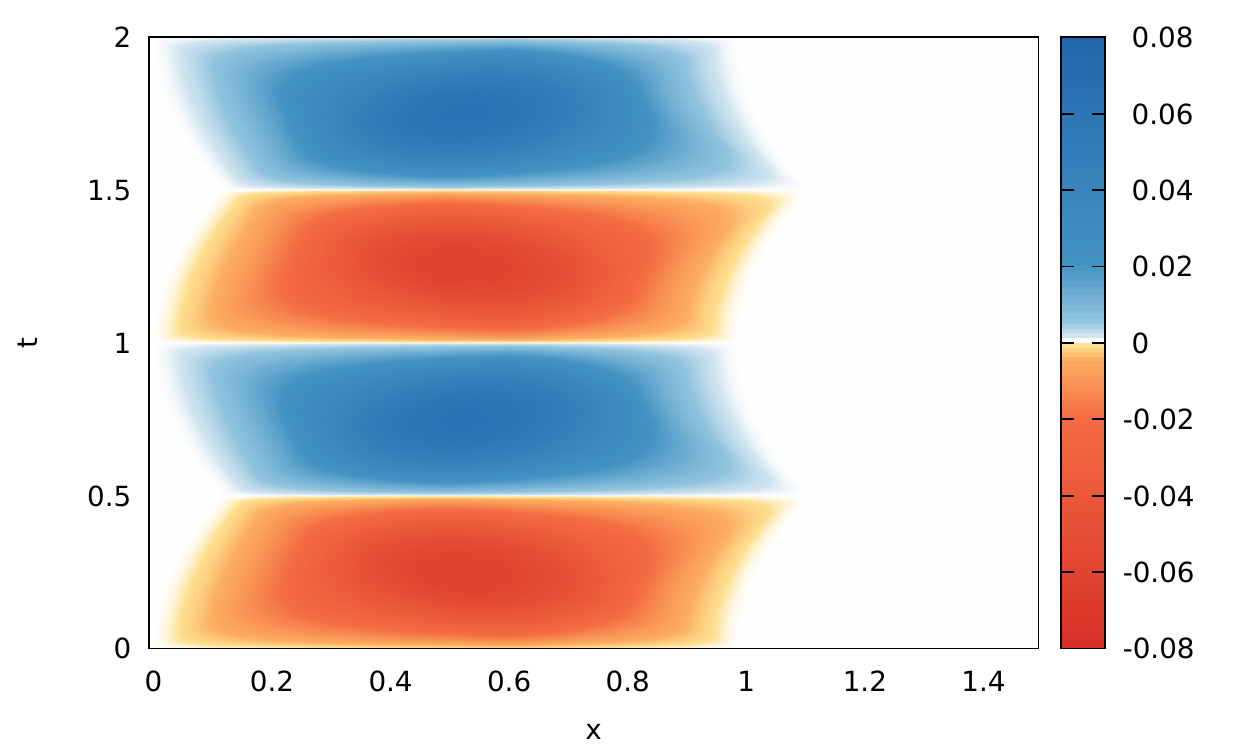}}
\subfigure[$\quad V=0,\,\, a=5$]{\includegraphics[width=0.45\textwidth,height=0.25\textwidth, angle =0]{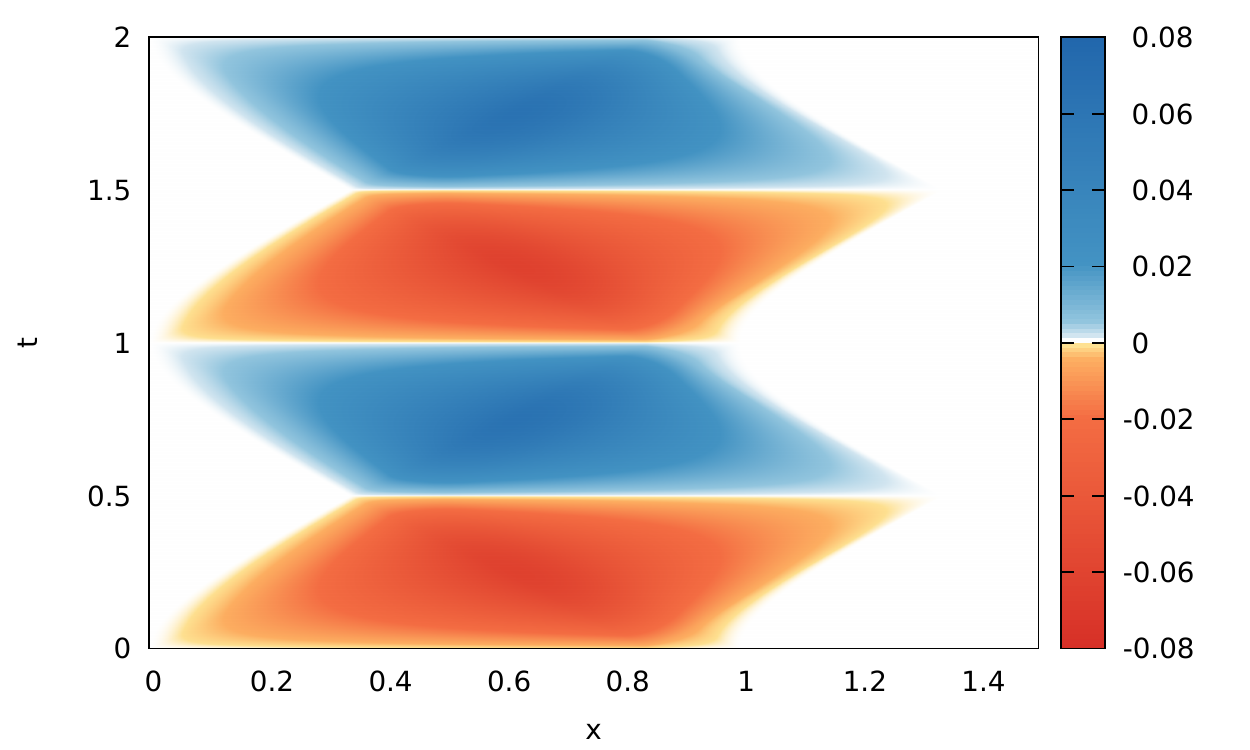}}
\subfigure[$\quad V=0.6,\,\, a=1$]{\includegraphics[width=0.45\textwidth,height=0.25\textwidth, angle =0]{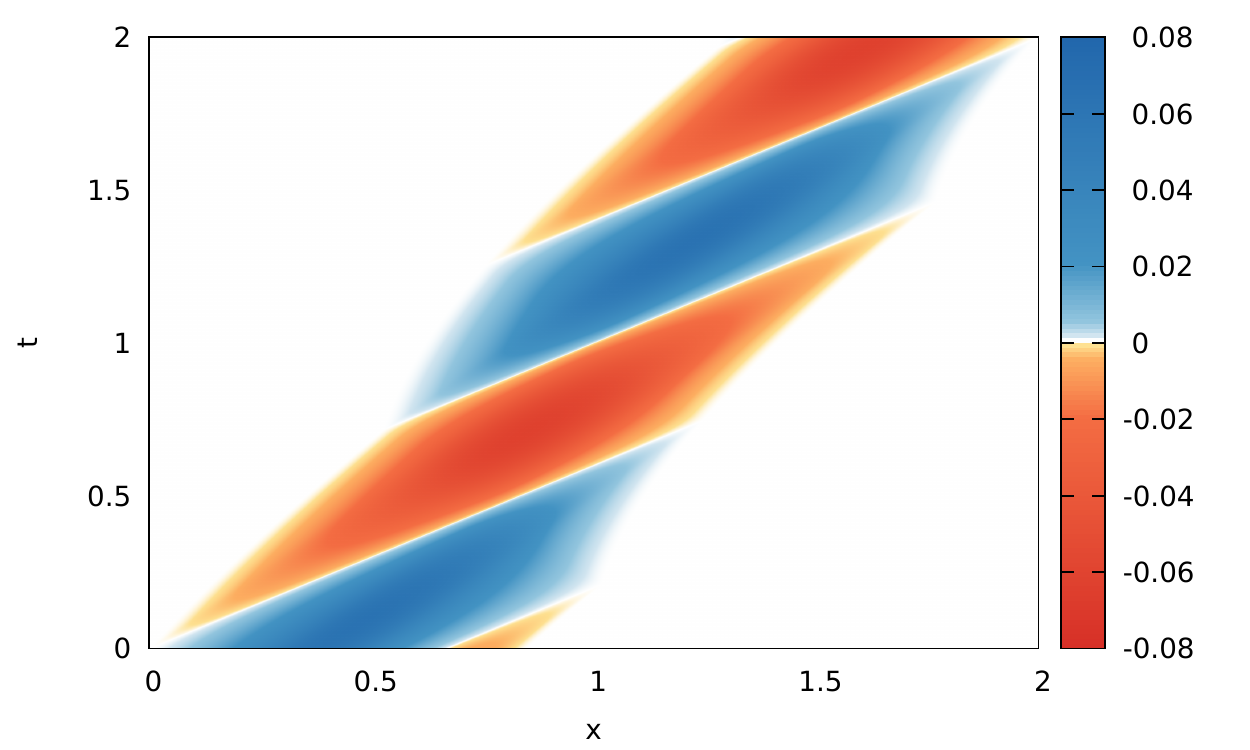}}
\subfigure[$\quad V=0.6,\,\, a=5$]{\includegraphics[width=0.45\textwidth,height=0.25\textwidth, angle =0]{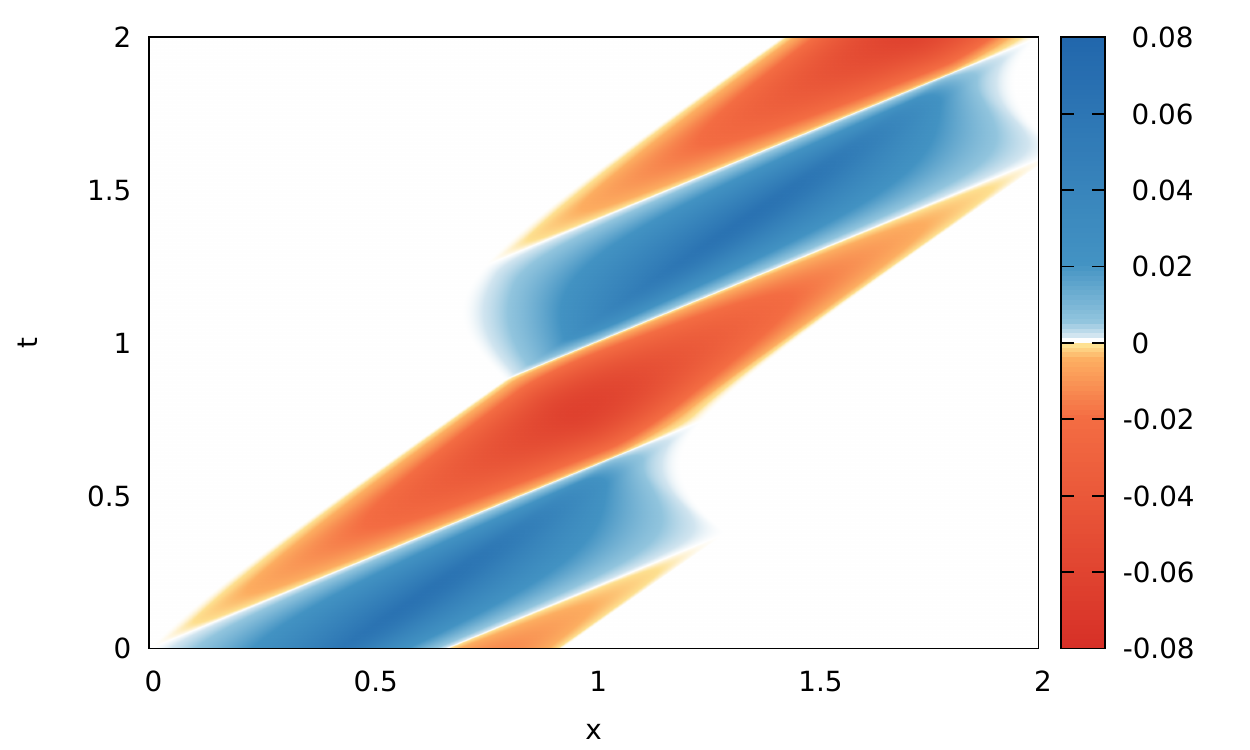}}
\subfigure[$\quad V=0.8,\,\, a=1$]{\includegraphics[width=0.45\textwidth,height=0.25\textwidth, angle =0]{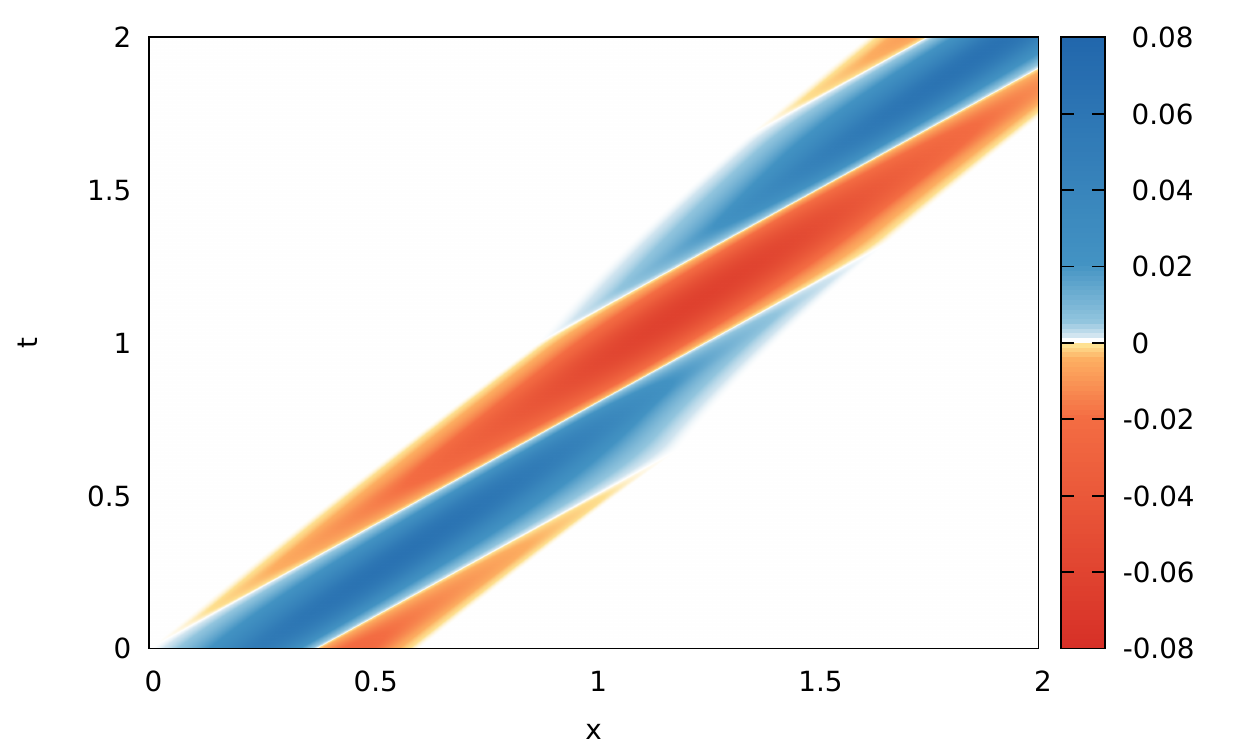}}
\subfigure[$\quad V=0.8,\,\, a=5$]{\includegraphics[width=0.45\textwidth,height=0.25\textwidth, angle =0]{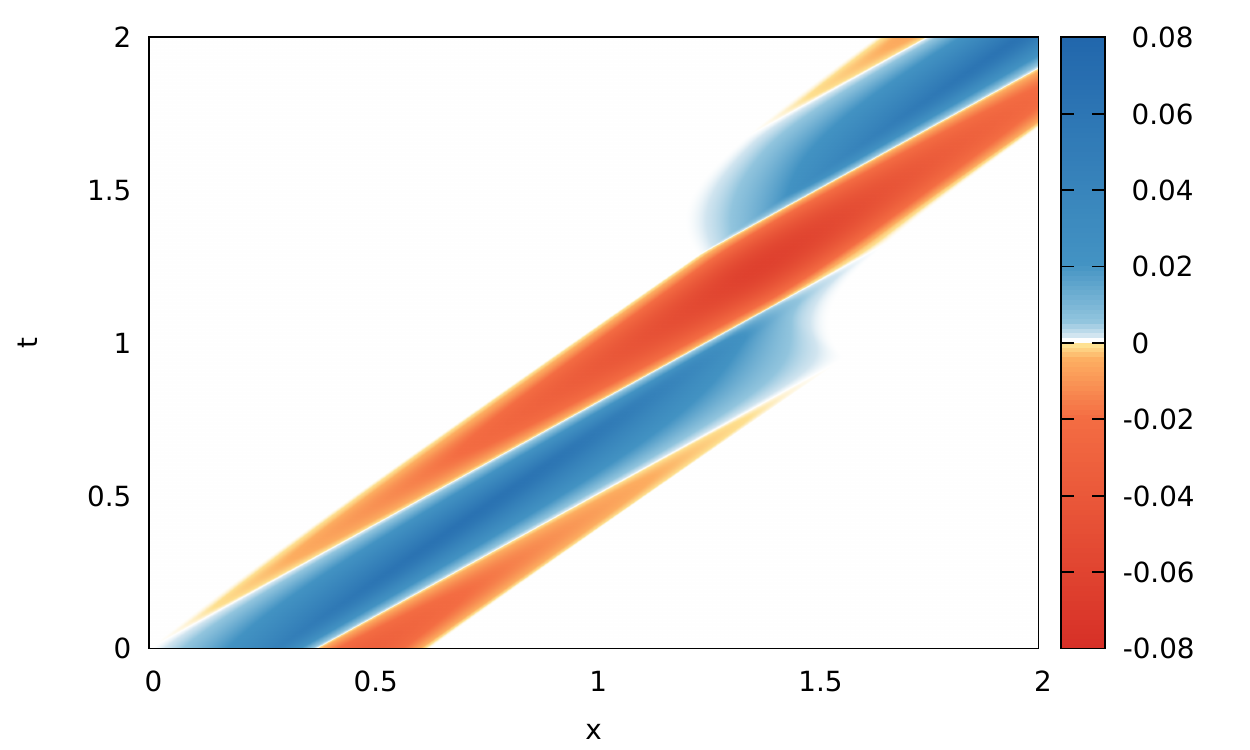}}
\caption{Generalized exact oscillon with uniformly accelerated border for different values of the boost velocity $V$ and acceleration $a$. }
\label{fig:gen2}
\end{figure}

Once again we can write the complete solution with the help of the step functions. We have
\begin{align}
	\Pi_C(t, x; a, v_0) &= \theta\qty(x + t - \frac{T}{2} - \Delta(a,v_0)) \theta\qty(x - t)\times\nonumber\\&\times\theta\qty(-x + t + \frac{T}{2} + \Delta(a,v_0)) \theta\qty(-x - t + T), \\
	\Pi_{L_1}(t, x; a, v_0) &= \theta\qty(x - t) \theta\qty(-x - t + \frac{T}{2} + \Delta(a,v_0)),
\end{align}
\begin{align}
	\Pi_{L_2}(t, x; a, v_0) &= \theta\qty(x - \frac{1}{a}\qty(\sqrt{1 + \qty(at + \gamma_0 v_0)^2} - \gamma_0)) \times\nonumber\\&\times\theta\qty(-x + t) \theta\qty(-x - t + \frac{T}{2} + \Delta(a,v_0)),
\end{align}
\begin{align}
	\Pi_{L_3}(t, x; a, v_0) &= \theta\qty(-x + t) \theta\qty(x + t - \frac{T}{2} - \Delta(a,v_0)), \\
	\Pi_{R_i}(t, x; a, v_0) &= \Pi_{L_i}(t, T - x; - a, - v_0),
\end{align}
where $ \Delta = x(T/2) - x_0 $ and so is given by 
\begin{equation}
	\Delta(a, v_0) = \frac{1}{a} \qty[ \sqrt{1 + \qty(a\frac{T}{2} + \gamma_0 v_0)^2} - \gamma_0].
\end{equation}

The periodicity of the solution can now be taken into account by involving the generalized forms of the functions $\tau(z)$ and $\sigma(z)$ for an arbitrary period $T$:
\begin{align}
	\tau(z)   &= \frac{T}{\pi} \arcsin\abs{\sin\qty(\frac{\pi z}{T})} \\
	\sigma(z) &= \sgn\qty(\sin\qty(\frac{2 \pi z}{T})).
\end{align}

Thus, the complete solution has the form:
\begin{equation}
	\varphi(t,x; a, v_0) = \sum_k \sigma(t) \Pi_k(\tau(t), x; a, v_0) \varphi_k\qty(\tau(t),x; A(a, v_0), B(a, v_0))
\end{equation}
where $ A(a, v_0) $ and $ B(a, v_0) $ are given by \eqref{eq:A} and \eqref{eq:B}.

 Comparing the generalized exact oscilons presented in Fig.\ref{fig:gen2} with numerical solutions presented in Fig.\ref{fig:scatvel} and Fig.\ref{fig:scatvelanti} we conclude that there is certain similarity between outgoing oscillon-like object produced in the scattering process and the exact oscillons with non-uniformly moving endpoints. This suggests that the process of the scattering of two exact oscillons  can lead to the production of field configurations corresponding to perturbed  generalized oscillons with non-uniformly moving endpoints. In some cases these perturbations can be almost zero and when this happens one can call the outgoing oscillon-like field configurations quasi-oscillons with non uniformly moving endpoints. The transformation of oscillons from one class into the oscillons belonging to another class during the scattering process  is an open problem which requires further investigations.

\subsection{Fractal nature of the radiation}
\label{sec:fractal}

In this section we discuss the interesting possibility that the radiation generated in the process of the scattering of oscillons has a fractal-like nature. There are two facts that suggest this. Firstly, many numerical  simulations show  that the radiation of the signum-Gordon model is dominated by oscillating structures that look like travelling oscillons. 

Such oscillon-like structures are generated as radiation during the scattering of oscillon-like objects.  Alternatively,  they are emitted from strongly perturbed oscillons. In fact, a production of small-size oscillons during the evolution of perturbed oscillons has been conjectured in Ref. \cite{oscillon}. Our work presented in this paper and also in \cite{kswz} suggests that this conjecture is true. Secondly, the oscillon-like field configurations exist at arbitrarily small scales. Although the numerical approach does not allow for the arbitrary good resolution we know that the existence of exact oscillons with any size is guaranteed by the  dilation symmetry of the signum-Gordon equation \eqref{EL}. This symmetry implies that for any real number $\lambda>0$ and any solution of the signum-Gordon equation $\phi_{(1)}(t,x)$
the function
\be
\phi_{(\lambda)}(t,x):=\lambda^2\phi_{(1)}\Big(\frac{t}{\lambda},\frac{x}{\lambda}\Big)\label{scalling1}
\ee
is also a solution of \eqref{EL}. Looking at energy of solutions we see that it scales according to
\be
E[\phi_{(\lambda)}]=\lambda^3E[\phi_{(1)}]\label{scalling2}
\ee
where
\be
E[\phi_{(1)}]
:=\int_{-\infty}^{\infty} d x\left[\frac{1}{2}(\partial_t\phi_{(1)})^2+\frac{1}{2}(\partial_x\phi_{(1)})^2+\left|\phi_{(1)}\right|\right].
\ee
Taking for $\phi_{(1)}(t,x)$ the generalized exact oscillon on a segment $x\in[0,1]$ with $E[\phi_{(1)}]=\frac{1}{24}$ and applying \eqref{scalling1} we obtain exact oscillons with arbitrarily small sizes.

Looking, for example, at Fig. \ref{fig:v02} (a) and (c) we see that the number of collisions between oscillon-like objects that form radiation  grows significantly with time. Each such a collision process is a source of  new smaller oscillon-like objects. Certainly, we do not expect that oscillon-like objects seen in our numerical simulations are exact oscillons.  On the other hand, many of them are surprisingly regular and sufficiently stable. They all possess characteristics necessary to be called quasi-oscillons. 
Less regular oscillon-like structures ``decay'' emitting smaller and more regular oscillating objects. The emission of smaller oscillons is a physical mechanism allowing  strongly perturbed oscillons to get rid of a surplus of their energy. Summarizing, we can say that interaction between individual oscillon-like objects (constituents of radiation) produces more and more  such objects during their evolution.

\begin{figure}[h!]
\centering
\subfigure[]{\includegraphics[width=0.48\textwidth,height=0.25\textwidth, angle =0]{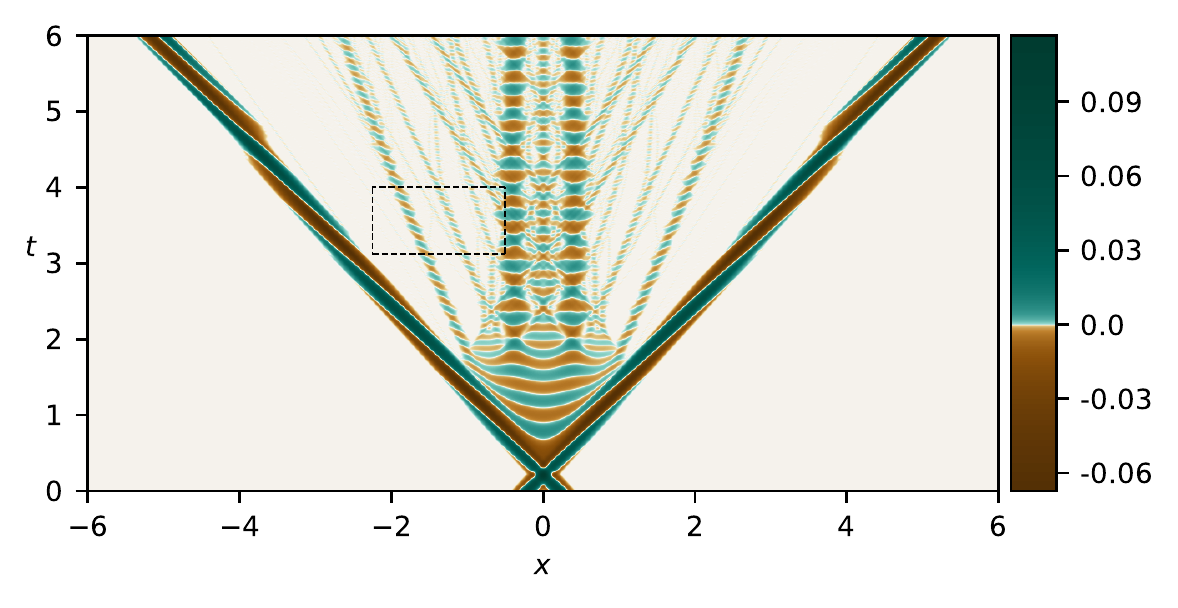}}\hskip0.5cm
\subfigure[]{\includegraphics[width=0.48\textwidth,height=0.25\textwidth, angle =0]{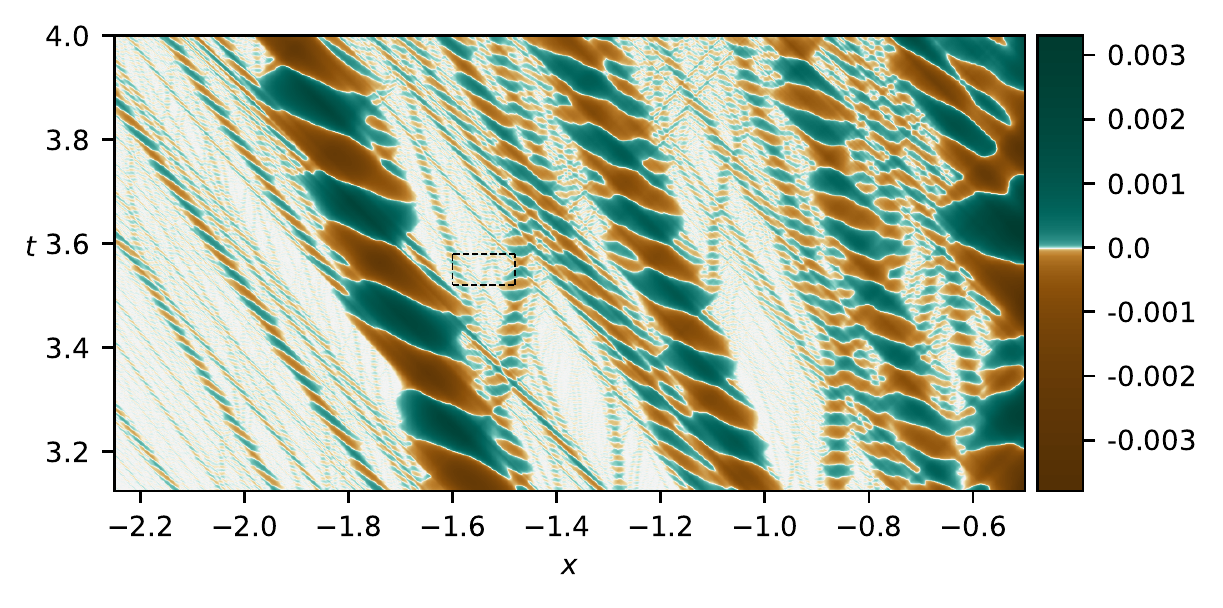}}
\subfigure[]{\includegraphics[width=0.48\textwidth,height=0.25\textwidth, angle =0]{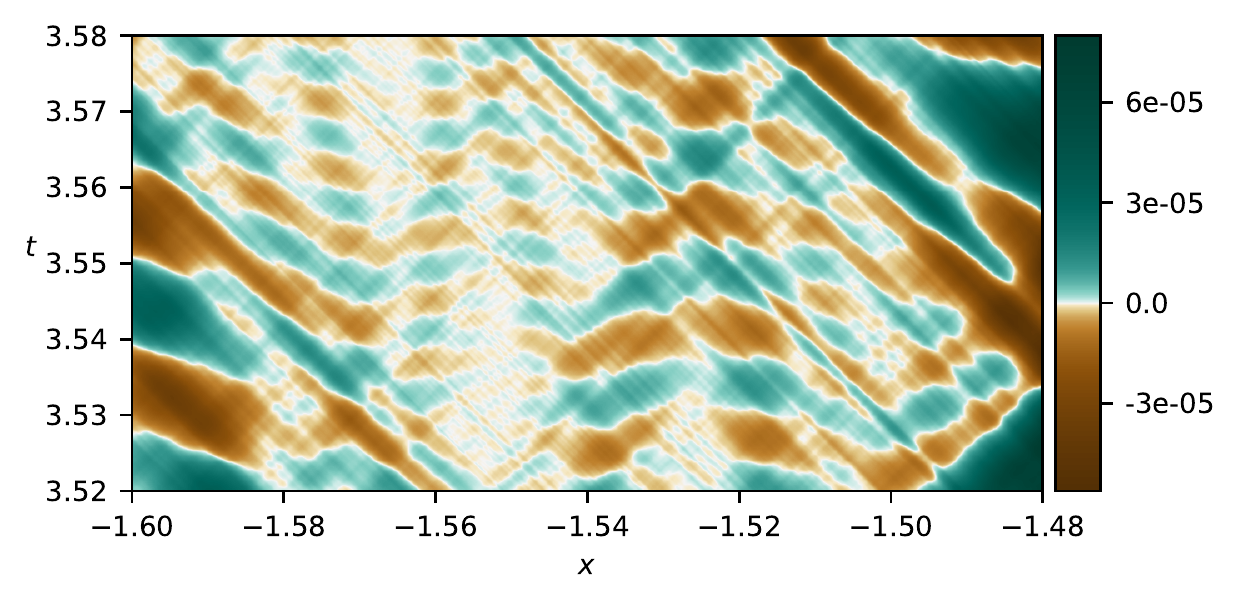}}
\caption{Scattering process of two oscillons: (b) blow-up of the rectangular region in (a); (c) blow-up of the rectangular region in (b).}
\label{fig:fractal}
\end{figure}

Finally, we note that the relation \eqref{scalling1} is quite general and it allows us to take as a solution $\phi_{(1)}(t,x)$ not only a single oscillon but the whole diagram.  Certainly, there is no qualitative difference between scattering processes involving two oscillons with $\lambda=1$ and a scattering  of smaller oscillons with $\lambda\ll1$. In principle, the whole diagram (like Fig.\ref{fig:v02} (a) and (c))  could repeat itself at any length scales. Such repetitions of structures involving oscillons at all length scales in the spacetime diagram suggest a possible fractal-like nature of the radiation. This statement still has a status of a {\it conjecture} and it certainly deserves further investigation. Below we present only some preliminary results  of a numerical study which reinforces this idea.

In order to check our hypothesis we have perform high resolution simulations of the scattering processes and then looked at the spacetime diagrams representing the result. In Fig.\ref{fig:fractal}(a) we plot the two main outcoming oscillons and the radiation in the central region between these oscillons. Looking in more detail at the region inside the rectangle in Fig.\ref{fig:fractal}(a) which we replot in  Fig.\ref{fig:fractal}(b) we see that there exist a huge number of smaller oscillons invisible in the previous picture. Choosing another rectangular region  of Fig.\ref{fig:fractal}(b) which we replot in Fig.\ref{fig:fractal}(c) we see that, again,  it contains many oscillating structures. This result supports our idea of the  fractal-like nature of the radiation of the signum-Gordon model.

\section{Conclusions}

In this paper we have reported our results on the scattering of compact oscillons in the signum-Gordon model in one spatial dimension. We have looked at two qualitatively distinct initial configurations -- symmetric and anti-symmetric one. In both cases the initial configurations consisted of exact compact solutions. Due to the compactness of oscillons there was no problem with their overlapping at $t=0$. In fact we also evolved oscillons whose supports touched each other but did not overlap at $t=0$. A time dependence of the shape of oscillons was responsible for the existence of an additional scattering parameter which we called the phase of the oscillon. This phase was a important quantity and the properties of the scattering process depended very strongly on it.

Looking at the results of the scattering of oscillons we have found that there was a significant qualitative difference between symmetric and anti-symmetric initial configurations. The emission of radiation for anti-symmetric configurations was restricted to situations where outgoing oscillons had irregular borders. Such irregular borders act as sources of radiation which has the form of showers of smaller oscillons sent out from the borders. The central region of the Minkowski diagrams just after emergence of outgoing oscillons was free of radiation. On the other hand, symmetric configurations produced much more radiation than the anti-symmetric ones. In this case the radiation was emitted mainly in the central part of the Minkowski diagram where structures similar to exact shock wave solutions of the signum-Gordon model were formed. These waves were not stable and, eventually, they decaysed into cascades of oscillons. We have spotted that there were special values of the phases of colliding oscillons for which there was almost no radiation. We suspect that this fact was associated with the absence of the shockwave-like structures between outgoing oscillons. The relation between the collapse (decay) of the shockwave-like solution and the appearance of a cascade of oscillons has been found to be a very interesting subject and it requires more thorough analysis than we could carry out in the present paper. We hope to report more on this subject in near future.

Comparing incoming oscillons with outgoing ones we have spotted that, in general, the later ones belong to a wider class of oscillons. This class is characterized by a non-uniform motion of the border of the oscillon in its own rest frame. In our numerical study  many of the outgoing oscillons had borders described by a segment of the worldline curve  whereas for incoming oscillons these borders were segments of straight lines. Thus the collision transformed the compact oscillons of a very special class into more general compact oscillons.

We have also looked at the radiation of the signum-Gordon model and have found that it possessed what looked like a self-similar structure. Since the model has the scalling symmetry one can show that an exact compact oscillon can have arbitrarily small support and energy. Our numerical studies have shown that small quasi-oscillons were emitted from perturbed oscillons or appeared in the scattering processes of two oscillons-like structures. Since, in general, they were also perturbed the process of emission repeated itself (in principle infinitely many times). This mechanism of emission of oscillons from perturbed objects and the fact that oscillons existed at arbitrarily small scales suggests to us the emergence of dynamical fractals {\it i.e.} of the fractal structures in the spacetime diagrams.

\subsection*{Final remarks}

\begin{enumerate}
\item
Our investigations of the scattering processes have been based primarily on the numerical integration of the signum-Gordon equation. The complexity of this process excludes any analytical approach to this problem. We have made many attempts to calculate analytically the evolution of the initial profile containing two exact oscillons. Unfortunately, even before the emergence of main oscillons we have encountered technical difficulties in the construction of partial solutions. Moreover, the number and localization of any partial solution depends on the initial data. In contradiction to the standard analytical non-linear models the perturbative approach cannot be used in the case of signum-Gordon model because of the non analytical character of the potential $V=|\phi|$ at $\phi=0$. Hence, the small perturbations of the vacuum solutions are always nonlinear.
\item
One can get some analytical results considering the decay of shock waves in a cascade of oscillons. This subject has been recently studied and the results have been reported in \cite{decay}. 
\item
Of course, we can also think of the comparisons of our results with those obtained in other models, such as the Sine-Gordon model of $\lambda\phi^4$ model. However, these models are basically very different as they do not possess compact solitons. Of course, their solitons are exponentially localised and some studies of such solitons have also been performed in much detail. The most comparable studies involved looking at the properties of Sine-Gordon kinks on scattering
on various obstructions (potential wells or barriers) and the effects of the obstructions on the properties of the basic solitonic structures. The obstructions generated the emission of kink-antikink pairs, either in the form of breathers or as invidual pairs. And for perturbed models, which still had solitonic solutions, one had emission of long lived breathers (basically oscillons) or annihilations.
An interested reader can look at papers \cite{extra} and references therein.

\end{enumerate}

\begin{appendices}
\section{Comments on the numerics}
The numeric results presented in this paper have been generated by the use of the standard $4^{\rm th}$-order Runge-Kutta method, integrating the system via the discrete timesteps $\Delta t$. The second-order equation in time has been decomposed into a coupled system of two first-order equations
\begin{align}
	\phi(x, t_i) &=\phi(x, t_{i-1}) + \Delta t\,\psi(x, t_{i-1}) \\
	\psi(x, t_i) &=\psi(x, t_{i-1}) + \Delta t\,\left( \dfrac{d^2\phi(x,t_{i-1})}{dx^2} - {\rm sign}\big(\phi(x,t_{i-1})\big) \right),
\end{align}
where $t_i=n\,\Delta t$ ($n=1,2,\cdots$) and $\psi(x,t) = \frac{\partial\phi}{\partial t}$. The spatial dimension was made discrete over $N$ sites of width $\Delta x$, so it had width $L=N\,\Delta x$. 

In most of our simulations we have used a spatial resolution of $N=2^{15}$ (for $L=6$ this corresponds to $\Delta x\simeq1.8\times10^{-4}$). There were two exceptions to this. The first one corresponded to the case of the fractal, which in order to generate and capture the small scale details we had to us $N=2^{20}$ (in this case, the simulation space length was $L=12$, leading to $\Delta x\simeq1.14\times10^{-5}$). The second exception corresponded to the generation of the diagrams of describing the percentage of energy lost to radiation (\textit{e.g.} Figs. \ref{fig:symphaseshift}, \ref{fig:symphaseshift2}) and the balance of energy after the interaction (Figs. \ref{fig:rad-balance}(a) and \ref{fig:rad-balance}(b)). Since the value of each pixel of these images is computed based on one entire simulation, in order to speed up the computations (in the case of Fig. \ref{fig:a-sym-phase-map}, we have used $350^2=122500$ simulations) we had performed lower resolution simulations, with $N=2^{12}$.

The timestep value, in all our simulations, was given by the relation $a = \frac{\Delta t}{\Delta x}$ and we have used in all our simulations $a=0.1$. 
 
Finally,  we would like to point out to the reader the need of some caution while dealing numerically with the very small scale structures appearing within the fractals. This is further explained in Sec. \ref{sec:num-caveats}.

\section{Caveats} \label{sec:num-caveats}
We would like to add a few comments on the difficulties associated with the numerical integration of the signum-Gordon equation. The main difficulty has origin in the fact that the radiation contains perturbed oscillons of arbitrarily small sizes. Certainly, oscillons smaller than the size of numerical domain cannot be seen in the simulation. An obvious solution involves increasing the number of points in the grid. We have run many simulations changing the number of points and comparing the results. Such tests have shown that very small oscillons are very sensitive to the number of points. In some cases, changing the number of points by a factor of two resulted in the appearance or even disappearance of some tiny structures whereas bigger structures remained stable under this procedure. A similar problem was spotted in simulations of a special type of self-similar solutions with an infinite number of zeros on a finite segment. In that case the numerical solution and the analytical one diverged after a very short time (the numerical solution was very unstable), and the increase of the simulation resolution resulted in only very small in stability. On the other hand, our numerical simulations of the exact oscillons did not lead to visible instability within intervals of time corresponding to many oscillon periods. Also, the simulations of exact shock waves were very consistent with the analytical solutions. Thus, in the regions dominated by radiation (or the special self-similar solutions) the solutions of the model were found to be  sensitive to the initial conditions. In this sense, the signum-Gordon model shares some properties with chaotic systems. This property is, for instance, one of the main difficulties in generating of high resolution fractals.
However, although some results were quite sensitive to the details of the details of the numerical procedures, most of them were not and we strongly believe in their validity.

\end{appendices}

\section*{Acknowledgements}
The authors are grateful to H. Arod\'z, A. Wereszczy\'nski and Z. \'Swierczy\'nski for discussion and comments. PK wants to thank Faculty of Physics, Astronomy and Applied of Computer Science of the Jagiellonian University for financial support and hospitality during the symposium {\it The nonperturbative world} in honor of Prof. Henryk Arod\'z on the occasion of his 70th birthday  where some preliminary result of this paper were reported. FMH is supported by  CNPq Scholarship. This study was financed in part by the Coordena\c c\~ao de Aperfei\c coamento de Pessoal de N\'i­vel Superior a Brasil (CAPES) â Finance Code 001. WJZ thanks the Leverhulme Trust for the award of Emeritus Fellowship EM-2016-007.

\end{document}